\documentstyle{mn}
\input {psfig.sty}
\def \Mpc {~h^{-1}~{\rm Mpc} }

\def \Omb {\Omega_{\rm b}}
\def \Omcdm {\Omega_{\rm CDM}}
\def \Omlam {\Omega_{\Lambda}}
\def \Omm {\Omega_{\rm m}}
\def \ho {H_0}

\def \kms {{\rm ~km~s}^{-1}}
\def \kmsmpc {{\rm ~km~s}^{-1}~{\rm Mpc}^{-1}}

\def \bj {b_{\rm J}}
\def \mb {M_{\rm b_{\rm J}}}
\def \qso {_{\rm Q}}
\def \gal {_{\rm gal}}
\def \xibar {\bar{\xi}}
\def \max {_{\rm max}}
\def \gsim { \lower .75ex \hbox{$\sim$} \llap{\raise .27ex \hbox{$>$}} }
\def \lsim { \lower .75ex \hbox{$\sim$} \llap{\raise .27ex \hbox{$<$}} }
\def \deg {^{\circ}}

\def \mbh  {M_{\rm BH}}
\def \mdh  {M_{\rm DH}}
\def \msun {M_{\odot}}
\def \z {_{\rm z}}
\def \edd {_{\rm Edd}}
\def \lin {_{\rm lin}}
\def \nonlin {_{\rm non-lin}}
\def \wrms {\langle w_{\rm z}^2\rangle^{1/2}}
\def \dc {\delta_{\rm c}}

\title[The 2QZ two-point correlation function]
      {The 2dF QSO Redshift Survey - XIV.  Structure and evolution
      from the two-point correlation function.}
\author[S.M. Croom et al.]
       {Scott M. Croom$^{1}$\thanks{scroom@aaoepp.aao.gov.au},
	B.~J. Boyle$^2$, T. Shanks$^3$, R.~J. Smith$^4$, L. Miller$^5$,
\newauthor P.~J. Outram$^3$, N.~S. Loaring$^6$, F. Hoyle$^7$, J. da \^{A}ngela$^{3,8}$\\ 
$^1$Anglo-Australian Observatory, PO Box 296, Epping, NSW 1710,
      Australia.\\
$^2$Australia Telescope National Facility, PO Box 76, Epping, NSW
      1710, Australia.\\
$^3$Physics Department, University of Durham, South Road, Durham, DH1 3LE,
UK.\\
$^4$Astrophysics Research Institute, Liverpool John Moores University,
Twelve Quays House, Egerton Wharf, Birkenhead,  CH41 1lD, UK\\
$^5$Department of Physics, Oxford University, Keble Road, Oxford, OX1
      3RH, UK.\\
$^6$Mullard Space Science Laboratory, Holmbury St. Mary, Dorking,
Surrey, RH5 6NT, UK\\
$^7$Department of Physics, Drexel University, 3141 Chestnut Street,
      Philadelphia, PA 19104, U.S.A.\\
$^8$Centro de Astrof\'{i}sica da Universidade do Porto, R. das
      Estrelas s/n, 4150-762 Porto, Portugal.
}
\begin{document}

\maketitle

\begin{abstract}

In this paper we present a clustering analysis of QSOs using over
20000 objects from the final catalogue of the 2dF QSO Redshift Survey
(2QZ), measuring the redshift-space two-point
correlation function, $\xi(s)$.  When averaged over the redshift range
$0.3<z<2.2$ we find that $\xi(s)$ is flat on small scales, steepening
on scales above $\sim25\Mpc$.  In a WMAP/2dF cosmology ($\Omm=0.27$,
$\Omlam=0.73$) we find a best fit power law with
$s_0=5.48_{-0.48}^{+0.42}\Mpc$ and $\gamma=1.20\pm0.10$ on
scales $s=1$ to $25\Mpc$.  We demonstrate that non-linear
redshift-space distortions have a significant effect on the QSO
$\xi(s)$ at scales less than $\sim10\Mpc$.  A cold dark matter model assuming
WMAP/2dF cosmological parameters is a good description of the QSO
$\xi(s)$ after accounting for non-linear clustering and redshift-space
distortions, and allowing for a linear bias at the mean redshift of
$b\qso(z=1.35)=2.02\pm0.07$.  

We subdivide the 2QZ into 10 redshift intervals with effective
redshifts from $z=0.53$ to $2.48$.  We find a significant increase
in clustering amplitude at high redshift in the WMAP/2dF cosmology.
The QSO clustering amplitude increases with redshift such that the
integrated correlation function, $\xibar(s)$, within $20\Mpc$ is
$\xibar(20,z=0.53)=0.26\pm0.08$ and $\xibar(20,z=2.48)=0.70\pm0.17$.
We derive the QSO bias and find it to be a strong function of redshift with 
$b\qso(z=0.53)=1.13\pm0.18$ and $b\qso(z=2.48)=4.24\pm0.53$.  We use
these bias values to derive the mean dark matter halo (DMH) mass
occupied by the QSOs.  At all redshifts 2QZ QSOs inhabit approximately
the same mass DMHs with $\mdh=(3.0\pm1.6)\times10^{12}h^{-1}\msun$,
which is close to the  characteristic mass in the Press-Schechter mass
function, $M^*$, at $z=0$.  These  results imply that $L\qso^*$ QSOs
at $z\sim0$ should be largely unbiased.  If the relation between black
hole (BH) mass and $\mdh$ or host velocity dispersion does not evolve,
then we find that the accretion efficiency ($L/L\edd$) for $L\qso^*$
QSOs is approximately constant with redshift.  Thus the fading of the
QSO population from $z\sim2$ to $\sim0$ appears to be due to less
massive BHs being active at low redshift.  We apply different methods
to estimate, $t\qso$, the active lifetime of QSOs and constrain
$t\qso$ to be in the range $4\times10^6-6\times10^8$ years at $z\sim2$.

We test for any luminosity dependence of QSO clustering by measuring
$\xi(s)$ as a function of apparent magnitude (equivalent to luminosity
relative to $L\qso^*$).  However, we find no significant evidence of
luminosity dependent clustering from this data set.       
\end{abstract}


\section{Introduction}

The question of how activity is triggered in the nucleus of galaxies
is vital to answer if we wish to have a full understanding of the
galaxy formation process.  It appears that a large fraction of galaxies
may have contained an active galactic nuclei (AGN) at some point in
their history.  When local galaxies are surveyed (including our own
Milky Way) most show evidence of a super-massive black hole (BH)
(e.g. Kormendy \& Richstone 1995).  The BHs tend to be found in
dynamically hot systems (i.e. spheroids - elliptical galaxies
or bulges), and the mass of the BHs is well correlated with the mass
of the spheroid.  The tightest correlation is found between BH mass,
$\mbh$, and spheroid velocity dispersion, $\sigma^*$
\cite{geb00a,fer00}.  At higher redshift it is not clear that this
correlation holds, or indeed in general, how high redshift BHs relate
to their host galaxies. However Shields et al. (2003) do suggest that
the same $\mbh-\sigma$ seems to be appropriate at high redshift.
 
It is the powerful evolution in luminosity of the AGN population which
allows them to be readily observed to high redshift.  Understanding
this evolution goes hand-in-hand with our understanding of the
relation between AGN and galaxies.  Croom et al. (2004a) (which we
will henceforth call Paper XII) find that optically selected QSOs are
well described by so called 'pure luminosity evolution' (PLE) with an
exponential increase in the typical luminosity $L\qso^*$ (e-folding
time of $\sim2$ Gyr) up to $z\sim2$.  Work at higher redshift
(e.g. Fan et al. 2001) find that at $z\sim4-6$ the number density of
QSOs is much lower than at $z\sim2$.  The X-ray luminosity function
(LF) appears to give a more complex picture \cite{ueda03} but still
shows the general trend of luminous AGN being more active, peaking at
$z\sim2-3$.

The question is then, how do we gain further information about the
physical processes of QSO formation at high redshift?  One approach
is to attempt to directly image QSO host galaxies at high resolution
\cite{kuk01,croom04}.  These analyses seem to show
that high redshift QSO hosts (at least for radio quiet sources) are no
brighter than low redshift hosts, after accounting for only passive
evolution of the stellar populations in the galaxies.  QSO clustering
measurements gives us an important second angle to study the hosts of
QSOs, as the clustering amplitude can be considered as a surrogate for
host mass or dark matter halo (DMH) mass, $\mdh$.  With large samples
such as the 2dF QSO Redshift Survey (2QZ; Paper XII) it is possible
to determine these host properties over a wide range in redshift.
With an estimate
of the host mass of these high redshift QSOs we can hope to determine
whether the host mass vs. BH mass correlation at low redshift
continues to high redshift.  We can also attempt to predict the masses
of the descendents of high redshift QSOs, and locate them in the local
universe.  

A number of authors (e.g. Martini \& Weinberg 2001; Haiman
\& Hui 2001; Kauffmann \& Haehnelt 2002) have constructed models for
QSO evolution including clustering, and these need to be tested
against accurate measurements.  One parameter that can be derived from
these models is a mean QSO lifetime, although the exact interpretation
of this is rather model dependent.

As well as being used for the study of QSO formation/evolution, QSOs
are also powerful probes of large-scale structure in their own right.
The large volumes probed ($\sim6\times10^9$h$^{-3}$Mpc$^3$ for the 2QZ
in a universe with $\Omm=0.3$ and $\Omlam=0.7$) and high redshift
sampled makes observations quite  complementary with lower redshift
galaxy observations and higher redshift CMB observations.  A number of
authors have attempted to detect high redshift QSO clustering
\cite{o81,s84,sfbp87,is88,ac92,mf93,sb94,cs96,lac98} and made some
preliminary measurements of clustering evolution, but these have all
been based on small samples of QSOs (typically a few hundred objects).
At low redshift, there have also been a number of recent analysis.
Grazian et al. (2004) find $s_0=8.6\pm2\Mpc$ for a sample of bright,
$B<15$, low redshift, $z<0.3$, QSOs.  Miller et al. (2004) show that
the AGN fraction in the SDSS galaxy survey is not dependent on
environment, while Croom et al. (2004c) and Wake et al. (2004) show
that low redshift, low luminosity AGN are clustered identically to
non-active galaxies.  The 2QZ provided the first large, deep sample with
which to perform detailed clustering analysis at high redshift.
Outram et al. (2003),  Outram et al. (2004), Miller et al. (2004) and
others have used the 2QZ to test cosmological  models.  The two-point
correlation function (the subject of this paper) has been discussed by
Croom et al. (2001a) for the preliminary, 10k, data release of the 2QZ
\cite{paper5}.  They found that the clustering of high redshift
($\bar{z}\simeq1.5$) QSOs to be very similar to the clustering of typical
galaxies at low redshift.  They also found that the amplitude of
clustering was approximately constant, or slightly increasing, with
redshift.

For comparison to the high redshift QSO clustering results, there are
now some measurements of galaxy clustering over similar redshift
intervals.  These suggest moderately high clustering amplitudes,
generally not inconsistent with that measured for QSOs.  E.g. Deep
wide-field ($\sim$ few degrees) imaging surveys used to measure the
angular correlation function of galaxies also suggest high clustering
amplitudes \cite{plso98}.  However, various differences are found,
depending on the magnitude limits and photometric bands used to define
the samples.  This is not surprising given that there is clearly
evidence that galaxy clustering
is a function of luminosity \cite{n2001}.  This may also be the
case for QSOs, although there has been no significant evidence for
this to date \cite{paper9}.  At $z\sim3$, galaxy
surveys using the drop-out technique (e.g. Steidel et al. 1998) have
found that $L\sim L^*$ galaxies also cluster similarly to local
galaxies on scales $\lsim10\Mpc$, with $r_0\simeq4-6\Mpc$ for a
cosmology with $\Omm=0.3$ and $\Omlam=0.7$
\cite{adelberger98,f03,adelberger03}.

In this paper we use the final data relase of the 2QZ (Paper XII)
to measure the QSO two-point correlation function over a wide range in
redshift.  The 2QZ is currently the best sample on which to perform
this analysis, being by far the largest QSO sample with a high
surface density ($\sim35$ deg$^{-2}$).  We focus in this paper on the
redshift-space correlation function $\xi(s)$ and attempt to account
for the effects of any $z$-space distortions.  The real-space
correlation function will be addressed in a further paper (da \^{A}ngela
et al. in preparation), and the cross-correlation of QSOs in different
luminosity intervals will be discussed by Loaring et al. (in
preparation).  In Section 2 we
introduce the 2QZ sample and the techniques used in our analysis.  In
Section 3 we use mock QSO catalogues \cite{hoyle} constructed from the
large simulations to test the reliability of our corrections for
variations in completeness in the 2QZ.  The redshift averaged,
redshift dependent and luminosity dependent 2QZ $\xi(s)$ measurements
are presented in Sections 4, 5 and 6 respectively.  We finally discuss
our conclusions in Section 7.

\begin{figure*}
\centering
\centerline{\psfig{file=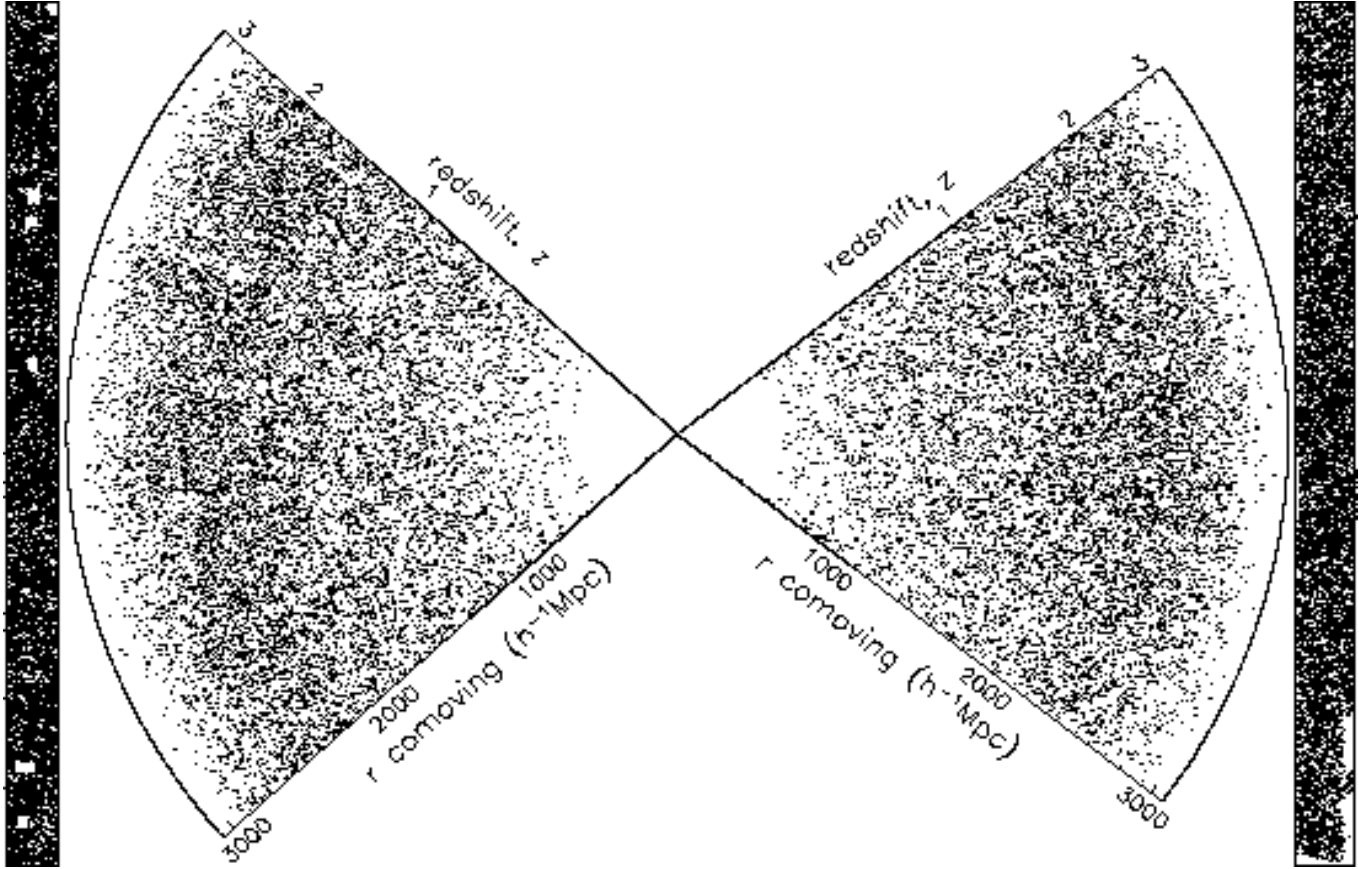,width=18.0cm,angle=270}}
\caption{The distribution of 2QZ QSOs from the final catalogue.  The
SGP strip is on the left, the equatorial strip on the right.  The
rectangular regions show the distributions projected onto the sky.  An
EdS cosmology is assumed in calculating the comoving distances to each
QSO.} 
\label{fig:wedge}
\end{figure*}

\section{Data and Techniques}\label{section_data}

\subsection{The 2dF QSO Redshift Survey}

There is a full description of the 2QZ in Paper XII.  Briefly, the
survey covers two
$75\deg \times 5\deg$ strips, one passing across the South
Galactic Cap centred on $\delta = -30\deg$ (the SGP strip) and the
other across the North Galactic Cap centred on $\delta = 0\deg$
(the NGP or equatorial strip).  The SGP strip extends from $\alpha$ =
21$^{\rm h}$40 to $\alpha$ = 3$^{\rm h}$15 and the equatorial strip
from $\alpha$ = 9$^{\rm h}$50 to $\alpha$ = 14$^{\rm h}$50 (B1950).
The total survey area is 721.6 deg$^2$, when allowance is made for
regions of sky excised around bright stars.  

2dF spectroscopic observations were carried out on colour selected
targets in the magnitude range $18.25< \bj < 20.85$.  This resulted in
the discovery of 23338 QSOs at redshifts less than $z\sim3$.  In this
paper we restrict our analysis to QSOs with quality 1 identifications
(see Paper XII), that is 22655 QSOs.  The distribution of
QSOs in the final sample is shown in Fig. \ref{fig:wedge}.

\subsection{Correlation function estimates}

\begin{figure*}
\centering
\centerline{\psfig{file=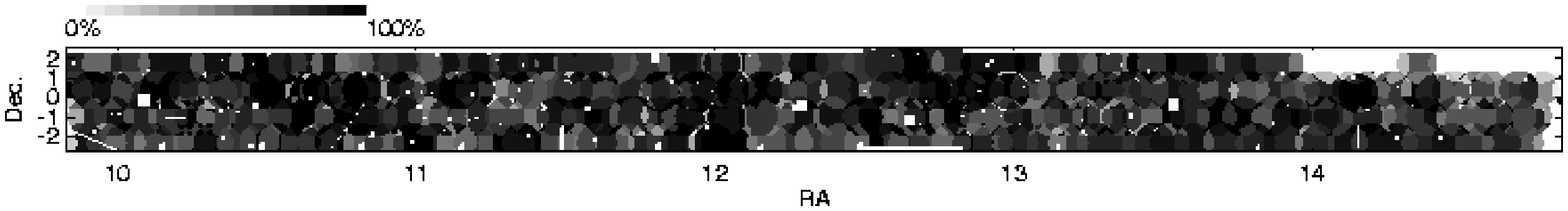,width=18cm,angle=90}}
\centerline{\psfig{file=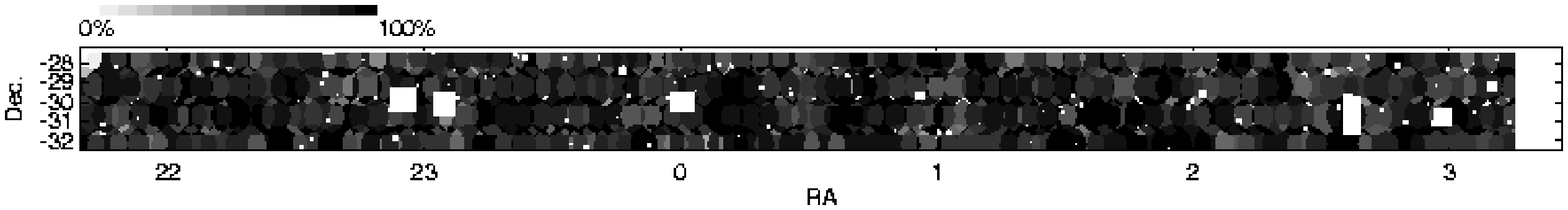,width=18cm,angle=90}}
\caption{The completeness map of the 2QZ catalogue for the equatorial
(top) and SGP (bottom) regions.  The grey-scale indicates the
percentage of all 2QZ targets that were both observed and positively
identified (quality 1) over the two survey strips.}
\label{fig:comp}
\end{figure*}

As the QSO correlation function, $\xi(s)$, probes high redshifts and
large scales, the measured values are highly dependent on the assumed
cosmology.  In determining the comoving separation of pairs of
QSOs we choose to calculate $\xi(s)$ for two representative
cosmological models.  The first uses the best fit cosmological
parameters derived from WMAP, 2dFGRS and other data
\cite{wmap,2dfgrspk02} with $(\Omm$,$\Omlam)=$ $(0.27,0.73)$, which we
will call the WMAP/2dF cosmology.  The second model assumed an
Einstein-de Sitter cosmology with $(\Omm$,$\Omlam)=$ $(1.0,0.0)$,
which we denote as the EdS cosmology.  We will quote distances in
terms of $\Mpc$, where $h$ is the dimensionless Hubble constant such
that $\ho=100h\kmsmpc$.

We have used the minimum variance estimator suggested by Landy \&
Szalay (1993)\nocite{ls93} to calculate $\xi(s)$, where $s$ is the
redshift-space (or $z$-space) separation of two QSOs (as opposed to
$r$, the real-space separation).  This estimator is
\begin{equation}
\xi(s)=\frac{QQ(s)-2QR(s)+RR(s)}{RR(s)},
\label{lseq}
\end{equation}
where $QQ$, $QR$ and $RR$ are the number of QSO-QSO, QSO-random and
random-random pairs counted at separation $s\pm\Delta s/2$.  $QR$ and
$RR$ are normalized to the total number of QSOs.  The density of
random points used was $50$ times the density of QSOs.

We calculate the errors on $\xi(s)$ using the Poisson estimate of
\begin{equation}
\Delta\xi(s)=\frac{1+\xi(s)}{\sqrt{QQ(s)}}.
\label{xierr}
\end{equation}
At small scales, $\lsim50\Mpc$, this estimate is accurate because each
QSO pair is independent (i.e. the QSOs are not generally part of
another pair at scales smaller than this).  On larger scales the QSO
pairs become more correlated and we use the approximation that
$\Delta\xi(s)=[1+\xi(s)]/\sqrt{N\qso}$, where $N\qso$ is the total
number of QSOs used in the analysis \cite{sb94,cs96}.  We
also derive field-to-field errors and compare these to the errors
found in simulations.  On small
scales, $\lsim2\Mpc$, the number of QSO-QSO pairs can be $\lsim10$.  In
this case simple {\it root-n} errors (Eq. \ref{xierr}) do not give the
correct upper and lower confidence limits for a Poisson distribution.
We use the formulae of Gehrels (1986) to estimate the Poisson
confidence intervals for one-sided 84\% upper and lower bounds
(corresponding to $1\sigma$ for Gaussian statistics).  These errors
are applied to our data for $QQ(s)<20$.  Above this number of pairs
root-n errors adequately describe the Poisson distribution.

In our analysis below we will also use the integrated correlation
function 
out to some pre-determined radius as a measure of clustering
amplitude.  This is commonly denoted by $\xibar$, where
\begin{equation}
\xibar(s\max)=\frac{3}{s\max^3}\int_0^{s\max} \xi(x)x^2{\rm d}x.
\label{eq:xibar}
\end{equation}
As in Paper II we will generally take $s\max=20\Mpc$ as this is on a
large enough scale that linear theory should apply.  The effect of
$z$-space distortions due to small-scale peculiar velocities or
redshift errors is also minimal on this scale.

\subsection{Selection functions and incompleteness}

The area of the survey is covered by a mosaic of 2dF pointings.  These
pointings overlap in order to obtain near complete coverage in all
areas, including regions of high galaxy and QSO density.  In order to
take into account the variable completeness between 2dF pointings, due
to variations in observational conditions, we use a mask that
specifies the completeness of each survey {\it sector}, where we
define a sector as the unique intersection of a number of circular 2dF
fields.  These masks are fully discussed in Paper XII.  The
completeness of each survey strip as a function of angular position on
the sky is shown in Fig. \ref{fig:comp}.  The distribution of random
points used in our correlation analysis is constructed to have an
identical distribution on the sky.  In order to minimize the influence
of low completeness fields, we restrict the analysis in this paper to
sectors for which the spectroscopic completeness is at least 70 per
cent.  This results in a sample of 20686 QSOs in the redshift range
$0.3<z<2.9$.

It is possible that on scales smaller than a 2dF field systematic
variations in completeness may exist (e.g. see Paper XII).  In
order to test the consequence of these, detailed simulations have been
carried out (see below).  On larger scales small residual calibration
errors in the relative magnitude zero-points of the UKST plates could add
spurious structure.  These are also assessed using simulations.

After generating random points according to the angular distribution
specified by the completeness masks, we then assign a random redshift
to each point.  This random redshift is draw from a distribution
defined by a polynomial fit to the observed $n(z)$ distribution (see
Fig. \ref{fig:nz}a and Section \ref{sec:nzest} below).

As a direct test of the effectiveness of the above corrections, we
also use random distributions generated by taking right ascensions
(RAs) and declinations (Decs.) from the QSO catalogue itself.  We then
assign a redshift based on either the fitted $n(z)$ (as above; this we
call the RA-Dec mixing method) or by assigning a random QSO redshift
taken from the catalogue (the RA-Dec-z mixing method).  These methods
will mimic the 2QZ QSO angular distributions exactly, but with the
effect of reducing the amount of structure measured (particularly on
larger scales).  We examine the reduction in large-scale power that
these estimates cause below.

These two alternative methods also demonstrate that the QSO
correlation function is not affected by the deficit of close ($<1'$)
pairs in the 2QZ.  The deficit is due to the fact that the 2dF
instrument cannot position two fibres closer than $\sim30''$.  It has
in large part been alleviated by the overlapping field arrangement in
the 2QZ strips, and the fact that the vast majority of QSO pairs which
are close in angular position have very different redshifts.  We
therefore make no further corrections for this effect in our analysis.

Extinction by galactic dust will also imprint a signal on the
angular distribution of the QSOs.  Primarily this changes the
effective magnitude limit in $\bj$ by $A_{\rm b_{\rm J}}=4.035\times
E(B-V)$ where we use the dust reddening $E(B-V)$ as a function of
position calculated by Schlegel, Finkbeiner \& Davis
(1998)\nocite{schlegel98}.  We then weight the random distribution
according to the reduction in number density caused by the extinction
such that
\begin{equation}
W_{\rm ext}(\alpha,\delta)=10^{-\beta A_{\rm b_{\rm
J}}(\alpha,\delta)},
\end{equation}
where $\beta$ is the slope of the QSO number counts at the magnitude
limit of the survey.  At $\bj=20.85$, the magnitude limit of the 2QZ,
the QSO number counts are flat, with $\beta\simeq0.3$. Applying this
correction we find that it only makes a significant difference to
$\xi(s)$ on scales of $\sim1000\Mpc$.  

\subsection{Making model comparisons to $\xi(s)$}

Below we make comparisons of the data to a number of models, both
simple  functional forms (power laws) and more complex, physically
motivated, models (e.g. cold dark matter; CDM).  We use the maximum
likelihood method to determine the best fit parameters.  The
likelihood estimator is based on the Poisson probability distribution
function, so that
\begin{equation}
L=\prod_{i=1}^{N}\frac{e^{-\mu}\mu^{\nu}}{\nu!}
\end{equation}
is the likelihood, where $\nu$ is the observed number of QSO-QSO
pairs, $\mu$ is the expectation value for a given model and $N$ is the
number of bins fitted.  We fit the data with bins $\Delta\log(s)=0.1$,
although we note that varying the bin size by a factor of two makes no
noticeable difference to the resultant fit.  In practice we minimize
the function $S=-2{\rm ln}(L)$, and determine the errors from the
distribution of $\Delta S$, where $\Delta S$ is assumed to be
distributed as $\chi^2$.  This procedure does not give us an absolute
measurement of the goodness-of-fit for a particular model.  We
therefore also derive a value of $\chi^2$ for each model fit in order
to confirm that it is a reasonable description of the data.  In
particular this is appropriate when fitting on moderate to large
scales ($\gsim5\Mpc$), where the pair counts are large enough that the
Poisson errors are well described by Gaussian statistics.

\section{Correlation function tests using mock QSO catalogues}

\subsection{Mock QSO catalogues}

To test both our correlation function estimation methods and the
effect of incompleteness we apply our analysis to mock QSO catalogues
produced from the large {\it Hubble Volume} simulations of the Virgo
Consortium \cite{frenk00,evrard02}.  In particular we make use of the
$\Lambda$CDM Hubble Volume simulation where data on each
particle has been output along the observer's past light cone to mimic
the 2QZ.  The simulation contains $10^9$ particles in a cube that is
$3000\Mpc$ on a side.  The cosmological parameters of the simulation
are $\Omb=0.04$, $\Omcdm=0.26$, $\Omlam=0.7$, $\ho=70\kmsmpc$ and
$\sigma_8=0.9$ (at $z=0$).  The light cone data was output in a
$75\deg\times15\deg$ wedge oriented along the maximal diagonal of the
cube, allowing the light cone to extend to a scale of $\sim5000\Mpc$
($z\sim4$).  These three slices are then split up into 3 largely
independent $75\deg\times5\deg$ slices, each one mimicing a single 2QZ
strip.  We note that there will be some correlation between the
largest structures in the different simulation strips, however, it was
not practical to generate simulations large enough to select many
completely independent volumes.

In order to create realistic mock QSO catalogues, the mass particles
are then biased to give a similar clustering amplitude to that
observed in the 2QZ (based on the results of Croom et al. 2001a).  The
biasing prescription is based on that of Cole et al. (1998) (their
model 2), but varying the parameters as a function of redshift to
match the Croom et al. (2001a) results and using a cell size of
$20\Mpc$ to determine the local density (Hoyle 2000).  In
our analysis below we consider mock catalogues with large numbers of
biased particles ($\sim100000$), almost a factor of 10 more than a
single real 2QZ strip.  This allows us to test for possible weak
systematic affects.  Full details of the Hubble Volume
simulation are given by Hoyle (2000).

\subsection{The effect of different correlation function estimates}

\begin{figure*}
\centering
\centerline{\psfig{file=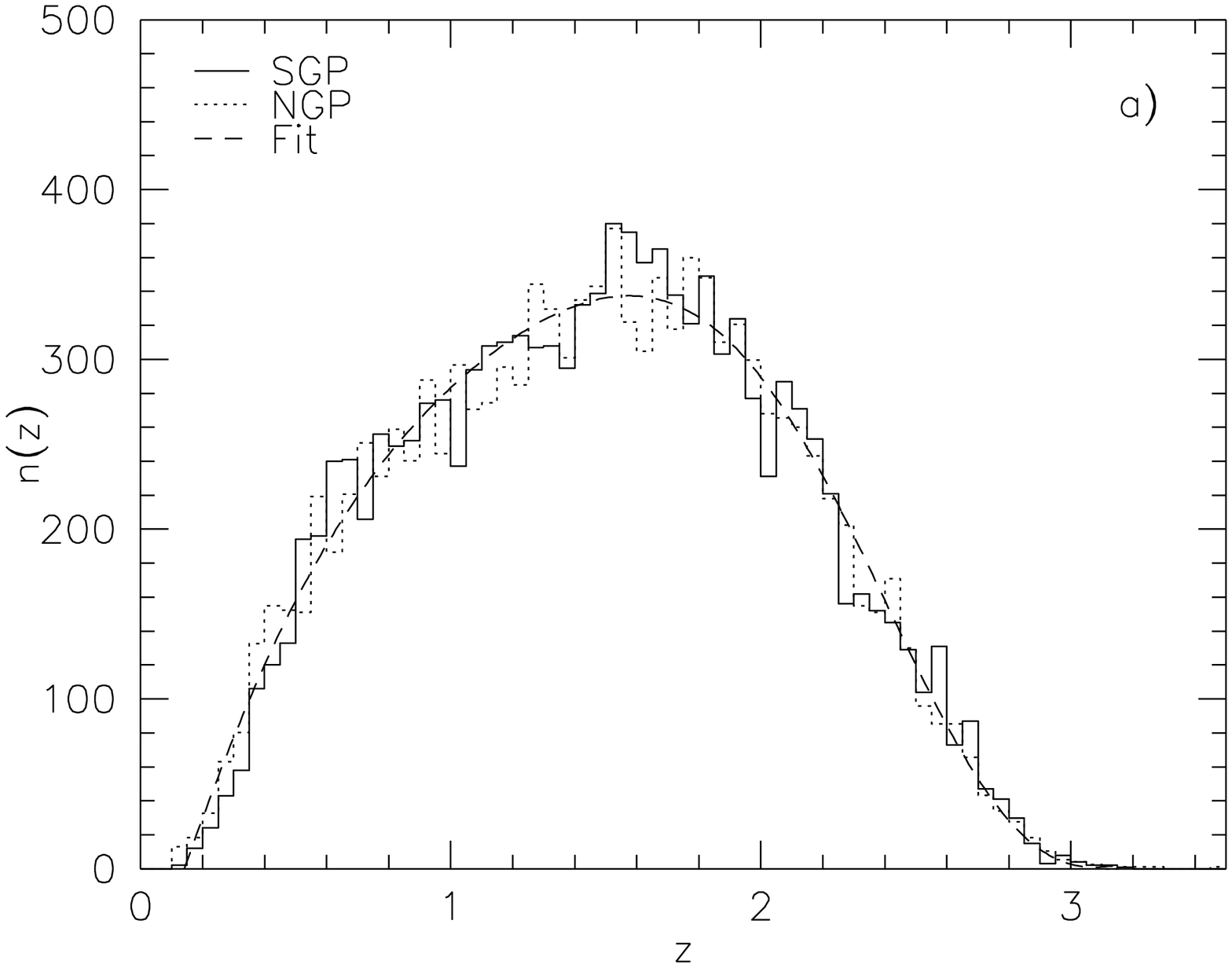,width=8cm}\psfig{file=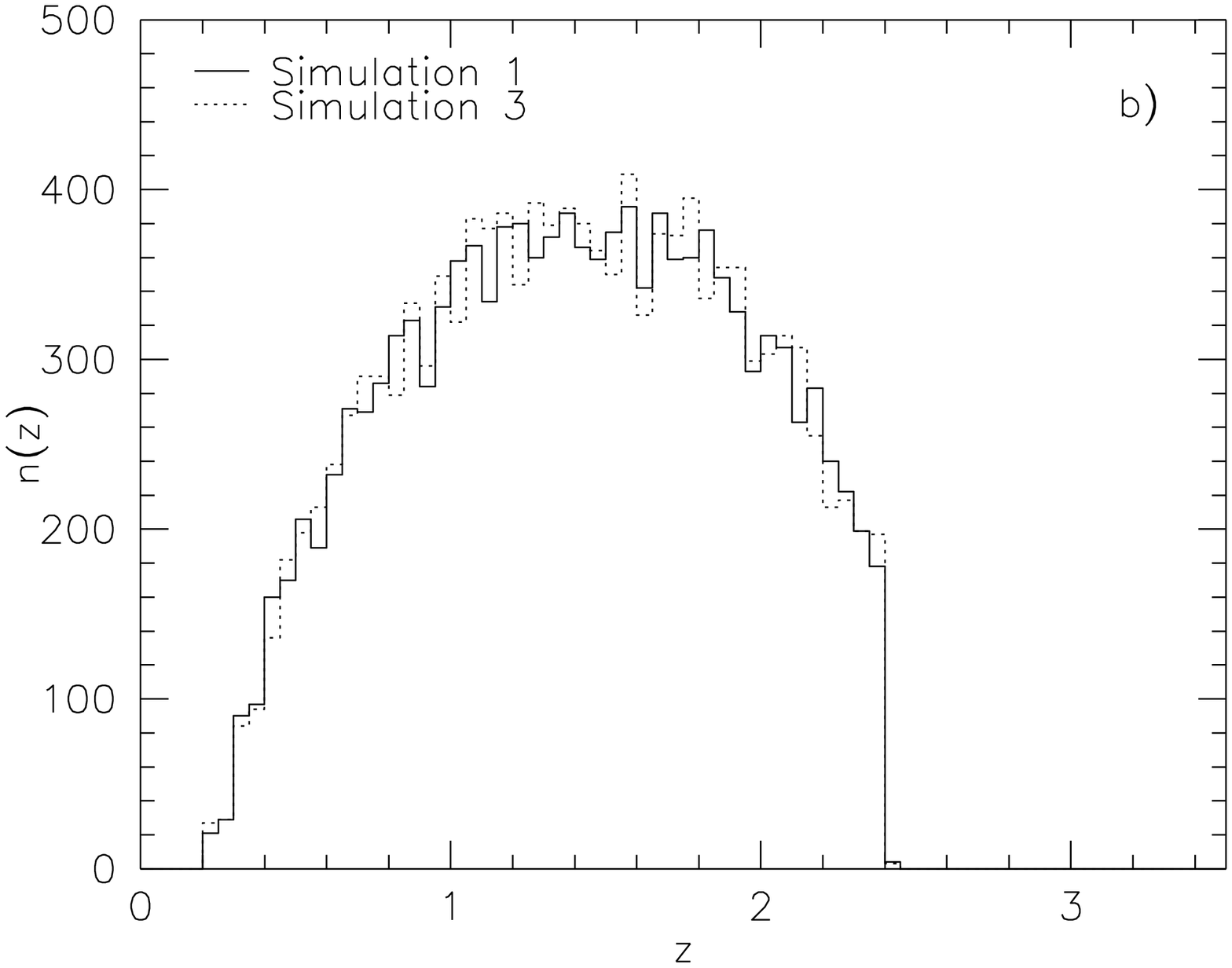,width=8cm}}
\caption{QSO and simulation $n(z)$ distributions. a) The $n(z)$
distributions in the two 2QZ slices, SGP (solid line) and NGP
(dotted line).  The NGP has been renormalized to the number of QSOs in
the SGP to aid comparison.  Also shown is the 12th order polynomial
fit to the combined $n(z)$ (dashed line).  b) The $n(z)$ distribution
of two Hubble Volume simulation slices each containing 12500
particles.}
\label{fig:nz}
\end{figure*}

There are several issues involved with accurately determining the
two-point correlation function.  We will investigate each of these in
turn. 

\subsubsection{Estimates of the QSO $n(z)$}\label{sec:nzest}

The redshift distributions, $n(z)$, of the two 2QZ slices are shown in
Fig. \ref{fig:nz}a.  In order to directly compare the two, we
renormalize the NGP $n(z)$ to contain the same total number as the
SGP.  The two strips have the same overall shape, however the we note
that they appear to have more structure that the $n(z)$
distributions of the Hubble Volume simulations shown in
Fig. \ref{fig:nz}b (note that the simulations have a cut off imposed
at $z=2.2$). By examining the spatial distribution of the QSOs
it is possible to see that the extra structure in the $n(z)$ is due to
a number of weak large-scale structures.  For example, the narrow peak in
the NGP $n(z)$ at $z=1.5$ is due to a wall-like feature (top
right of Fig. \ref{fig:wedge}).  We must therefore be careful not to
remove any excess large-scale power by fitting the $n(z)$ on too fine
a scale.  A detailed discussion of structure on very large scales is
given by Miller et al. (2004).  In Fig. \ref{fig:nz}a we plot the
polynomial fit (12th 
order) to the QSO $n(z)$ distribution used to generate the random
distributions.  Tests using higher and lower order polynomial fits
(8th -- 16th order) showed no significant differences between the
resultant $\xi(s)$ estimates.  We also
found that different methods of fitting the $n(z)$ of the simulations
(e.g. spline vs. polynomial) only caused differences at the $\sim0.1$
per cent level, much smaller than the random errors in the
measurements of $\xi(s)$ from the 2QZ.

\subsubsection{Masks vs. randomizing}\label{sec:maskrand}

\begin{figure*}
\centering
\centerline{\psfig{file=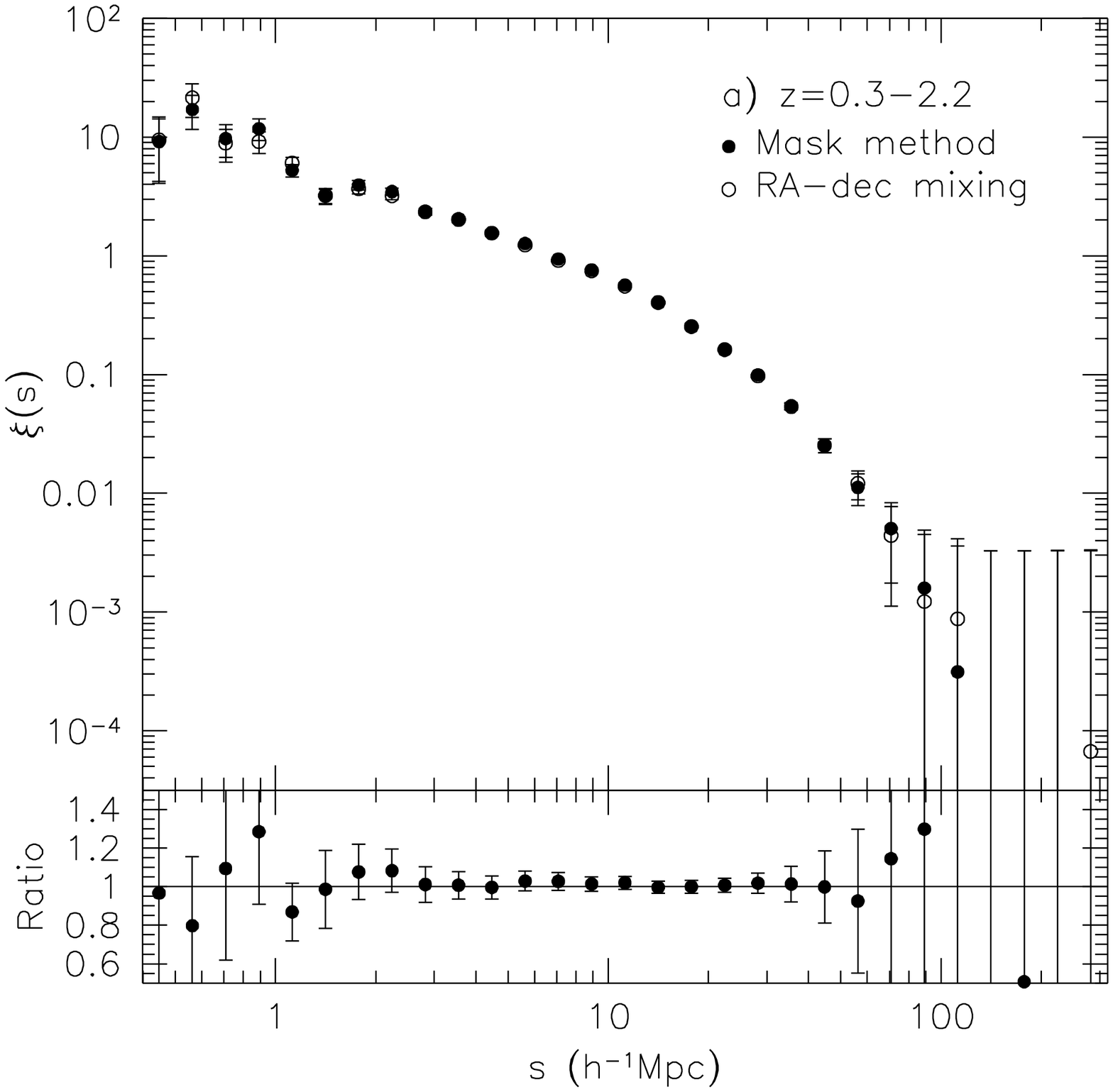,width=8cm}\psfig{file=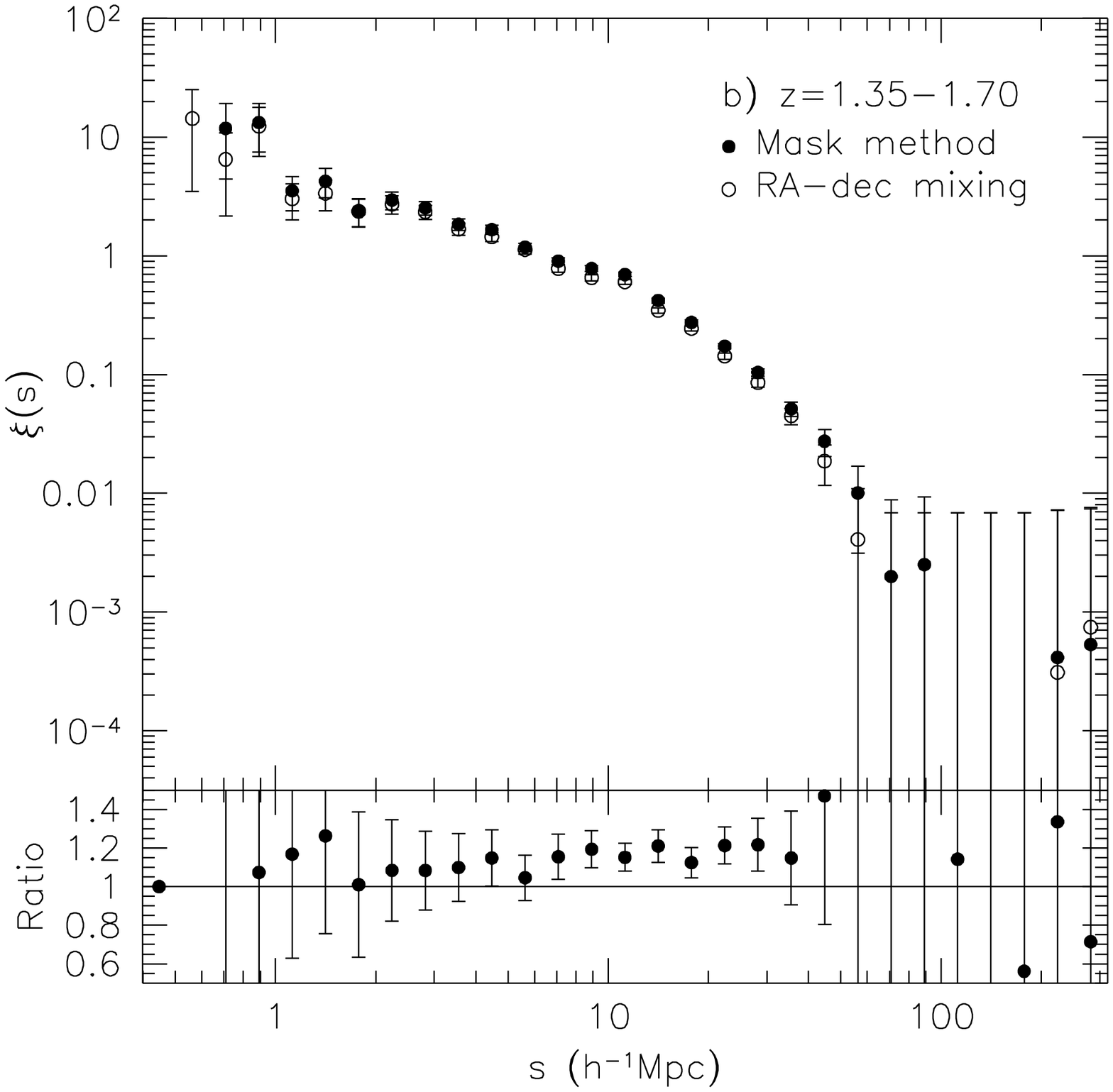,width=8cm}}
\caption{Comparison of masking (filled points) and RA-Dec mixing (open
points) methods for the Hubble Volume simulations.  Beneath each plot
we show the ratio of the two correlation function measures,
$\xi(s)_{\rm mask}/\xi(s)_{\rm mixing}$.  a) $\xi(s)$ measured over a
broad redshift range, $z=0.3-2.2$.  There is no significant difference
between the two estimates. b) $\xi(s)$ measured over a narrow 
redshift range, $z=1.35-1.70$.  In this case the RA-Dec mixing method
produces a correlation function which is $\sim10-20$ per cent lower than
the masking method.}
\label{fig:mix}
\end{figure*}

\begin{figure*}
\centering
\centerline{\psfig{file=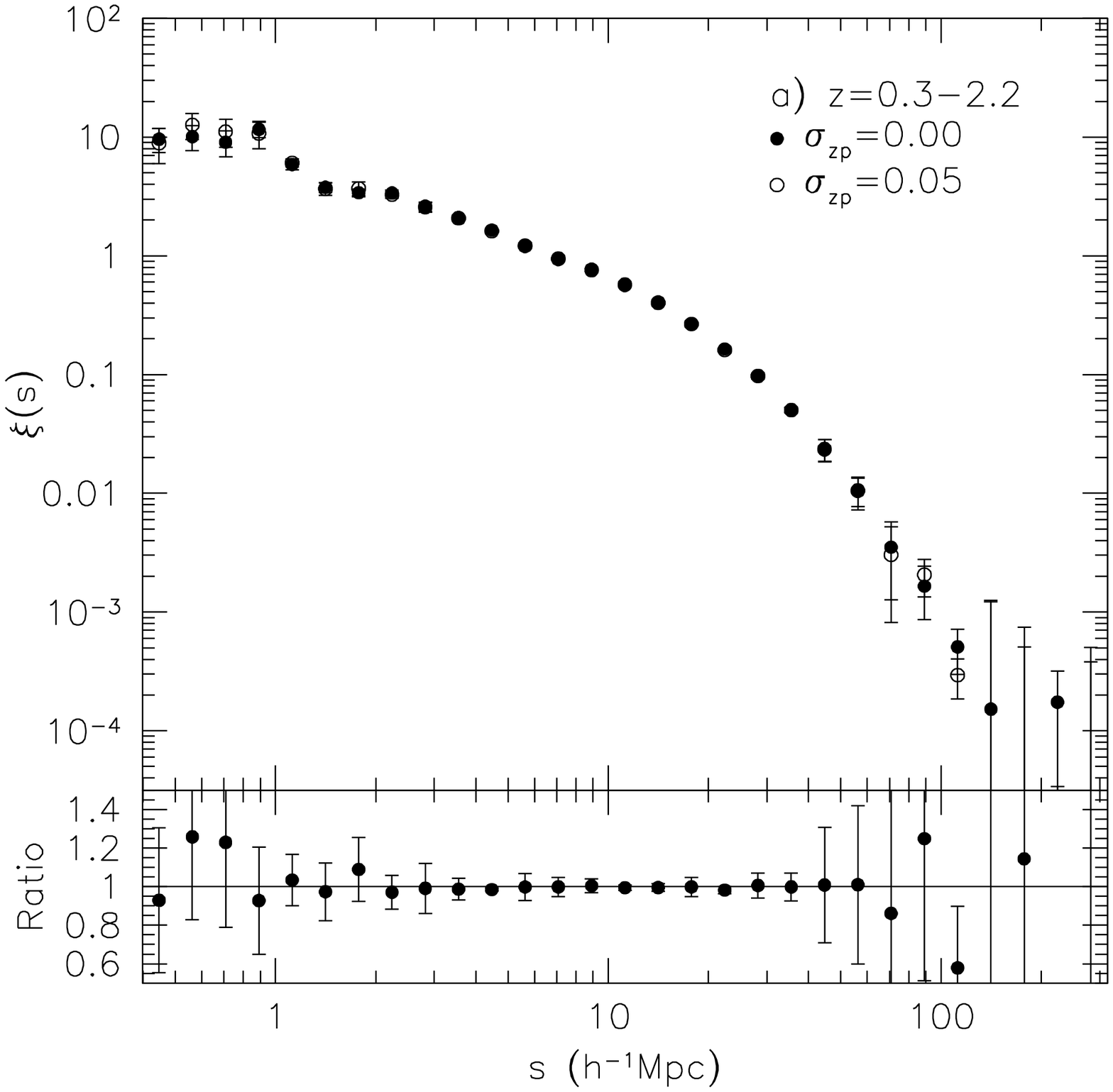,width=8cm}\psfig{file=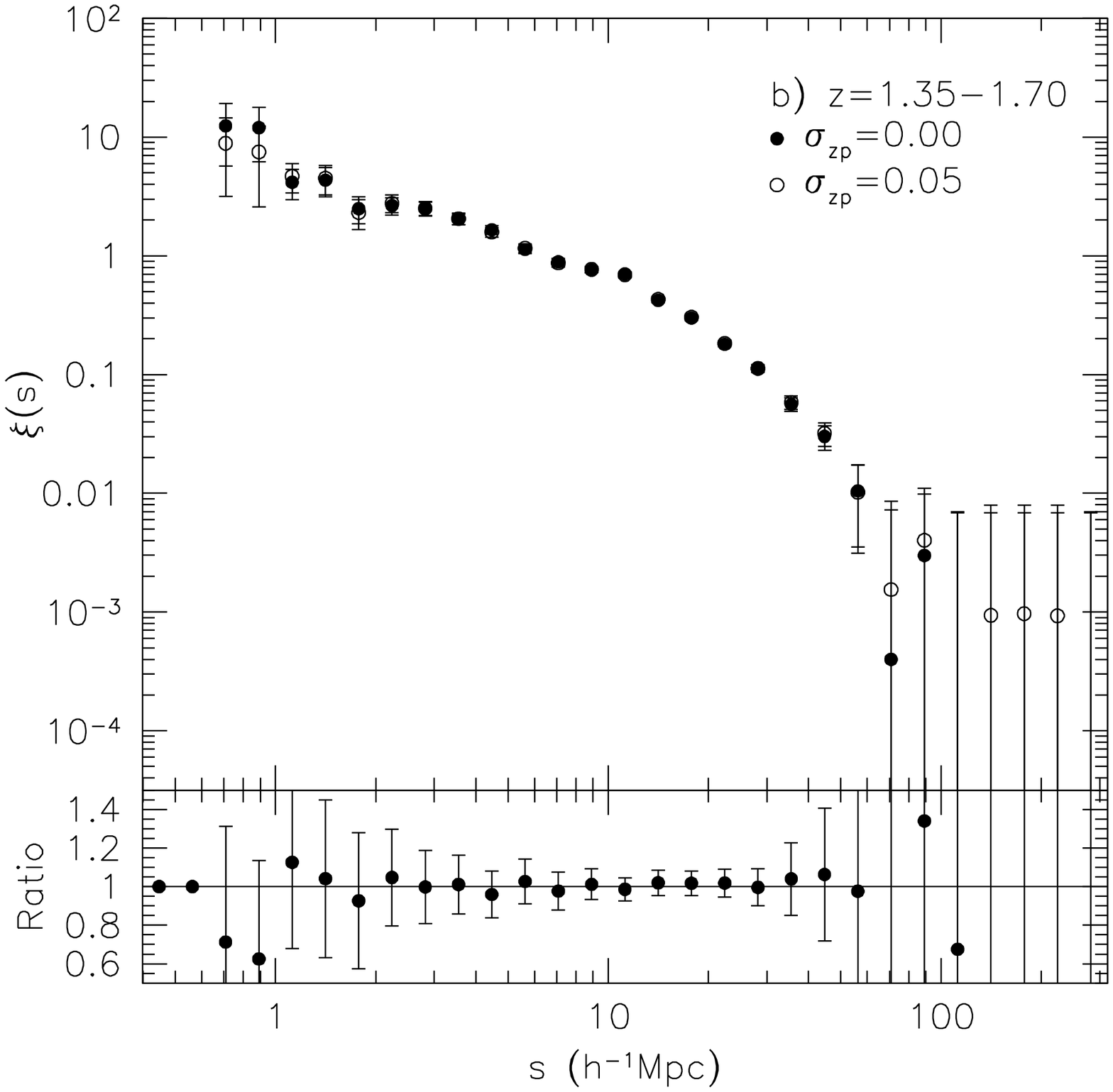,width=8cm}}
\caption{Comparison of simulated correlation functions with (open
points) and without (filled points) zero-point errors for a) the full
redshift range and b) a narrow redshift range with $z=1.35-1.70$.  The
ratio of the points with and without zero-point errors,
$\xi(s,\sigma_{\rm zp}=0.05)/\xi(s,\sigma_{\rm zp}=0.000)$, is shown
below each plot.}
\label{fig:ukstzp}
\end{figure*}

\begin{figure*}
\centering
\centerline{\psfig{file=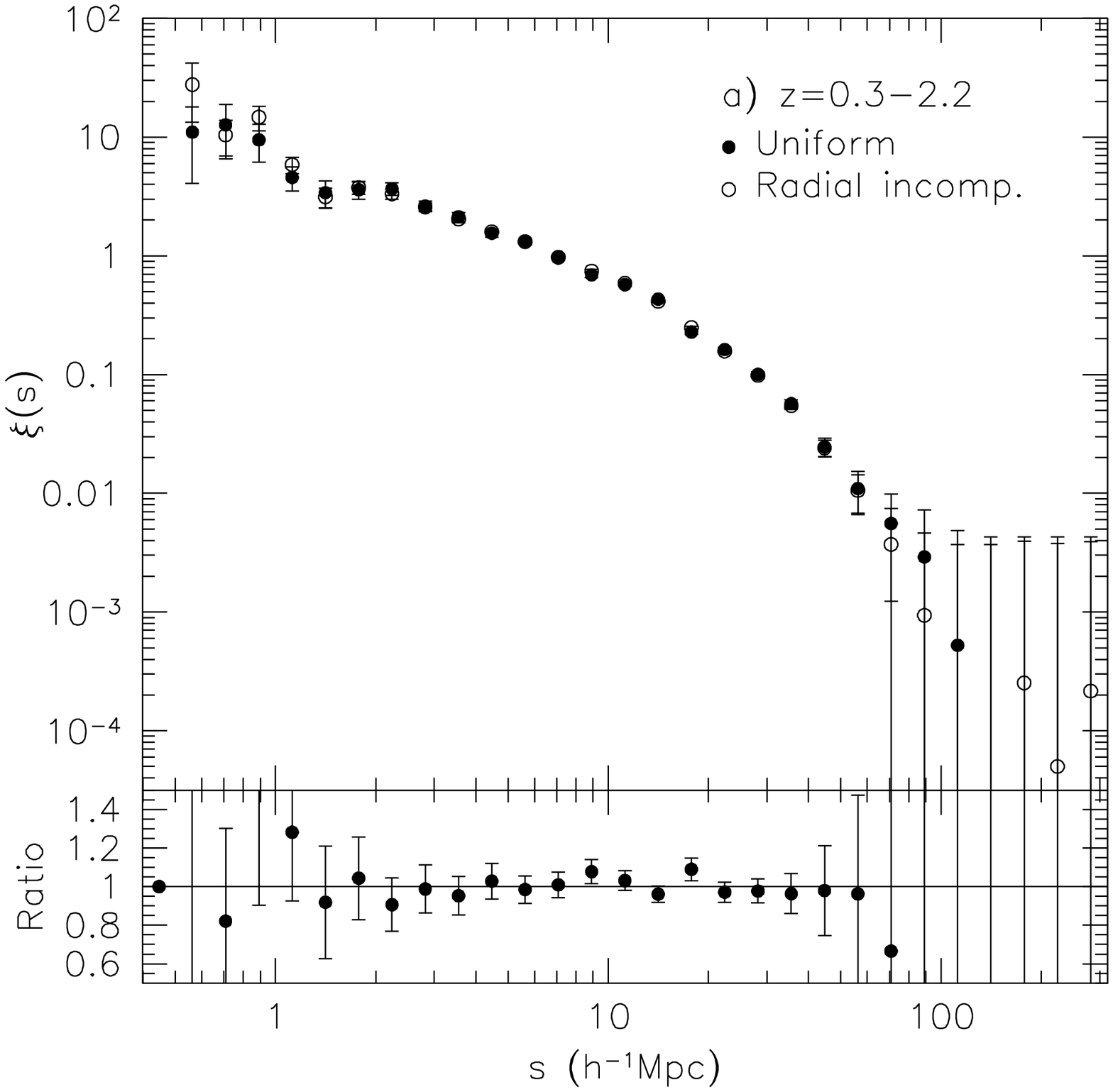,width=8cm}\psfig{file=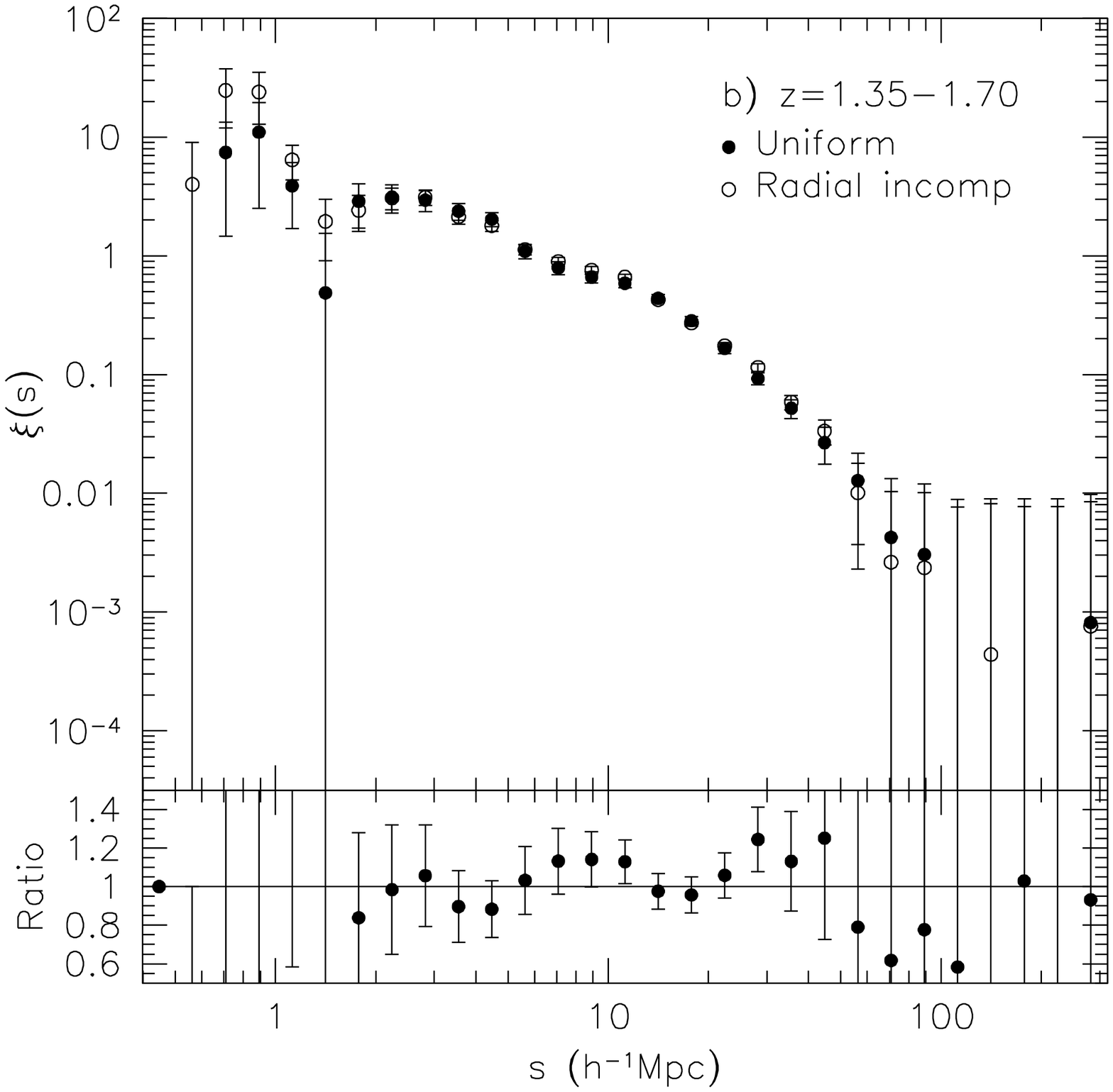,width=8cm}}
\caption{Comparison of simulated correlation functions with (open
points) and without (filled points) radially dependent incompleteness
within 2dF fields for a) the full redshift range and b) a narrow
redshift range with $z=1.35-1.70$.  The ratio of the points with and
without radial dependent incompleteness, $\xi(s)_{\rm rad}/\xi(s)$, is
shown below each plot.} 
\label{fig:radial}
\end{figure*}

We next investigate differences between the methods described above to
produce the random distributions.  In particular, although the RA-Dec
and RA-Dec-z mixing methods are effective at removing any variations
in completeness, we also need to assess whether they also remove
significant amounts of large-scale structure.  To do this we determine
the clustering in our simulations using these different methods.  In
Fig. \ref{fig:mix} we show a comparison of the masking and RA-Dec
mixing methods for a single Hubble Volume simulation slice.  When the
redshift range is broad (Fig. \ref{fig:mix}a) there is no significant
difference between the two methods and the ratio of the two
(bottom of Fig. \ref{fig:mix}a) is consistent with 1 at all scales.
However if we take a narrower redshift interval, as in
Fig. \ref{fig:mix}b, we do see significant depression of the
clustering strength in the RA-Dec mixing method.  This is because in a
narrow redshift interval, the angular clustering of QSOs will be
greater, due to the reduced amount of projection.  Therefore we
conclude that while the RA-Dec mixing method is a useful check of the
clustering amplitude averaged over the full survey, it is not an
accurate estimate when measuring QSO clustering evolution in narrow
redshift slices.  The same results were found for the RA-Dec-z mixing
method. 

\subsection{The effect of the survey selection function and incompleteness}

We now assess the effect of errors in the survey selection function on
our estimates of $\xi(s)$.  All these tests are carried out using the
masking method.  Errors in the zero-points of the UKST
photographic plates are a possible source of excess large-scale
power.  To mimic this effect we divide the simulated survey strips into 15
$5\deg\times5\deg$ regions and apply to each a Gaussian random
zero-point error $\Delta m$, with a $\sigma=0.05$ mag.
We then modulate the density of sources in that region by a factor of
$10^{-0.3\Delta m}$, as the faint end slope of the QSO number counts is
$\sim0.3$.  This equates to an error in the QSO density of 7 per cent
for a zero-point error of 0.1 mag.  With $\sigma=0.05$ the full range
of zero-point errors used was $\simeq0.15$ mag.  We do not expect there to
be real zero-point errors in the survey larger than this.  A
comparison of simulated correlation functions with and without
zero-point errors is shown in Fig. \ref{fig:ukstzp}.  We see no
systematic differences caused by the zero-point errors in either the
full redshift interval (Fig. \ref{fig:ukstzp}a), or narrower redshift
intervals (Fig. \ref{fig:ukstzp}b). We note that if the zero-point
errors are increased (to values greater than the likely photometric
errors in the survey) then significant differences can be seen.  With
$\sigma=0.1$ mag there are systematic offsets in $\xi(s)$ at the level of
$\sim1$ per cent which become significant on scales greater than
$\sim40\Mpc$.

\begin{figure}
\centering
\centerline{\psfig{file=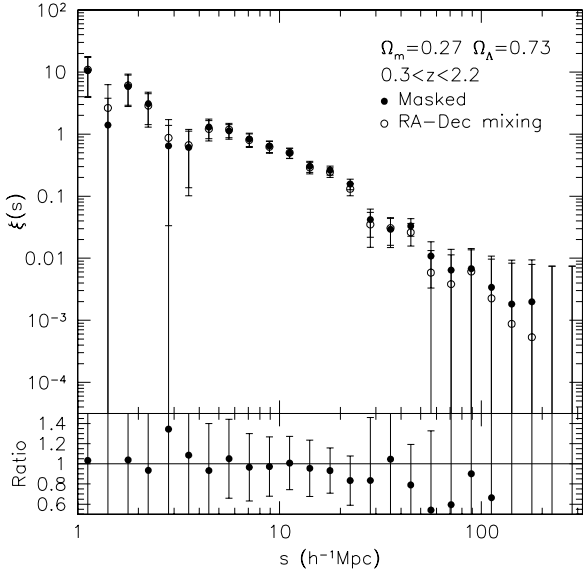,width=8cm}}
\caption{The QSO $\xi(s)$ from the 2QZ using the masking method
  (filled points) and and RA-Dec mixing method (open points).  A
  WMAP/2dF cosmology is assumed.  Below we show the ratio of the two,
  $\xi(s)_{\rm mixing}/\xi(s)_{\rm mask}$.}
\label{fig:ximethcomp}
\end{figure}

\begin{figure}
\centering
\centerline{\psfig{file=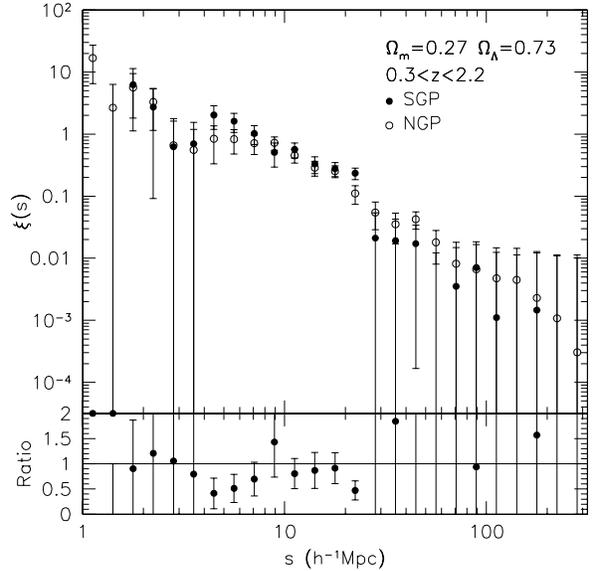,width=8cm}}
\caption{The QSO $\xi(s)$ from the 2QZ, plotting the results from the
  SGP (filled points) and NGP (open points) separately.  A WMAP/2dF
  cosmology is assumed.  Below we show the ratio of the two, 
  $\xi(s)_{\rm NGP}/\xi(s)_{\rm SGP}$.  Note that the scale of the
  ratio plot is broader than the previous similar plots.}
\label{fig:xins}
\end{figure}

Another possible cause of systematic errors in $\xi(s)$ is the
variations in completeness within 2dF fields.  These can be caused by
systematic errors in astrometry or field rotation which will be worse
at the edges of a field, or atmospheric refraction effects, if a field
was observed at a different hour angle to that which it was configured
for.  Paper XII showed that although radially dependent completeness
is noticeable when observations of many individual fields are
averaged together, if the overlap between fields and repeat
observations are taken into account there is no systematic decline in
completeness towards the edge of 2dF fields.  In order to confirm that
completeness variations within 2dF fields will not impact on our
clustering analysis we perform detailed tests.  We first position our
2dF field centres along the simulation strips, and then apply
spectroscopic completenesses selected randomly from the actual field
completenesses found in the survey.  A mask is also generated to
correct for this variable incompleteness.  We then modulate the
completeness within each simulated 2dF field such that it mimics the
radial decrease seen in Paper XII (filled points in their
Fig. 18).  We then calculated $\xi(s)$ from these simulations, using a
completeness mask which corrects for all effects apart from the
variation in completeness within the 2dF fields.  This is a worst case
scenario, as in the simulations we allocate an object to only one
field, and then derive the radial completeness variation from the
centre of that field.  In the actual survey, objects without IDs could
be observed in overlapping fields.  We compare the results to $\xi(s)$
measured without the radial completeness variations in
Fig. \ref{fig:radial}.  We find that the radial completeness
variations have no significant impact on $\xi(s)$ for either the whole
redshift range or in narrower redshift intervals.  We also determine
the effect of radial incompleteness on $\xibar(s)$ in narrow redshift
intervals (which is used extensively in Section \ref{sec:evol}).  The
radial incompleteness typically only changes $\xibar(s)$ by $2-5$
per cent, with the worst case being 10 per cent.  Given that the
radial selection model is a worst case scenario, and that the
measurement errors in $\xibar(s)$ are at least 20 per cent, any radial
dependence of completeness within 2dF fields will not impact on our
conclusions presented below.

\section{The redshift averaged QSO correlation function} 

\begin{figure*}
\centering
\centerline{\psfig{file=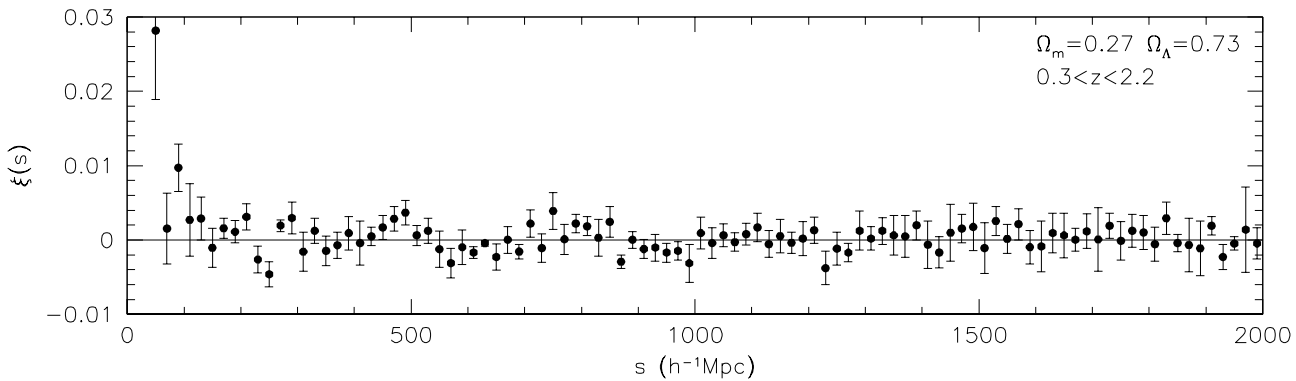,width=18cm}}
\caption{The QSO $\xi(s)$ from the 2QZ on scales $0-2000\Mpc$, plotted
  on a linear scale.  Error bars are derived from the
  field-to-field variance between six sub-samples of the data set.} 
\label{fig:xilargescales}
\end{figure*}

The above simulations confirm that our methods of correlation
analysis, and any residual systematic errors in the 2QZ should not
significantly bias our estimates of $\xi(s)$.  We now present the
results of applying our correlation analysis to the final 2QZ
sample, beginning with $\xi(s)$ averaged over the redshift range
$0.3<z<2.2$, for the most part, assuming a WMAP/2dF cosmology.  We
note that here we restrict the redshift range to regions of high
completeness, and do not include QSOs above $z=2.2$.  This is because
the mean QSO colours move progressively further into the stellar locus
above this redshift making the sample increasingly sensitive to small
systematic errors in selection.  This sample contains 18066 QSOs and
has a mean redshift of $\bar{z}=1.35$.

\subsection{Results}

We first plot a comparison between the masking method and the RA-Dec
mixing method for the redshift averaged QSO $\xi(s)$.  This is shown
in Fig. \ref{fig:ximethcomp}.  Note that we only plot $\xi(s)$ on
scales greater than $1\Mpc$ as we find no QSO-QSO pairs on scales
smaller than this (in a WMAP/2dF cosmology).  Also, for any other bins
without QSO-QSO pairs we plot a point on the bottom x-axis without an
error bar.  We see that on all scales the two 
estimates are consistent within the Poisson measurement errors.  There
is some indication that the RA-Dec mixing method is slightly
systematically lower than the mask method on scales $>20\Mpc$, which
could be an indication of a weak systematic error in the mask method,
but this is not a significant deviation.  Given the consistency of the
two methods, unless we state so explicitly, we will use the mask
method for all of our $\xi(s)$ estimates.

\begin{figure*}
\centering
\centerline{\psfig{file=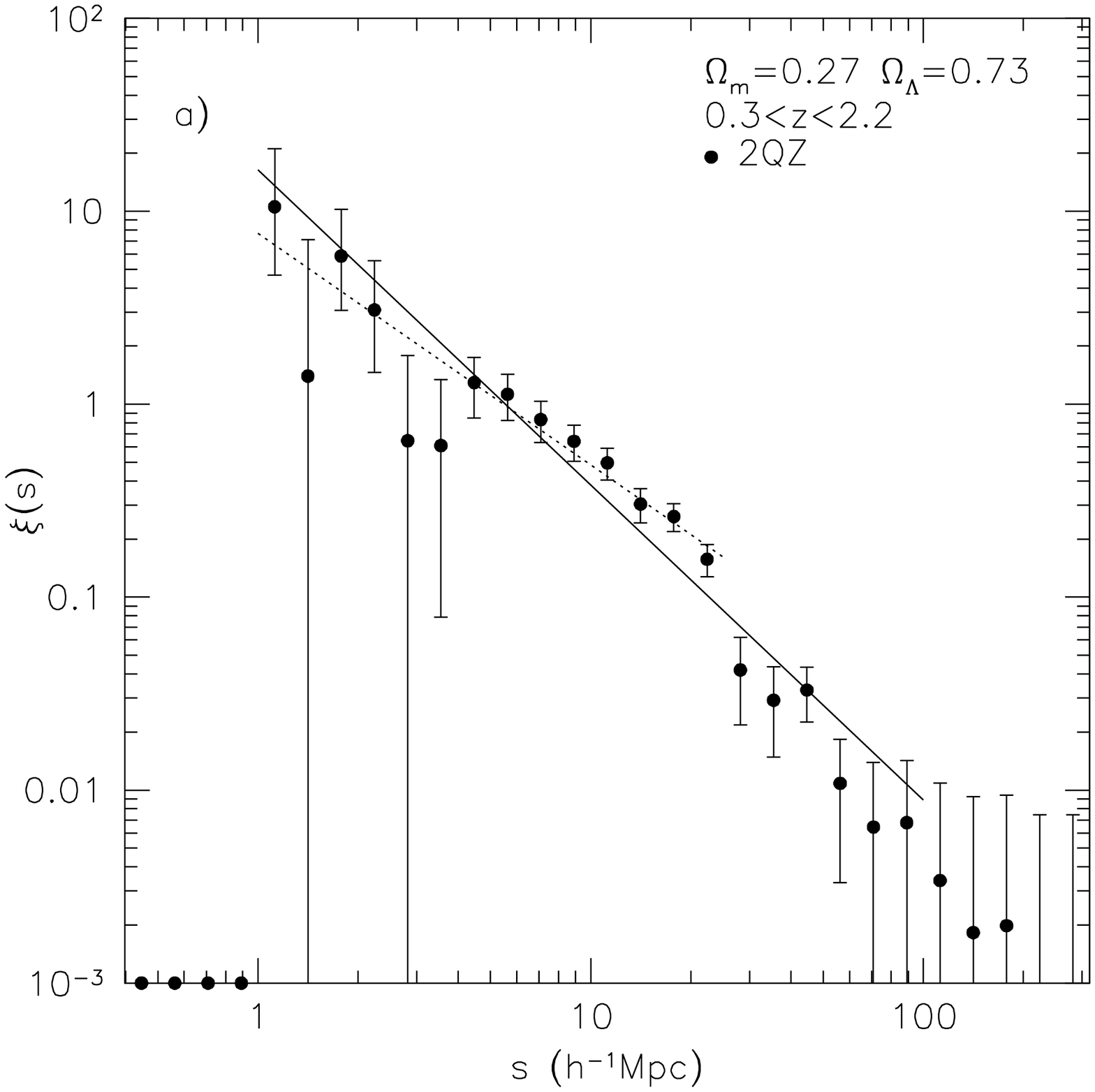,width=8cm}\psfig{file=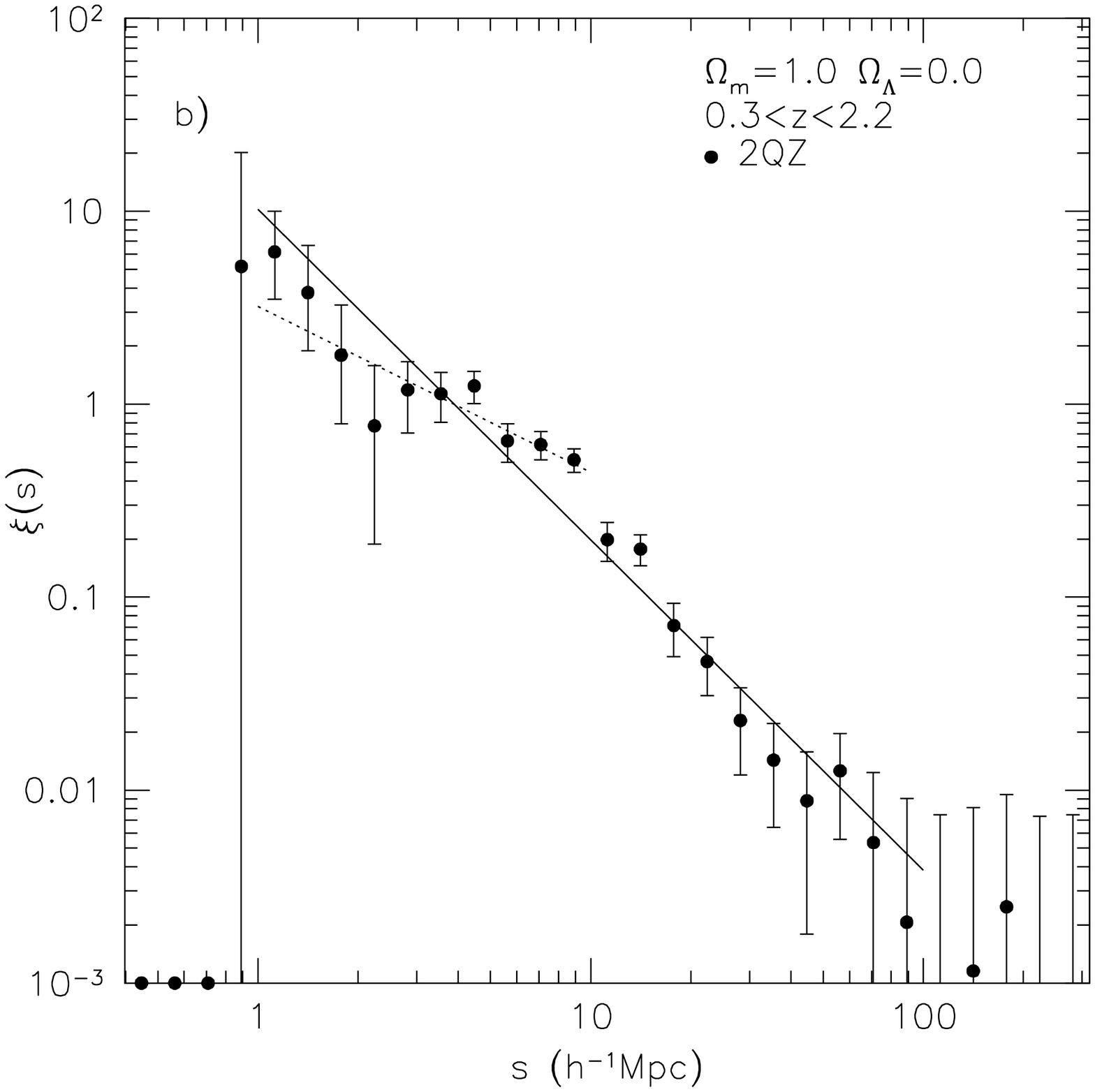,width=8cm}}
\caption{The QSO $\xi(s)$ from the 2QZ (filled points) compared to the
  best fit power laws over a range of scales: $s=1-100\Mpc$ (solid
  line) and a) $s=1-25\Mpc$ for a WMAP/2dF cosmology or b) $s=1-10\Mpc$
  for an EdS cosmology (dotted lines).}
\label{fig:xipowlawfit}
\end{figure*}

In a second check of the consistency of our results we plot a
comparison of the measured $\xi(s)$ in each of the NGP and SGP
strips (Fig. \ref{fig:xins}).  Although, the $\xi(s)$ measured from
the two strips is in broad agreement, the NGP strip shows slightly
stronger clustering on scales $>20\Mpc$.  Comparing the estimates of
$\xibar(s)$ on different scales in the two strips we find that they
are consistent ($0.9\sigma$, $1.3\sigma$ and $0.6\sigma$
differences for $s=20$, 30 and $50\Mpc$ respectively).

The large volume probed by the 2QZ allows $\xi(s)$ to be probed on
very large scales, in excess of $\sim1000\Mpc$.  Most models do not
predict any signal in $\xi(s)$ at large scales, however, there have
been some claims of features in the QSO $\xi(s)$ (including using data
from the 2QZ).  E.g. Roukema, Mamon \& Bajtlik (2002) claimed to see
several features, including a positive feature at the level of $\sim4$
per cent on a scale of $\sim240\Mpc$ in the $\xi(s)$ of
$\sim2300$ QSOs from the initial release of 2QZ catalogue
\cite{paper5}.  To test these claims we make an estimate of the 2QZ
$\xi(s)$ to the maximum scales probed by the sample.  The results of
this are shown in Fig. \ref{fig:xilargescales} for the WMAP/2dF
cosmology (Roukema et al. assume $\Omm=0.3$ and $\Omlam=0.7$, but our
results are similar for both cosmologies).  As
Fig. \ref{fig:xilargescales} probes very large scales, where
QSO pairs could be correlated, we determine errors by measuring the
variance between six subs-regions of the full data set (three
$5^{\circ}\times25^{\circ}$ regions in each 2QZ strip).  The errors
plotted is the measured rms between the six subsamples divided by
$\sqrt{6}$ to account for the greater volume of the full sample.  We
note that on the largest scales even these field-to-field errors will
somewhat inaccurate.  By comparing the QSO-QSO pair counts for the
full region and the six sub-regions we find that at $\sim200\Mpc$
$\sim10$ per cent of pairs come from correlations between different
sub-regions.  By $\sim1000\Mpc$ this number has risen so that
approximately half of all QSO-QSO pairs are from QSOs in different
sub-regions.  This means that on large scales there will be
significant correlation between the sub-regions, but the reduction of
pairs in each sub-regions will also increase the Poisson noise.

There is little evidence of any strong deviation from zero on any
scale larger that $\sim100\Mpc$ and the QSO $\xi(s)$ is zero to within
0.5 per cent over a broad range of scales.  One point (at $90\Mpc$)
deviates from zero by $\sim1$ per cent.  There is no evidence for a
feature at $\sim240\Mpc$.  At various different scales there are some
points that are greater than $1\sigma$ from zero.  A $\chi^2$ test
comparing the data to $\xi(s)=0$ at $s=100-1000\Mpc$ gives
$\chi^2=76.1$ with 45 degrees of freedom (dof), which implies
significant deviations at the 99.7 per cent level.  The rms scatter
over this scale range is $\pm0.002$.  The level of deviations away
from zero at large scales is so small that we cannot be confident that
they are real features and not due to low level residual systematics.
However, residual systematic effects at this level will not affect any
of our conclusions and we can have confidence that the masks used to
define the selection function are removing structure not due to QSO
clustering.

\subsection{Fitting models to the QSO $\xi(s)$}\label{sec:modelfits}

\begin{table*}
\begin{center}
\caption{The results of power law fits to the 2QZ $\xi(s)$ averaged
  over the redshift range $0.3<z<2.2$.  Model fits assuming a power
  law in $z$-space [$(s/s_0)^{-\gamma}$] and a power law in real-space
  [$(r/r_0)^{-\gamma}$] are presented (the second for a WMAP/2dF
  cosmology only).  The real-space power law is corrected for the
  effects of linear and non-linear $z$-space distortion.  We list the cosmology
  assumed, the scales fit over, the best fit parameters and associated
  errors, the measured $\chi^2$ values, number of dof, $\nu$ and
  probability of acceptance, $P(<\chi^2)$.} 
\setlength{\tabcolsep}{4pt}
\begin{tabular}{cccccccc}
\hline
Model &$\Omm$,$\Omlam$ & $s_{\rm min}$,$s_{\rm max}$ &
$s_0$/$r_0$ & $\gamma$ & $\chi^2$ & $\nu$ & $P(<\chi^2)$ \\
\hline
$(s/s_0)^{-\gamma}$ & 0.27,0.73 &   1.0,100.0 & $5.55_{-0.29}^{+0.29}$ & $1.633_{-0.054}^{+0.054}$ & 37.7 & 18 & 4.6e-3\\
$(s/s_0)^{-\gamma}$ & 0.27,0.73 &   1.0,25.0  & $5.48_{-0.48}^{+0.42}$ & $1.20_{-0.10}^{+0.10}$ & 8.1  & 12 & 7.8e-1\\
$(s/s_0)^{-\gamma}$ & 1.00,0.00 &   1.0,100.0 & $3.89_{-0.18}^{+0.18}$ & $1.713_{-0.052}^{+0.052}$ & 42.6 & 18 & 9.2e-4\\
$(s/s_0)^{-\gamma}$ & 1.00,0.00 &   1.0,10.0  & $3.88_{-0.53}^{+0.43}$ & $0.86_{-0.17}^{+0.16}$ & 5.6  & 8  & 7.0e-1\\
$(r/r_0)^{-\gamma}$ & 0.27,0.73 &   1.0,100.0 & $5.81_{-0.29}^{+0.29}$ & $1.866_{-0.060}^{+0.060}$ & 20.4 & 18 & 3.1e-1\\
$(r/r_0)^{-\gamma}$ & 0.27,0.73 &   1.0,25.0  & $5.84_{-0.33}^{+0.33}$ & $1.647_{-0.047}^{+0.047}$ & 7.2  & 12 & 8.4e-1\\
\hline
\label{tab:fitpowpar}
\end{tabular}
\end{center}
\end{table*}

We now attempt to fit a variety of models to the data.  The simplest
model traditionally fitted to correlation function estimates is a
power law of the form
\begin{equation}
\xi(s)=\left(\frac{s}{s_0}\right)^{-\gamma},
\label{eq:xipowerlaw}
\end{equation}
where $s_0$ is the comoving correlation length, in units of $\Mpc$.
We first fit a power law over the full range of scales where
significant clustering is detected, from 1 to $100\Mpc$, using the
maximum likelihood technique.  For the WMAP/2dF cosmology, this
resulted in best fit parameters
$(s_0,\gamma)=(5.55\pm0.29,1.633\pm0.054)$, however this fit is
unacceptable at the 99.5 per cent level (see Table
\ref{tab:fitpowpar}).  This best fit power law (solid line) is
compared to the data in Fig. \ref{fig:xipowlawfit}a and it can be seen
that the data are flatter on small scales and steeper on large scales
than model.  We then vary the maximum scale that we fit.  Only by
reducing this to $\sim25\Mpc$ is an acceptable power law fit achieved.
Over the range $1-25\Mpc$ we find best fit values
$(s_0,\gamma)=(5.48_{-0.48}^{+0.42},1.20_{-0.10}^{+0.10})$.  The
power law slope is significantly flatter when the fit is performed on
these smaller scales, but the scale length, $s_0$ is largely
unaffected.  This shows that the shape of the QSO $\xi(s)$ changes
with scale and does not follow a single pure power law, but steepens
at large scales.  We also fit similar power law models to $\xi(s)$
estimated assuming an EdS cosmology.  Over the range $s=1-100\Mpc$ we
find $(s_0,\gamma)=(3.89\pm0.18,1.713\pm0.052)$, but as for the
WMAP/2dF cosmology, this is clearly rejected (at the 99.9 per cent
level) (see Fig. \ref{fig:xipowlawfit}b).  As above, fitting on a more
restricted range of scales allows acceptable fits.  We find an
acceptable power law fit on scales $s=1-10\Mpc$ with
$(s_0,\gamma)=(3.88_{-0.53}^{+0.43},0.86_{-0.17}^{+0.16})$ (see
Fig. \ref{fig:xipowlawfit}b).  The apparent break in the QSO $\xi(s)$
is unsurprising given that we generally only expect power law
clustering in the regime where clustering is non-linear.  Similar
breaks have been seen in the clustering of low redshift galaxies
(e.g. Hawkins et al. 2003).  On scales
$\gsim10\Mpc$ where $\xi(s)<1$ clustering should be close to linear.
Other affects, such as $z$-space distortions could also distort the
measured $\xi(s)$ away from a power law.

\begin{figure}
\centering
\centerline{\psfig{file=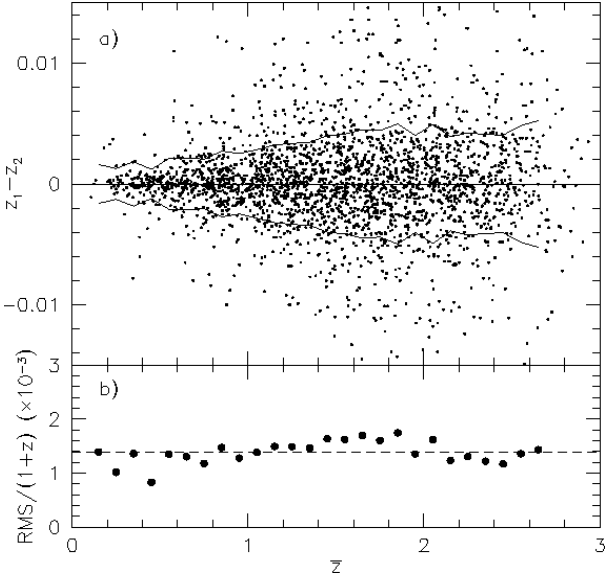,width=8cm}}
\caption{a) The difference between redshift measurements for repeated
  QSO observations in the 2QZ (using only quality 1 identifications
  and redshifts) as a function of mean redshift (points), also shown
  is the calculated rms in $\Delta z=0.1$ bins (solid lines). b) The
  rms redshift difference divided by $1+z$ as a function of mean
  redshift.  The mean $\sigma_z/(1+z)$ is shown by the dashed line.}
\label{fig:zerror}
\end{figure}

We assess the impact of $z$-space distortions on a power law.  Small
scale peculiar velocities will tend to reduce $\xi(s)$ on small
scales.  Both intrinsic peculiar velocities and redshift measurement
errors will generate a similar effect.  If due to intrinsic peculiar
velocities, this should be best described by an exponential distribution
\cite{rspf98,paper7,hawk03} such that
\begin{equation}
f_{\rm exp}(w\z)=\frac{1}{\sqrt{2}\wrms}\exp\left(-\sqrt{2}\frac{|w\z|}{\wrms}\right), 
\label{eq:expdisp}
\end{equation}
where $\wrms$ is the rms pairwise line-of-sight
velocity dispersion.  If it is the redshift measurement errors which
dominate, then the distribution may be better described by a Gaussian,
\begin{equation}
f_{\rm norm}(w\z)=\frac{1}{\wrms\sqrt{2\pi}}
\exp\left(-\frac{w\z^2}{2\langle w\z^2\rangle}\right).
\label{eq:gaussdisp}
\end{equation}
The rms
pairwise redshift error measured from repeat observations of 2QZ QSOs
is given as $\sigma\z=0.0027z$ in Paper XII.  We have re-assessed this
redshift error using the same data as Paper XII
(Fig. \ref{fig:zerror}) and find that a better estimate of the
pairwise redshift error is $\sigma\z=0.0014(1+z)$ (the dashed line in
Fig. \ref{fig:zerror}b).  Thus the pairwise velocity
error [$\delta v = c\delta z/(1+z)$] corresponding to this redshift
error is $\delta v\z = 416\kms$ largely independent of redshift.  To
this we need to add the intrinsic velocity dispersion of the QSOs,
$\delta v_{\rm i}$.  At low redshift the typical intrinsic galaxy
pairwise velocity dispersion is  $\simeq500\kms$ (e.g. Hawkins et
al. 2003) at $z\simeq0.15$.  We note that Hawkins et al. did not
include the factor of $1+z$ in Eq. \ref{eq:xisigmapi} (see below).
Correcting for this, the pairwise velocity is actually
$\simeq430\kms$.  It is uncertain
whether this will decline with redshift.  While the dark matter
velocity dispersion should decline, as QSOs are biased tracers of
large-scale structure, their pairwise velocity may not decline.  Zhao,
Jing \& Borner (2002) predict that the pairwise velocity dispersion of
Lyman-break galaxies at $z\sim3$ could be $\sim200-400\kms$.  Given
the uncertainty in the evolution of $\delta v_{\rm i}$ we will assume
a fixed value of $\simeq430\kms$ at all redshifts, noting that any
evolution is likely to reduce this value.  A final issue that needs to
be considered is the velocity error due to intrinsic emission-line
shifts in QSOs, $\delta v_{\rm l}$.  The UV emission lines in QSO
spectra typically show blue-shifts relative to their systemic
velocity, this is particular so of lines such as C{\small IV}.
Richards et al. (2002) demonstrated that the dispersion between
centroids of C{\small IV} and Mg{\small II} lines was $511\kms$, while
the dispersion between Mg{\small II} and [O{\small III}] was a somewhat
smaller $269\kms$.  This dispersion will cause an extra dispersion in
our redshift estimates which is not taken into account by the repeat
observations (as they are repeats of the same QSO spectrum).  Thus
$\delta v_{\rm l}$ should take values in the range $200-450\kms$ allowing
for measurement errors \cite{r02}.  Combining the three components of
velocity dispersion together in quadrature results in
$\wrms\simeq630-750\kms$.  In our analysis below we will
assume a value of $690\kms$ which lies in the middle of this range.
As a combination of $\delta v_{\rm l}$ and  $\delta v\z$ dominates the
total pairwise velocity dispersion, we use Eq. \ref{eq:gaussdisp} to
model the effects of $z$-space distortions on small scales.  We note
that other authors (e.g. Outram et al. 2004; Hoyle et al. 2002) used a
similar value of $\wrms\simeq800\kms$ (however they miss the
factor of $1+z$ in Eq. \ref{eq:xisigmapi} below).

\begin{figure}
\centering
\centerline{\psfig{file=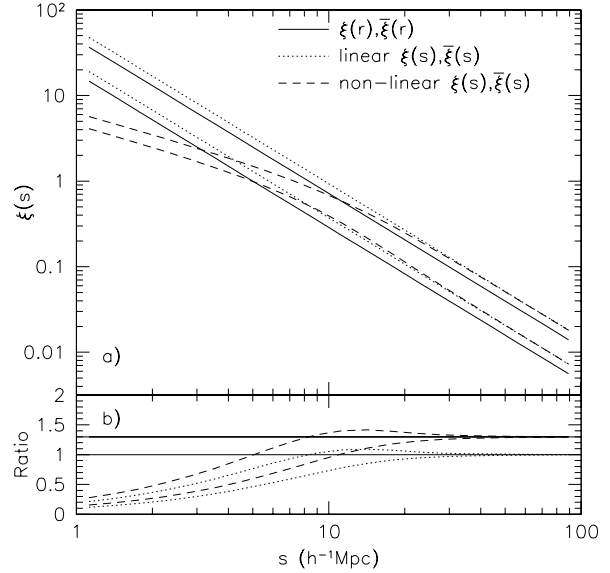,width=8cm}}
\caption{a) model power law correlations functions demonstrating the
  effects of $z$-space distortions, assuming a real space
  $\xi(r)=(r/5)^{-1.8}$.  In each case $\xi$ and $\xibar$ are plotted
  with $\xibar$ being the upper line.  We show $\xi(r)$ (solid line),
  $\xi_{\rm lin}(s)$  (dotted line) and and $\xi_{\rm non-lin}(s)$
  (dashed line).  For the redshift-space distortion model we assume
  a WMAP/2dF cosmology, at a mean redshift of 1.35 with $\beta(z)=0.4$
  and $\wrms=690\kms$.  b) the ratio of different models
  comparing the ratios of $\xi_{\rm non-lin}(s)/\xi(r)$ (upper dashed
  line) and $\xi_{\rm non-lin}(s)/\xi_{\rm lin}(s)$ (upper dotted
  line).  The other two dashed and dotted lines are the $\xibar$
  equivalents.  The two solid lines are set at 1.0 and at
  $(1+2\beta/3+\beta^2/5)=1.30$ for $\beta=0.4$.}
\label{fig:xizspacemodels}
\end{figure}

We should also take into account the effect of linear $z$-space
distortions.  Kaiser (1987) showed that 
\begin{equation}
\xi(s)=\xi(r)(1+\frac{2}{3}\beta+\frac{1}{5}\beta^2),
\label{eq:kaisereq}
\end{equation}  
where $\xi(r)$ is the real-space correlation function and 
$\beta\simeq\Omm^{0.6}/b$.  More generally, $\xi(\sigma,\pi)$, the
correlation function across (the $\sigma$ direction) and along (the
$\pi$ direction) the line of sight is distorted, such that 
\begin{equation}
\xi(\sigma,\pi)=\left[1+\frac{2(1-\gamma\mu^2)}{3-\gamma}\beta+\frac{3-6\gamma\mu^2+\gamma(2+\gamma)\mu^4}{(3-\gamma)(5-\gamma)}\beta^2\right]\xi(r),
\label{eq:xisigmapilin}
\end{equation}         
assuming that $\xi(r)$ is a power law \cite{ms96}. $\mu$ is the cosine
of the angle between $r$ and $\pi$ (the distance along the line of
sight), and $\gamma$ is slope of the power law.  Then including the
effects of non-linear $z$-space distortions, the full model for
$\xi(\sigma,\pi)$  is given by
\begin{equation}
\xi(\sigma,\pi)=\int^\infty_{-\infty}\xi'[\sigma,\pi-(1+z)w\z/H(z)]f_{\rm
  norm}(w\z)\rm{ d}w\z,
\label{eq:xisigmapi}
\end{equation}
where $\xi'[\sigma,\pi-(1+z)w\z/H(z)]$ is given by
Eq. \ref{eq:xisigmapilin}, $f_{\rm norm}(w\z)$ is given by
Eq. \ref{eq:gaussdisp} and $H(z)$ is Hubble's constant at a redshift,
$z$.  Finally, we carry out a spherical integral over the model
$\xi(\sigma,\pi)$ to derive the model $\xi(s)$ which we then fit to the
data.  We note that there is an extra factor of $1+z$ in
Eq. \ref{eq:xisigmapi} compared to previous works (e.g. Hawkins et
al. 2003; Hoyle et al. 2002).  This is because the velocity
dispersions are generally given in {\it proper} coordinates, rather
than comoving coordinates.  At low redshift this has a minimal affect,
however, at high redshift this extra term boosts the effective scale
corresponding to a given proper velocity by $1+z$ (in fact it
approximately cancels out the increase of $H(z)$ with redshift, so
that the proper velocity dispersion corresponds to a similar comoving
scale at every redshift).   It is therefore critical to incorporate
this term. In this paper we are not specifically focussing on
$\xi(\sigma,\pi)$ and $z$-space distortions, but only wish to
determine their affect in shaping the measured $\xi(s)$.  Detailed
investigation of $\xi(\sigma,\pi)$ is discussed by da \^{A}ngela et
al. (in preparation).

\begin{figure}
\centering
\centerline{\psfig{file=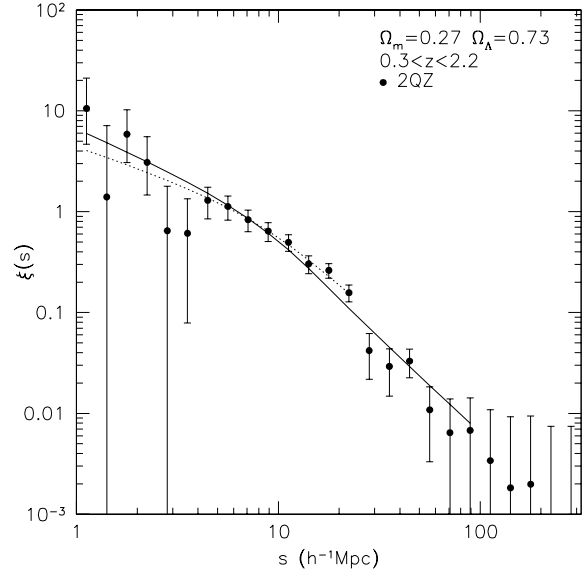,width=8cm}}
\caption{The QSO $\xi(s)$ from the 2QZ (filled points) compared to the
  best fit power law model incorporating the affects of linear and
  non-linear redshift-space distortions.  A WMAP/2dF cosmology is
  assumed.  The fits are carried out on scales $s=1-100\Mpc$ (solid
  line) and $s=1-25\Mpc$ (dotted line).} 
\label{fig:xizspacefit}
\end{figure}

Estimates of the
strength of $z$-space distortions via the QSO power spectrum have been
made by Outram et al. (2004).  They find that at $z=1.4$, the mean
redshift of the sample used, $\beta=0.4\pm0.1$.  We assume this
value for $\beta$ and a small-scale velocity dispersion of $690\kms$.
We then produce a grid of model real-space correlation functions which
are adjusted for these $z$-space distortions and fitted to our
observed $\xi(s)$ using the maximum likelihood technique. 

In Fig. \ref{fig:xizspacemodels}a we show a comparison of models with
and without $z$-space distortions. Assuming a real-space correlation
function of $\xi(r)=(r/5)^{-1.8}$ in a WMAP/2dF cosmology, and the
above values of  $\beta=0.4$ and $\wrms=690\kms$.  The solid
lines  show the real-space $\xi(r)$ and $\xibar(r)$ (see
Eq. \ref{eq:xibar}).  The model $\xibar(r)$ is a factor of
$3/(3-\gamma)=2.5$ above $\xi(r)$.  The dotted lines show the model
$\xi(s)$ and $\xibar(s)$ for linear $z$-space distortions only
($\xi\lin(s)$ i.e. $\beta=0.4$ and $\wrms=0.0$), while
the dashed lines show the full model with linear and non-linear
$z$-space distortions ($\xi\nonlin(s)$ i.e. $\beta=0.4$ and
$\wrms=690\kms$).  On scales less than $10\Mpc$ the
non-linear $z$-space distortions cause a significant suppression of
$\xi$.  In Fig. \ref{fig:xizspacemodels}b we plot the ratio of these
various models.  The dashed lines are $\xi\nonlin(s)$ (top) and
$\xibar\nonlin(s)$ (bottom) divided by $\xi(r)$ and $\xibar(r)$
respectively.  The dotted lines are $\xi\nonlin(s)$ (top) and
$\xibar\nonlin(s)$ (bottom) divided by $\xi\lin(s)$ and
$\xibar\lin(s)$ respectively.  The solid lines are set at 1 and at
$(1+2\beta/3+\beta^2/5)=1.30$ (for $\beta=0.4$).  From this it can be
seen that on scales $\sim20-30\Mpc$ and larger the affect of
non-linear $z$-space distortion is small, while the linear term
affects $\xi$ on all scales.  For the above power law, we find that
$\xibar\nonlin(s)/\xibar\lin(s)=0.93$, 0.97 and 0.99 for $s=20$, 30
and $50\Mpc$ respectively.

To begin with we assume a power law model for $\xi(r)$
(Eq. \ref{eq:xipowerlaw}).  We generate a grid of models with
different power law slopes ($\gamma$), and fit these models to the
data using the maximum likelihood technique over the range
$s=1-100\Mpc$.  The resulting best fit  model with  $\beta=0.4$ and
$\wrms=690\kms$ is shown by the solid line in
Fig. \ref{fig:xizspacefit}.  We find a power law slope of
$\gamma=1.866\pm0.060$ and a real-space scale length
$r_0=5.81\pm0.29\Mpc$.  This provides an acceptable fit to the data
with $\chi^2=20.4$ (18 dof) and an acceptance
probability of 31 per cent.  If we fit over a more restricted range of
scales, noting that we expect deviations from a pure power law in real
space on large scales, then we find best fit values of
$\gamma=1.647\pm0.047$ and $r_0=5.84\pm0.33\Mpc$ for $s=1-25\Mpc$.
Both fits are compared to the data in Fig. \ref{fig:xizspacefit} (see
also Table \ref{tab:fitpowpar}).
When fitting on smaller scales the power law slope is flatter,
however, $r_0$ is unchanged.  It can be seen that the affect of small
scale $z$-space distortions has a significant impact on scales less
than $\sim10\Mpc$.

\begin{figure}
\centering
\centerline{\psfig{file=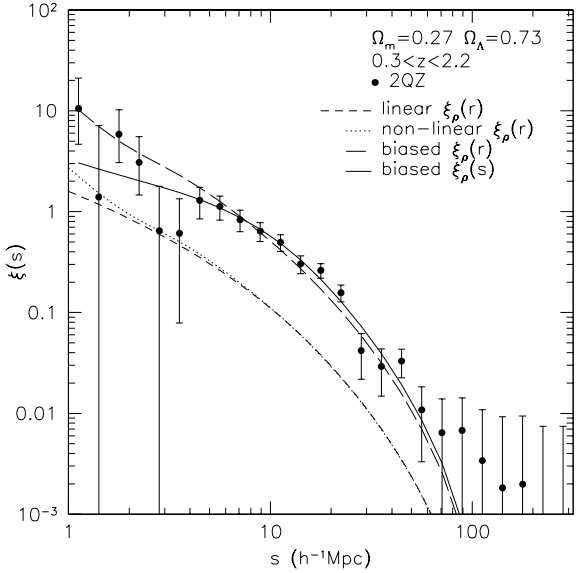,width=8cm}}
\caption{The QSO $\xi(s)$ from the 2QZ (filled points) compared to
  a WMAP/2dF CDM model.  The models shown are the linear real-space mass
  correlation function, $\xi_\rho(r)$ (short dashed line), the
  non-linear $\xi_\rho(r)$ (dotted line).  The non-linear
  $\xi_\rho(r)$ scaled by the best fit bias value (long dashed line)
  and the non-linear mass correlation function corrected for $z$-space
  distortions, $\xi_\rho(s)$, scaled by the best fit bias (solid
  line).}
\label{fig:xizspacefitcdm}
\end{figure}

More generally we should fit
a model where the shape of  $\xi(r)$ is governed by the underlying
physics of the dark-matter distribution (e.g. CDM).  In particular,
Hamilton et al. (1991,1995) provide an analytic description of the
generic linear CDM $\xi(r)$.  The input parameters for the CDM model
are taken from the now standard WMAP/2dF cosmological model
\cite{wmap,2dfgrspk02} with $\Omm=0.27$, $\Omlam=0.73$, $\Omb=0.04$,
$H_0=71\kms~{\rm Mpc}^{-1}$, $\sigma_8=0.84$ (at $z=0$).  We calculate
the model $\xi(s)$ at the mean redshift of the 2QZ sample ($\bar{z}=1.35$),
and correct for the affects of non-linear clustering
\cite{hklm91,jmw95}.  Linear and non-linear $z$-space effects are
accounted for as above, but using the more general prescription
of Hamilton (1992) rather than Eq. \ref{eq:xisigmapilin} for the
linear distortions.  For the $z$-space distortions we assume
$\beta=0.4$ and $\wrms=690\kms$.  We then perform a maximum likelihood
fit for a single parameter, a scale independent QSO bias, over the
scale range $s=1-100\Mpc$.  QSO bias is defined as
\begin{equation}
b\qso(z)=\sqrt{\frac{\xi\qso(r)}{\xi_\rho(r)}},
\label{eq:bias}
\end{equation}
where $\xi\qso(r)$ and $\xi_\rho(r)$ are the real-space QSO and mass
correlation functions respectively.  We note that our assumed value of
$\beta$ includes an implicit assumption of QSO bias.  If we substitute
the $\xi\qso(r)$ in Eq. \ref{eq:bias} with that from
Eq. \ref{eq:kaisereq} and solve the resultant quadratic in $b\qso(z)$
we find that 
\begin{equation}
b\qso(z)=\sqrt{\frac{\xi\qso(s)}{\xi_\rho(r)}-\frac{4\Omm^{1.2}(z)}{45}}-\frac{\Omm^{0.6}(z)}{3}.
\label{eq:bz}
\end{equation}
This relation thus directly gives us the QSO bias at a redshift $z$,
but is only strictly true if non-linear $z$-space distortions, which
affect the shape of $\xi(s)$, are not present.  The linear distortions
do not affect the shape of $\xi(s)$ (this is exactly the case when
there are no non-linear effects, and correct to first order in the
presence of non-linear effects), so we fit a model $\xi_\rho(s)$
divided by $(1+2\beta/3+\beta^2/5)$ (using the same $\beta=0.4$ value
used above) to obtain the ratio $\xi\qso(s)/\xi_\rho(r)$ seen in
Eq. \ref{eq:bz}.  Assuming $\Omm(z=0)=0.27$ [implying
$\Omm(z=1.35)=0.83$] we find a best fit QSO bias of
$b\qso(z=1.35)=2.02\pm0.07$.  This model is fully
consistent with the data, with a $\chi^2=14.3$ from 19 dof (acceptable
at the 76 per cent level, see the solid line in Fig.
\ref{fig:xizspacefitcdm}).  The implied values of $\beta$ for this
best fit bias is $\beta=0.44\pm0.02$.  This is close to our assumed
value of $\beta=0.4$ and within the errors estimated by Outram et
al. (2004) of $\pm0.1$.  To test the impact of making the 
$z$-space corrections to our model, we also fit the non-linear real space
model to the data.  This results in a best fit bias of
$2.12\pm0.09$ (long dashed line in Fig. \ref{fig:xizspacefitcdm}),
however, this is a slightly worse fit with a $\chi^2=25.2$ (19
dof) acceptable at the 15 per cent level.  From
Fig. \ref{fig:xizspacefitcdm} we see that the real-space model does
not have a strong enough break at $\sim10-20\Mpc$ to match the data.
We conclude that the 2QZ QSO $\xi(s)$ averaged over redshift is
fully consistent with the WMAP/2dF cosmology once allowance is made
for the affects of $z$-space distortions.

\subsection{Comparisons to other results}

\begin{figure}
\centering
\centerline{\psfig{file=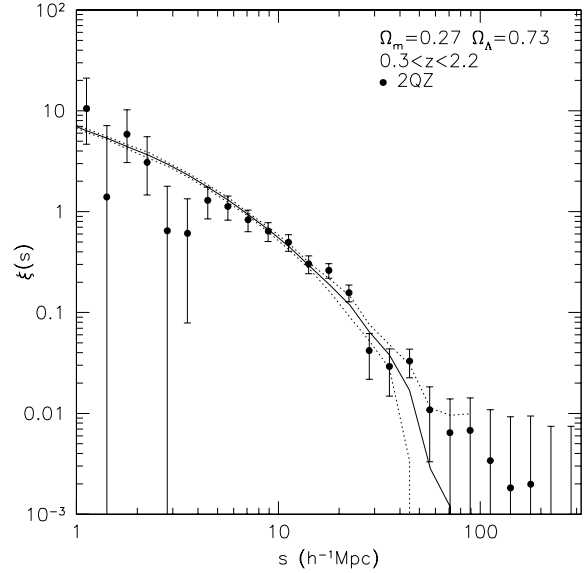,width=8cm}}
\caption{The QSO $\xi(s)$ from the 2QZ (filled points) compared to the
 the 2dFGRS $\xi(s)$ of Hawkins et al. (2003) (solid line, with
 $\pm1\sigma$ errors shown by the dotted lines).}
\label{fig:xisgalcomp}
\end{figure}

\begin{figure}
\centering
\centerline{\psfig{file=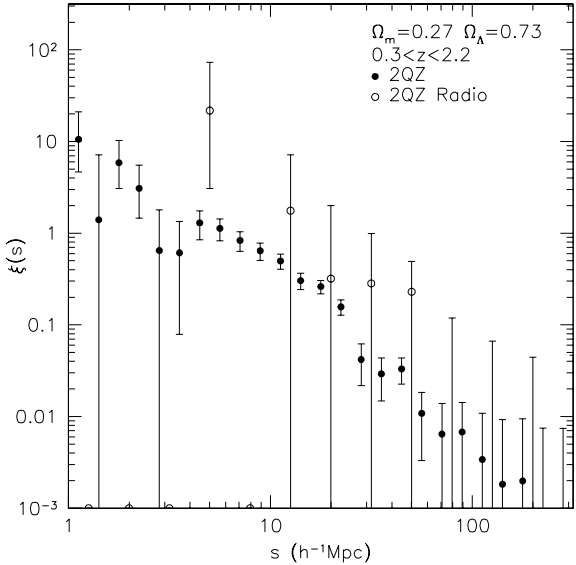,width=8cm}}
\caption{The QSO $\xi(s)$ from the 2QZ (filled points) compared to
 that for NVSS detected 2QZ QSOs (open points).  The radio detected
 $\xi(s)$ uses broader bins of $\Delta \log(s)=0.2$.}
\label{fig:xiradio}
\end{figure}

The redshift averaged QSO $\xi(s)$ from the
2QZ is consistent with the current best fit cosmological model, after
allowing for a linear bias  $b\qso(z=1.35)=2.02\pm0.07$.  We now
compare our results to those from other estimates of $\xi(s)$.  We
find that there is very good agreement between the 2QZ $\xi(s)$ and
2dF Galaxy Redshift Survey (2dFGRS; Hawkins et al. 2003) $\xi(s)$ both
in the shape and amplitude (see Fig. \ref{fig:xisgalcomp}).  We note
that the 2QZ $\xi(s)$ may be slightly flatter than that of the 2dFGRS
on small scales, as would be expected given the smaller influence of
non-linear clustering at high redshift together with the larger impact
of non-linear $z$-space distortions.  However this is not significant.
While the agreement in shape is not particularly surprising, the
impressive match in amplitude is more surprising.  This was also found
in the preliminary 2QZ data release \cite{paper2}.  Considering the
evolution of clustering seen (see Section \ref{sec:evol} below), this must be
considered as something of a coincidence.

A number of authors have measured the spatial clustering of radio
galaxies over a range of redshifts.  Overzier et al. (2003) finds a
real-space clustering scale-length $r_0=14\pm3\Mpc$ at $z\sim1$ for
powerful radio galaxies, while weaker radio sources appear less
clustered, with $r_0\sim4-6\Mpc$.  The clustering of 2QZ QSOs (which
are largely radio quiet) is more similar to the radio weak
sources.  The 2QZ contains a small fraction of sources
detected in the radio.  There are 428 2QZ QSOs in the redshift range
$0.3<z<2.2$ that are detected by the NRAO VLA Radio Survey (NVSS;
Condon et al. 1998).  The $\xi(s)$ we measure for this radio-detected
population is shown in Fig. \ref{fig:xiradio} (open circles).  The
small number of sources and their low surface density means that there
is barely a detection of clustering, with only 2 QSO pairs detected
vs. 1.15 expected at $s<20\Mpc$.  The clustering of radio-detected
QSOs in the 2QZ does not therefore impact on the clustering
measurements of the full sample.  There is a clear difference between
the clustering of radio-quiet QSOs, as sampled by the 2QZ, and
powerful radio galaxies, implying that radio galaxies must exist in
more massive dark matter halos that radio-quiet QSOs.

The low redshift galaxy cluster correlation function has a much higher
amplitude with $s_0$ typically $12-25\Mpc$ depending on the richness
of the clusters \cite{bah03}.  There are few measurements of the
cluster correlation length at high redshift.  Gonzalex, Zaritsky \&
Wechsler (2002) find that approximately velocity dispersion limited
samples of clusters at $z=0.35-0.575$ have similar clustering scale
lengths to local clusters.  For a WMAP/2dF cosmology, linear theory
predicts that the amplitude of mass clustering between
$z=1.35$ and $z=0$ will increase by a factor of $\simeq3.4$, which is
equivalent to an increase in $s_0$ by a factor of 2.0 (assuming
$\gamma=-1.8$).  Hence, even if QSO clustering at a mean redshift of
$z=1.35$ evolved as strongly as linear theory evolution allows (making
no allowance for evolution of bias), the descendents of objects that
contained QSOs at $z\sim1.4$ could not be clustered any more
strongly than poor clusters at low redshift.  Below we make a more
detailed analysis of the evolution of QSO clustering to extend this
analysis.   

\section{The evolution of QSO clustering}\label{sec:evol}

\begin{table*}
\baselineskip=20pt
\begin{center}
\caption{2QZ clustering results as a function of redshift for a
  WMAP/2dF cosmology. All fits are on scales $s=1-25\Mpc$.  We
  list the redshift interval, and mean redshift, apparent magnitude
  and absolute magnitude (assuming $h=0.71$) for each bin together
  with the number of QSOs used.  The best fit values of $s_0$ (in
  comoving units of $\Mpc$) and
  $\gamma$ are given with their $\chi^2$ values, number of dof,
  $\nu$ and probability of acceptance, $P(<\chi^2)$.  Lastly
  we also list the measured values of $\xibar(s)$ for $s=20$, 30 and
  $50\Mpc$.}
\setlength{\tabcolsep}{4pt}
\begin{tabular}{ccccccccccccc}
\hline
$z$ interval & $\overline{z}$ & $\overline{b}_{\rm J}$ & $\overline{M}_{\rm b_{\rm J}}$ & $N\qso$ &
$s_0$ & $\gamma$ & $\chi^2$ & $\nu$ & $P(<\chi^2)$
& $\xibar(20)$ & $\xibar(30)$ & $\xibar(50)$ \\
\hline
0.30,0.68 & 0.526 & 19.85 & --22.16 & 2119 & $ 5.73^{+0.79}_{-0.94}$ & $-1.49^{+0.25}_{-0.25}$ & 15.9 & 10 & 1.02e-01 & $0.263\pm0.075$ & $0.162\pm0.041$ & $0.071\pm0.023$\\
0.68,0.92 & 0.804 & 19.93 & --23.23 & 2067 & $ 3.94^{+1.00}_{-0.98}$ & $-1.15^{+0.24}_{-0.25}$ &  7.2 &  9 & 6.12e-01 & $0.332\pm0.085$ & $0.118\pm0.044$ & $0.020\pm0.022$\\
0.92,1.13 & 1.026 & 19.95 & --23.86 & 2012 & $ 4.76^{+0.97}_{-1.02}$ & $-1.23^{+0.25}_{-0.25}$ &  6.7 &  9 & 6.71e-01 & $0.353\pm0.094$ & $0.146\pm0.048$ & $0.063\pm0.024$\\
1.13,1.32 & 1.225 & 19.97 & --24.27 & 2066 & $ 5.52^{+0.98}_{-1.00}$ & $-1.04^{+0.25}_{-0.25}$ &  8.0 &  8 & 4.29e-01 & $0.511\pm0.100$ & $0.226\pm0.050$ & $0.082\pm0.024$\\
1.32,1.50 & 1.413 & 20.02 & --24.57 & 2063 & $ 5.28^{+0.98}_{-1.00}$ & $-1.04^{+0.25}_{-0.25}$ &  3.4 &  7 & 8.51e-01 & $0.452\pm0.099$ & $0.211\pm0.050$ & $0.064\pm0.023$\\
1.50,1.66 & 1.579 & 20.02 & --24.82 & 2011 & $ 4.87^{+0.95}_{-1.02}$ & $-0.94^{+0.25}_{-0.24}$ &  4.3 &  7 & 7.43e-01 & $0.379\pm0.096$ & $0.205\pm0.050$ & $0.066\pm0.024$\\
1.66,1.83 & 1.745 & 20.03 & --25.06 & 2044 & $ 6.25^{+0.83}_{-0.85}$ & $-1.80^{+0.24}_{-0.25}$ &  3.5 & 10 & 9.66e-01 & $0.321\pm0.098$ & $0.096\pm0.049$ & $0.045\pm0.023$\\
1.83,2.02 & 1.921 & 20.05 & --25.29 & 2020 & $ 6.39^{+0.98}_{-1.00}$ & $-1.09^{+0.25}_{-0.25}$ &  3.7 &  9 & 9.29e-01 & $0.483\pm0.111$ & $0.260\pm0.057$ & $0.100\pm0.026$\\
2.02,2.25 & 2.131 & 20.07 & --25.51 & 2049 & $ 8.00^{+0.99}_{-1.00}$ & $-1.17^{+0.25}_{-0.25}$ &  5.6 &  9 & 7.82e-01 & $0.607\pm0.128$ & $0.249\pm0.063$ & $0.074\pm0.028$\\
2.25,2.90 & 2.475 & 20.09 & --25.86 & 2235 & $ 8.81^{+0.98}_{-1.01}$ & $-1.24^{+0.25}_{-0.25}$ &  5.9 &  7 & 5.56e-01 & $0.701\pm0.174$ & $0.289\pm0.086$ & $0.144\pm0.039$\\
\hline
\label{tab:fitpar}
\end{tabular}
\end{center}
\end{table*}

Above we have calculated $\xi(s)$ average over a broad redshift
range.  Under the assumption that QSO bias is largely scale
independent (at least compared to the uncertainties in the clustering
measurements) this should preserve the correct underlying shape of
$\xi(s)$, particularly on large scales.  However, according to the
standard picture of gravitational growth of structure, the mass
distribution should evolve with redshift.  Croom et al. (2001a) showed
that QSO clustering was constant or slightly increasing with redshift,
with $s_0\simeq5\Mpc$ up to $z\sim2.5$.  This demonstrated that QSOs must
be biased tracers of the matter distribution, and that the amount of
bias must evolve with redshift.  Below we repeat this analysis with
the final 2QZ data set, and discuss in detail the implications for QSO
formation models.  We will assume a WMAP/2dF cosmology unless stated
otherwise. 

\subsection{Measurements of $\xi(s,z)$}

\begin{figure*}
\centering
\centerline{\psfig{file=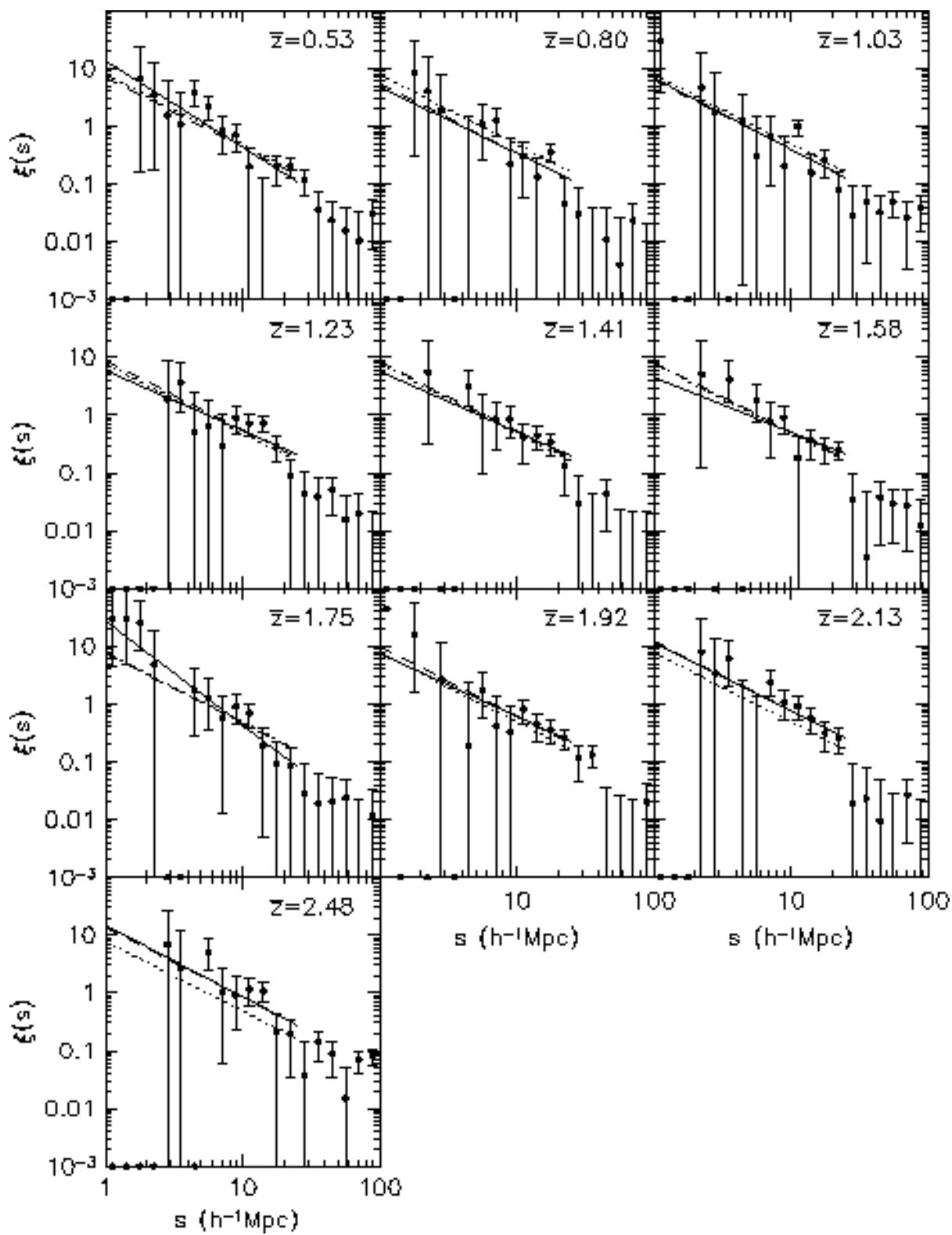,width=16cm}}
\caption{The QSO $\xi(s)$ from the 2QZ (filled points) as a function
  of redshift in 10 redshift bins containing approximately 2000 QSOs
  each.  The best fit power law is shown in each case (solid line), as
  well as the best fit fixing $\gamma$ to be 1.20 (dashed lines).  We
  also show the best fit power law for the full redshift range
  ($0.3<z<2.2$) for comparison (dotted line).  A WMAP/2dF cosmology
  is assumed.}
\label{fig:xi_zbins}
\end{figure*}

We split the QSOs up into 10 redshift intervals, such that there are
approximately equal numbers of QSOs ($\sim2000$) in each bin.  Here we
sample the redshift range $0.3<z<2.9$ and note that the final redshift
interval $z=2.25-2.90$ could be affected by systematic variations in
completeness on large scales. We
perform the correlation analysis as described above on each of these
sub-samples.  In particular we use the mask method to correct for
incompleteness, as the RA-Dec mixing method was shown to significantly
suppress clustering measurements in narrow redshift intervals (see
Section \ref{sec:maskrand}).  We do, however, perform tests with the
RA-Dec and RA-Dec-z mixing methods to confirm that there are no
obvious unaccounted for systematic errors in our analysis.  The
resulting correlation functions are plotted in
Fig. \ref{fig:xi_zbins}.

\begin{figure*}
\centering
\centerline{\psfig{file=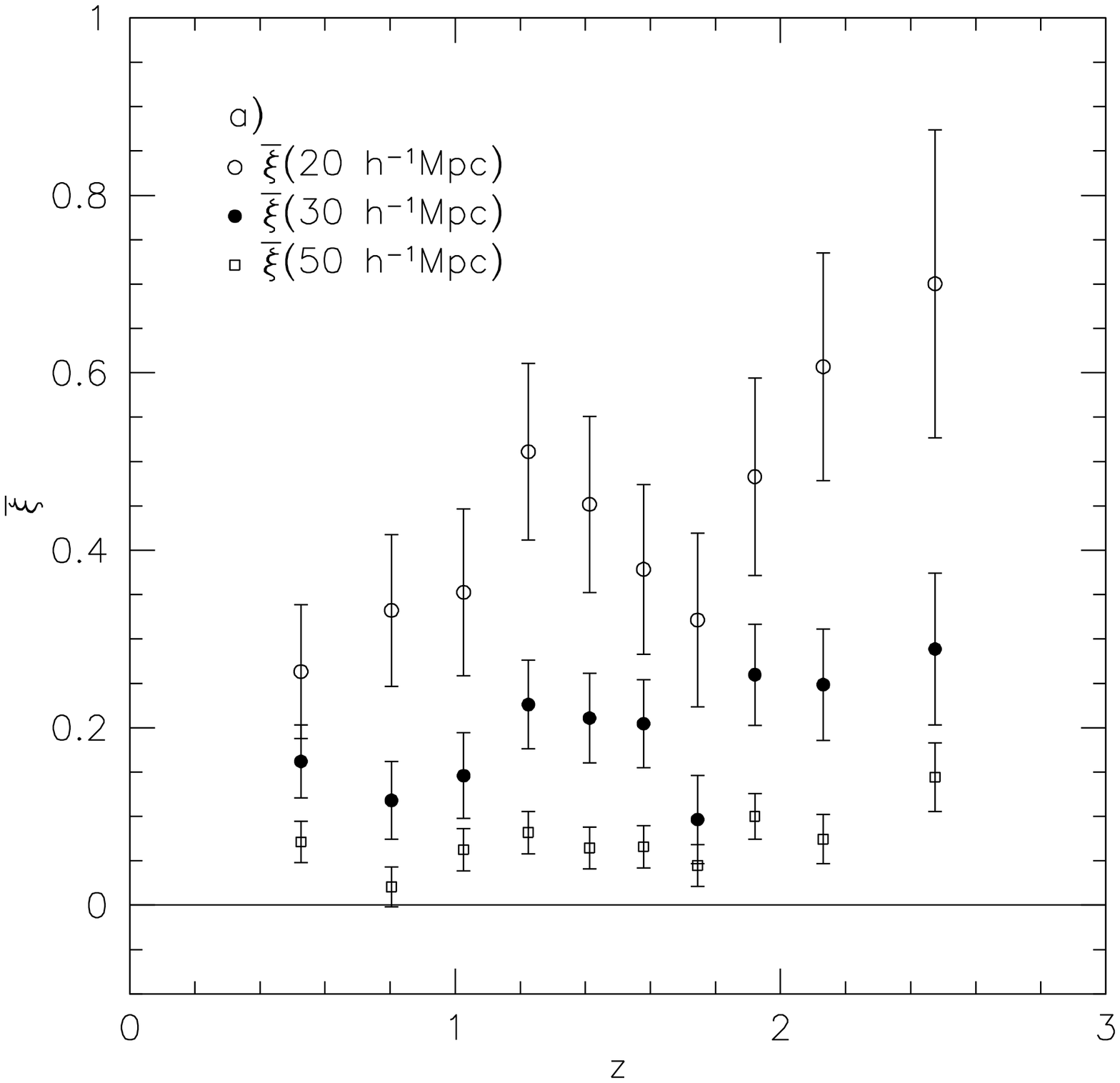,width=8cm}
\psfig{file=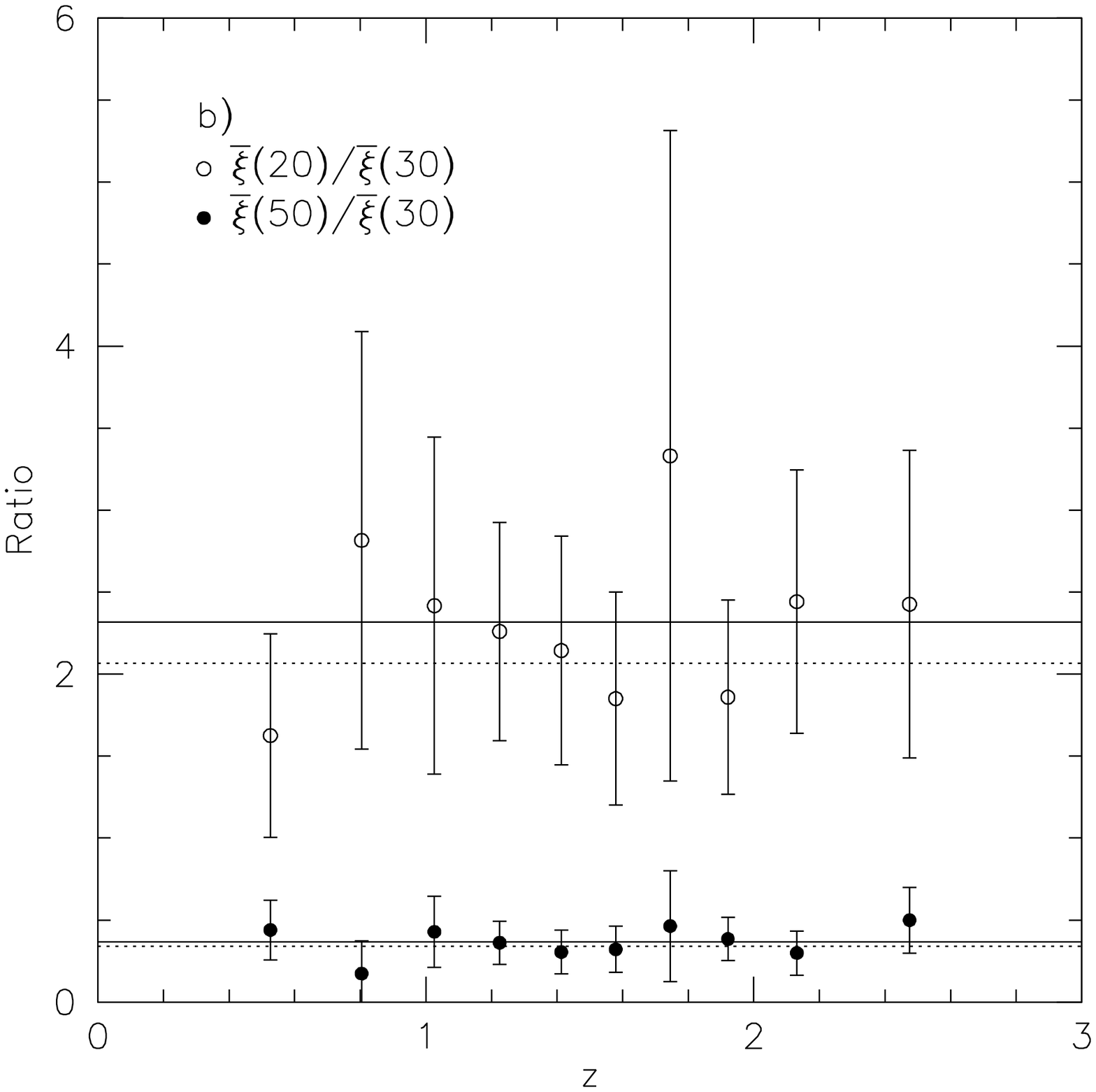,width=8cm}}
\caption{a) The evolution of $\xibar(s)$ for three different values of
  $s=20$, 30 and $50\Mpc$ (open circles, filled circles and open
  squares respectively). There is evidence for an increase in
  $\xibar$ with increasing redshift in all cases.  b) The ratios of
  $\xibar$ as a function of redshift for $\xibar(20)/\xibar(30)$ (open
  circles) and $\xibar(50)/\xibar(30)$ (filled circles).  The redshift
  averaged mean values for the ratios are indicated by the solid lines.
  The ratios are consistent with an unchanging shape for $\xi(s)$.
  The WMAP/2dF cosmology is assumed.  Also plotted are the expected
  ratios for a CDM model with WMAP/2dF parameters (dotted lines).}
\label{fig:xibarevol}
\end{figure*}

In order to make quantitative measure of the clustering properties we
calculate $\xibar(20)$ (Eq. \ref{eq:xibar}) for each redshift
interval.  To test for any evidence of a change in shape of $\xi(s)$
we also calculate $\xibar$ using radii of $30$ and $50\Mpc$.  The
evolution of $\xibar$ is plotted in Fig. \ref{fig:xibarevol}a using
all three scales (the values are also listed in Table
\ref{tab:fitpar}).  In each case there is a general trend for $\xibar$ 
to increase with redshift.  To assess the significance of the
evolution we perform a Spearman rank correlation test on the $\xibar$
values.  We find Spearman rank-order correlation coefficients, $\rho=
0.721$, $0.648$ and $0.552$ for $\xibar$ determined at a radius of 20,
30 and $50\Mpc$ respectively.  These correspond to correlation
significances of $98.1$, $95.7$ and $90.2$ per cent.  We note, of
course, that as these are integral measures they are not independent of
each other.  The above test implies a significant correlation with
redshift, however the data are still found to be consistent (via a
$\chi^2$ test) with a single parameter model which is constant with
redshift (only rejected at the 81, 77 and 75 per cent levels for
$\xibar(20)$, $\xibar(30)$ and $\xibar(50)$ respectively). 

\begin{figure}
\centering
\centerline{\psfig{file=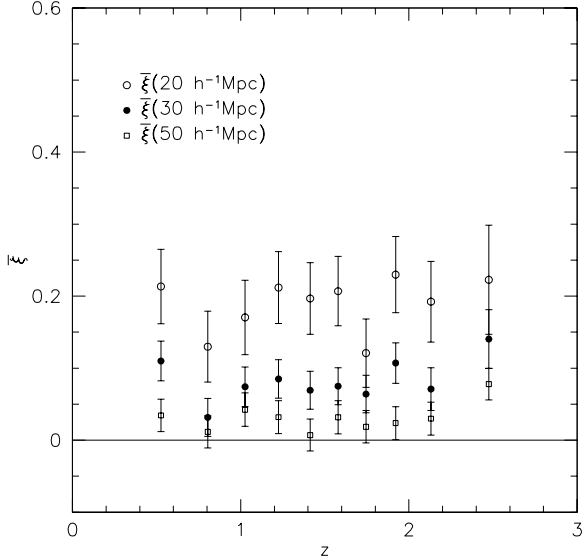,width=8cm}}
\caption{The evolution of $\xibar(s)$ for three different values of
  $s=20$, 30 and $50\Mpc$ (open circles, filled circles and open
  squares respectively) in an EdS cosmology.}
\label{fig:xibarevoleds}
\end{figure}

In Fig. \ref{fig:xibarevol}b we show the
ratio of $\xibar(20)/\xibar(30)$ and $\xibar(50)/\xibar(30)$ to
provide a simple test for any evidence of a change in the shape of
$\xi(s)$ with redshift.  These ratios are consistent with being
constant over the full redshift range of the data set, suggesting that
the shape of $\xi(s)$ does not change significantly with redshift.  We
also compare the $\xibar$ ratios to those assuming a CDM power
spectrum in a WMAP/2dF cosmology (dotted lines in
Fig. \ref{fig:xibarevol}b).  These are fully consistent with the
observed ratios.  In 
Fig. \ref{fig:xibarevoleds} we show the evolution of $\xibar(s)$ for
an EdS cosmology.  In this cosmology clustering is completely constant
as a function of redshift, a Spearman rank correlation test shows no
significant correlation.

\begin{figure}
\centering
\centerline{\psfig{file=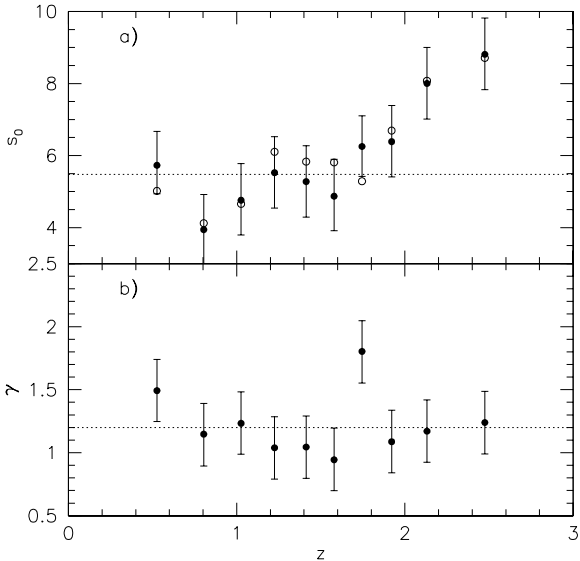,width=8cm}}
\caption{a) The fitted values of $s_0$ with freely varying $\gamma$
  (filled points) and fixed $\gamma$ (open points) as a function of
  redshift. b) The best fit values of $\gamma$ as a function of
  redshift.  The dotted lines indicate the best fit values to the full
  redshift range.  A WMAP/2dF cosmology is assumed.}   
\label{fig:s0gamevol}
\end{figure}

We next fit a simple power law model (Eq. \ref{eq:xipowerlaw}).  In
Section \ref{sec:modelfits} we find that a power law is an acceptable
fit to the redshift averaged QSO $\xi(s)$ on scales $s=1-25\Mpc$.  We
therefore fit the data sub-divided into redshift intervals over the
same range of scales.  The best fit $s_0$ and $\gamma$ values are
shown in Fig. \ref{fig:s0gamevol} (and listed in Table \ref{tab:fitpar}).
We carry out a Spearman rank test on both $s_0$ and $\gamma$
vs. redshift.  For $s_0$ we find $\rho=0.770$ (99 per cent
significant), while for $\gamma$ we find $\rho=-0.030$ (7 per cent
significant).  The measured values of $s_0$ are inconsistent with a
constant value at 98 per cent significance.  Given the lack of evolution in
$\gamma$ we now fix its value and re-perform the fitting.  For this we use the
best fit power law slope  of $\gamma=1.20$.  The $s_0$ values
derived are plotted in Fig. \ref{fig:s0gamevol} (open points).  These
are similar to those found when allowing $\gamma$ to vary freely.  A
Spearman rank correlation test confirms that the correlation is still
present with $\rho=0.842$ significant at the 99.8 per cent level.

Examining the highest redshift bin in Fig. \ref{fig:xi_zbins} we see
that there is significant signal at scales $\sim70-100\Mpc$.  This
redshift interval at $2.25<z<2.90$ has a large
variation in completeness with redshift, as the absorption due to the
Lyman-$\alpha$ forest quickly moves the mean QSO colours into the
stellar locus (see Paper XII).  We do not need to calculate the
absolute completeness in each redshift interval, as we rely on fitting
to the observed shape of the QSO $n(z)$ relation.  However, if this
fit is not accurate enough over a given redshift interval, or there are
systematic differences in the $n(z)$ covering different regions of the
2QZ survey, extra spurious large-scale structure could be added.  We
test for the presence of any such systematic affect by first
calculating the $\xi(s,z)$ using RA-Dec-z mixing.  This produces
estimates of $\xi(s)$ which are systematically biased low (see Section
\ref{sec:maskrand}), however any broad trends should still be
present.  We find that the highest redshift bin still has the largest
best fit value of $s_0$ using these mixing methods.  As a second test
we calculate 
$\xi(s)$ for the $2.25<z<2.90$ interval by normalizing the total
number and the redshift distribution of the random points within each
UKST field.  This would remove the affects of any UKST photometric 
zero-points errors or the differential affects of variability on
completeness in different fields.  The results of this analysis are
indistinguishable from those using masking and the full 2QZ strips.
While it is possible that this excess large-scale structure is still
caused by systematic error, its size does not influence any of our
main results below.  Infact the final redshift bin could be completely
ignored without changing our basic conclusions.

\subsection{Comparison to simple models}

\begin{figure}
\centering
\centerline{\psfig{file=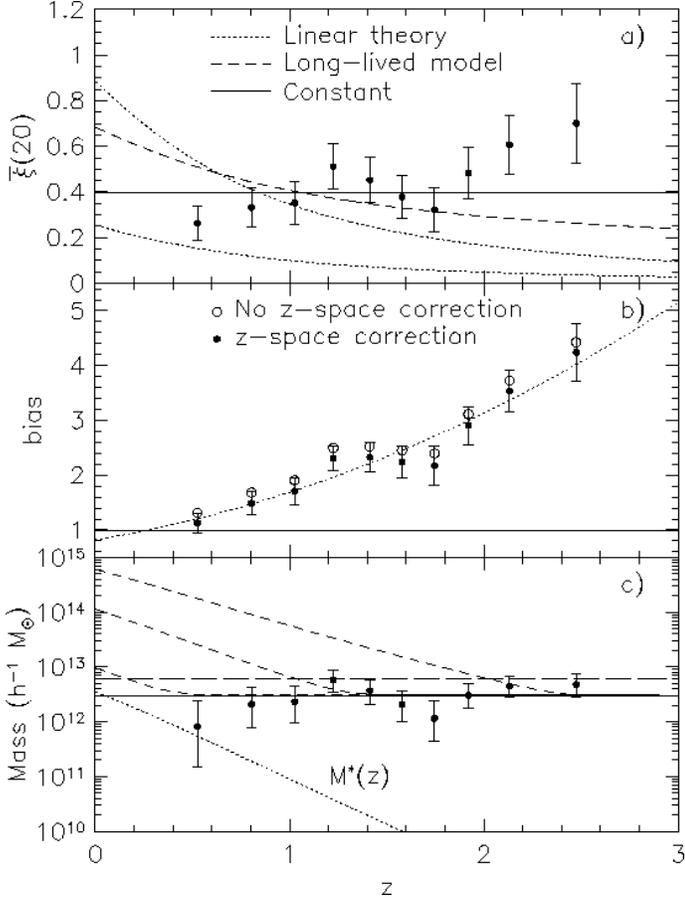,width=9cm}}
\caption{a) Our measurement of $\xibar(20)$ for 2QZ QSOs as a function
  of redshift (filled points).  The data are compared to linear theory
  gravitational evolution (dotted lines) for two normalizations, one
  normalized to a WMAP/2dF cosmology (lower dotted line) and a second
  normalized to provide a 'best fit' to the data points (upper dotted
  line).  We also compare to the best fit for a constant $\xibar(20)$
  (solid line) and a long-lived model (short dashed line).  b) The QSO
  bias, $b\qso(z)$ as
  a function of redshift derived from a comparison of $\xibar(20)$ for
  QSOs to that expected for the WMAP/2dF cosmology.  The open points
  are the raw bias values [i.e. $\xibar\qso(s)/\xibar_\rho(r)$] while
  the filled points with error bars are the values after making a
  consistent correction for $z$-space distortions.  A simple
  empirical model is also shown (dotted line).  c) The mean
  mass of DMHs containing QSOs derived from the measured
  bias (filled points).  We also show
  the mean mass averaged over redshift (solid line) and the mean plus
  twice the rms of the points (long dashed line).  $M^*(z)$, the
  characteristic mass which is just collapsing at a given redshift is
  denoted by a dotted line.  The short dashed lines show the median
  expected growth in DMH mass from the mean DMH mass of QSO hosts at
  $z=0.53$, 1.41 and 2.48.}    
\label{fig:xibiasmass}
\end{figure}

Following Paper II we
test a number of simple models against the observed data.  To be
conservative we use the $\xibar(20)$ measurements, rather than the
best fit $s_0$ values which are dependent on the range of scales fit
and assumptions concerning the slope, $\gamma$.  We note that removing the
highest redshift point does not remove the detected correlation
between $\xibar(20)$ and redshift, although it does reduce its
significance ($\rho=0.617$, significant at the 92 per cent level).
The significance of the correlations of $\xibar(30)$ and $\xibar(50)$
with redshift are also reduced removing when the highest redshift
point is removed (to 85 and 69 per cent respectively).

We compare our results to the expected growth in density perturbations
from linear theory, which should be applicable on the scales we are
probing.  For an EdS universe, the linear growth rate, $D(z)$, is
given by $D(z)=1/(1+z)$, and for other cosmologies we use the accurate
fitting formula of Carroll, Press \& Turner (1992).  In
Fig. \ref{fig:xibiasmass}a we plot the measured $\xibar(20)$ for QSOs
vs. linear theory models (dotted lines).  We assume a CDM model with
WMAP/2dF parameters.  In this model the values of $\xibar(r,z=0)$ for the
mass distribution are 0.254, 0.123 and 0.042 for $r=20$, 30 and
$50\Mpc$ respectively.  We
plot two linear theory lines, the first (lower dotted line) assumes
the above normalization given by WMAP/2dF, which is
significantly below the points at all redshifts.  The second (upper
dotted line) is the linear theory model re-normalized by a constant
bias to a 'best fit' value for the data points.  As in Croom et
al. (2001a) we
find linear theory evolution with a fixed bias to be in clear
disagreement with the data (the probability of acceptance is formally
$3.6\times10^{-9}$).  Assuming an EdS cosmology, we also get a
rejection of QSOs following linear theory evolution (rejected at the
99.98 per cent level).  We next fit the long-lived QSO model discussed
by Croom et al. (2001a) which has the form 
\begin{equation}
b\qso(z)=1+[b\qso(z=0)-1]/D(z).
\end{equation}
This model is equivalent to assuming that QSOs have ages of order the
Hubble time, and after formation at some arbitrarily high redshift
subsequent evolution is governed by their motion within the
gravitational potential \cite{f96}.  It is also equivalent to QSOs
forming in density peaks above a constant threshold \cite{cs96}.  The
best fit value of $b\qso(z=0)=1.64\pm0.05$ (short dashed
lines in Fig. \ref{fig:xibiasmass}a), however, while Croom et
al. (2001a) found this model was marginally acceptable in a cosmology
with $\Omm=0.3$ and $\Omlam=0.7$ we find that the extra signal in the
final 2QZ data set rejects the long-lived model at a significance
level of 99.97 per cent in the WMAP/2dF cosmology.  Fitting this model
in the EdS Universe gives $b\qso(z=0)=1.40\pm0.04$, and is marginally
acceptable (rejected at the 89 per cent level).

\subsection{Bias, dark matter halo mass and the evolution of QSOs}

\begin{figure}
\centering
\centerline{\psfig{file=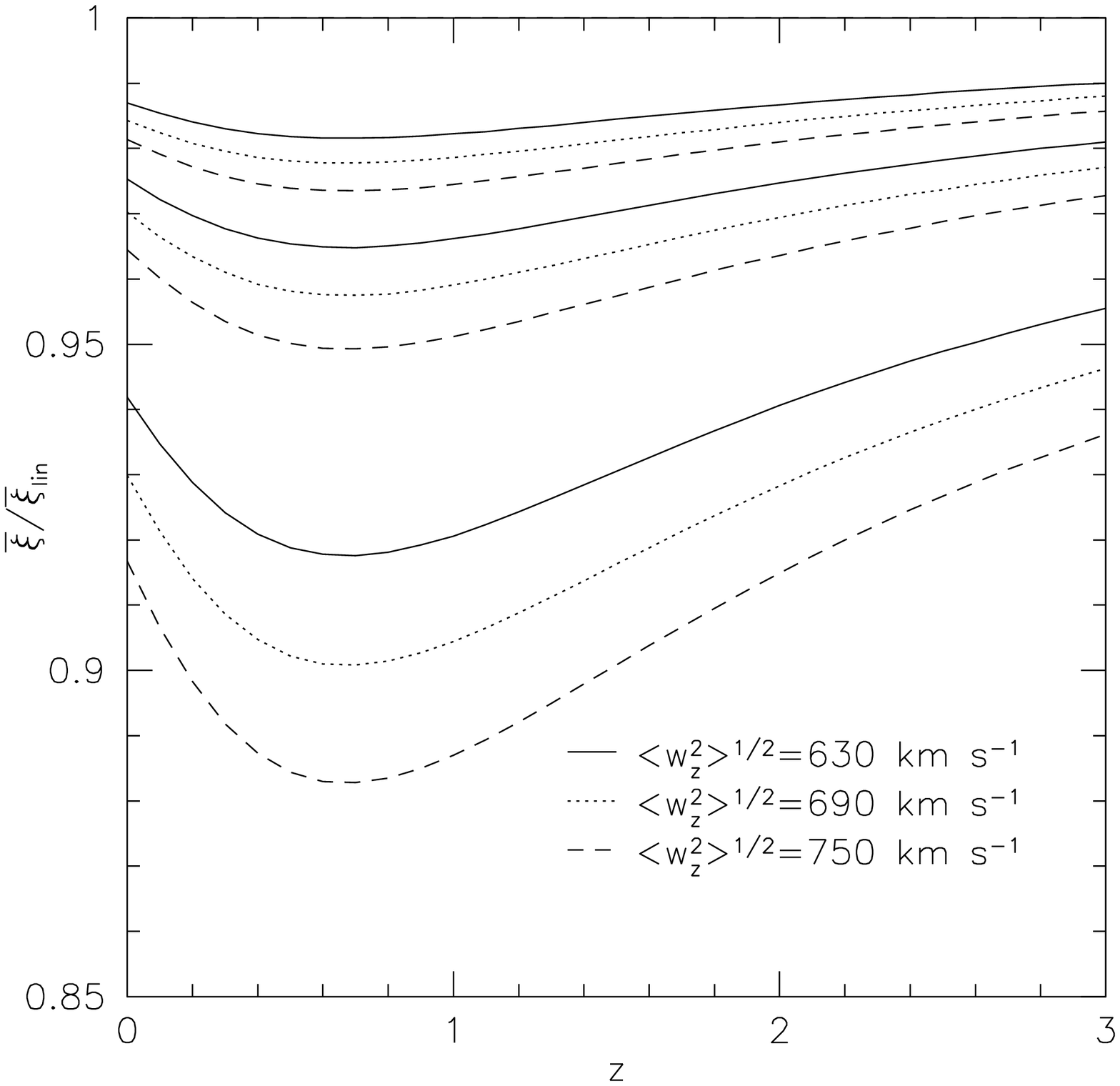,width=8cm}}
\caption{The ratio of $\xibar\nonlin(s)/\xibar\lin(s)$ ($\xibar$
  including and not including non-linear $z$-space distortions) as a
  function of redshift for $\wrms=630$, 690 and $750\kms$
  (solid, dotted and dashed respectively) at scales $s=20$, 30 and
  $50\Mpc$ (bottom to top).} 
\label{fig:xibarmodeldiff}
\end{figure}

By assuming an underlying cosmological model we are able to convert
the measured values of $\xibar$ to an effective bias by making
comparisons to linear theory evolution.  This allows us to directly
determine QSO bias as a function of redshift.  In doing so, we need to
account for the affect of $z$-space distortions on the measured values
of $\xibar(s)$.  The non-linear $z$-space distortions have a small
affect on the scales we are examining here (see Section
\ref{sec:modelfits}).  To determine their affect on $\xibar(s)$ we
derive the ratio of $\xibar(s)$ with linear and non-linear $z$-space
distortions to that including only the linear distortions,
$\xibar\nonlin(s)/\xibar\lin(s)$.  This is plotted for the CDM model
with WMAP/2dF parameters as a function of redshift for $\wrms=630$,
690 and $750\kms$ in  Fig. \ref{fig:xibarmodeldiff} (solid,
dotted and dashed lines respectively).  In constructing the models we
assume values of $\beta$ that are consistent with the
$\beta(z=1.4)=0.4\pm0.1$ of Outram et al. (2004) and also account for 
the evolution of bias we find below.  This assumption of $\beta$ only
influences the shape of $\xi(\sigma,\pi)$ that is convolved with
Eq. \ref{eq:gaussdisp} to determine the non-linear $z$-space
distortions. Varying the assumed $\beta$ within reasonable limits
results in negligible difference in the
$\xibar\nonlin(s)/\xibar\lin(s)$ ratio (less than 0.5 per cent).
We plot the ratio for $s=20$, 30 and
$50\Mpc$ (top to bottom) and see that even at $s=20\Mpc$ the worst
correction is only 12 per cent.  Assuming $\wrms=690\kms$, the
range of reasonable values for $\wrms$ results in a scatter of
only $\sim2$ per cent at $s=20\Mpc$ and less at larger scales.  This
is considerably smaller than the measurement errors in $\xibar$, and we
therefore use the derived ratio for $\wrms=690\kms$ to correct
our results for non-linear $z$-space affects (dotted lines in
Fig. \ref{fig:xibarmodeldiff}).  Linear $z$-space distortions
(Eq. \ref{eq:kaisereq}) have a more significant affect, (e.g. a factor
of $\sim1.3$ at $z\sim1.4$).  We use Eq. \ref{eq:bz} to
self-consistently determine the QSO bias at a given redshift.

\begin{table}
\baselineskip=20pt
\begin{center}
\caption{The derived QSO bias, $b\qso$ and DMH mass, $\mdh$ as for
  2QZ QSOs at a function of redshift in a WMAP/2dF cosmology.  We also
  list the mean redshift and absolute magnitude of each redshift
  interval, as well as the value of $M_{\rm b_{\rm J}}^*$ derived from
  the polynomial evolution model of Paper XII and the space density of
  QSOs, $\Phi$, found by integrating the QSO luminosity function
  between the apparent magnitude limits of the 2QZ.} 
\setlength{\tabcolsep}{3pt}
\begin{tabular}{cccccc}
\hline
$\overline{z}$ & $\overline{M}_{\rm b_{\rm J}}$ & $M_{\rm b_{\rm J}}^*$ &
$\Phi$ $h^3{\rm Mpc}^{-3}$ & $b\qso$ & $\mdh$ $h^{-1}\msun$\\
\hline
 0.526 & --22.16 & --23.24 & $9.6\times10^{-6}$ & $ 1.13\pm 0.18$ & $ 0.82_{- 0.67}^{+ 1.55}\times10^{12}$\\
 0.804 & --23.23 & --23.94 & $7.6\times10^{-6}$ & $ 1.49\pm 0.21$ & $ 2.09_{- 1.30}^{+ 2.18}\times10^{12}$\\
 1.026 & --23.86 & --24.41 & $6.8\times10^{-6}$ & $ 1.71\pm 0.24$ & $ 2.31_{- 1.37}^{+ 2.23}\times10^{12}$\\
 1.225 & --24.27 & --24.78 & $6.6\times10^{-6}$ & $ 2.31\pm 0.23$ & $ 5.76_{- 2.21}^{+ 2.90}\times10^{12}$\\
 1.413 & --24.57 & --25.07 & $6.3\times10^{-6}$ & $ 2.32\pm 0.27$ & $ 3.69_{- 1.62}^{+ 2.24}\times10^{12}$\\
 1.579 & --24.82 & --25.29 & $6.1\times10^{-6}$ & $ 2.24\pm 0.30$ & $ 2.05_{- 1.07}^{+ 1.61}\times10^{12}$\\
 1.745 & --25.06 & --25.47 & $5.8\times10^{-6}$ & $ 2.17\pm 0.35$ & $ 1.15_{- 0.72}^{+ 1.24}\times10^{12}$\\
 1.921 & --25.29 & --25.61 & $5.3\times10^{-6}$ & $ 2.91\pm 0.35$ & $ 3.05_{- 1.34}^{+ 1.85}\times10^{12}$\\
 2.131 & --25.51 & --25.72 & $4.8\times10^{-6}$ & $ 3.53\pm 0.38$ & $ 4.46_{- 1.68}^{+ 2.20}\times10^{12}$\\
 2.475 & --25.86 & --25.76 & $3.5\times10^{-6}$ & $ 4.24\pm 0.53$ & $ 4.78_{- 1.99}^{+ 2.68}\times10^{12}$\\
\hline
\label{tab:biasmass}
\end{tabular}
\end{center}
\end{table}

Fig. \ref{fig:xibiasmass}b shows the derived bias of 2QZ QSOs as a
function of redshift (filled points).  The open points are the values
found without accounting for $z$-space distortions.  Here we see that
QSO bias is strongly evolving with redshift, from
$b\qso(z=0.53)=1.13\pm0.18$ to $b\qso(z=2.48)=4.24\pm0.53$ (see Table
\ref{tab:biasmass}).  A simple empirical description of the bias
evolution found is
\begin{equation}
b\qso(z)=(0.53\pm0.19)+(0.289\pm0.035)(1+z)^2,
\end{equation}
which is shown in Fig. \ref{fig:xibiasmass}b (dotted line).  At
$z\sim0.5$ the value of $b\qso$ is already close to 1, and
a simple extrapolation of the trend observed would predict that the
bias would at or below 1 at $z=0$.  We note at this point that because
of the apparent magnitude limit of the 2QZ, the mean absolute
magnitude in each interval increases with redshift (see Table
\ref{tab:fitpar}).  However, the 2QZ selects QSOs that are close to 
$\sim L\qso^*$ (the characteristic luminosity of the QSO optical
luminosity function) at every redshift, and the space density of
objects in each of the redshift slices is also approximately equal.
Table \ref{tab:biasmass} lists the values of $M_{\rm b_{\rm J}}^*(z)$
assuming the polynomial evolution model of Paper XII (which is an
uncertain extrapolation beyond $z=2.1$).  Although the actual values of
$M_{\rm b_{\rm J}}^*(z)$ should be used with caution as the fitted
value of $M_{\rm b_{\rm J}}^*(0)$ is correlated with the bright and
faint end slopes of the LF, it can be seen that there is little change
in the relative difference between $M_{\rm b_{\rm J}}^*(z)$ and
$\overline{M}_{\rm b_{\rm J}}(z)$ (less than 1 mag at $z<2.2$).  Also
listed is the space density found by integrating the observed
luminosity function over the apparent magnitude range of the 2QZ for
each redshift.  Between $z=0.5$ and $z=2.1$ there is only a factor of
2 change in space density (increasing to a factor of 2.7 if we include
the highest redshift bin). 
Paper XII found that the extrapolated $\mb^*$ (the absolute
magnitude equivalent of $L\qso^*$) at $z=0$ has in the range
$-20.5$ to $-21.6$ (where the large range is due to correlation
between the value of $\mb^*$ and the bright/faint slopes of the QSO
LF, and uncertainty in the exact model to extrapolate
to zero redshift).  Thus we would expect that at these moderate
luminosities, QSOs (or more properly AGN) would be close to unbiased
at $z=0$.  It has been shown \cite{hawk03,verdi02} that $\sim L\gal^*$
galaxies at low redshift are largely unbiased.  This implies
that typical low redshift AGN (which are much less luminous than those
at high redshift) are clustered similarly to $\sim L\gal^*$ galaxies.
There is some direct evidence that this is the case, as Croom et
al. (2004c) have shown that the cross-correlation between low redshift
2QZ QSOs and 2dFGRS galaxies is equal to the auto-correlation of the
galaxies.

Once the bias is derived it is possible to relate this to the mean
mass of the DMHs that the QSOs reside in.  Halos of a
given mass, $M$, are expected to be clustered differently to the
underlying mass distribution.  Mo \& White (1996) developed the
formalism for relating mass to bias.  This was extended to low mass
halos by Jing (1998).  Both of these works were based on the spherical
collapse model.  Sheth, Mo \& Tormen (2001) extend the formalism to
account for ellipsoidal collapse, to provide an improved relation
between bias and mass.  It is this relation that we will use in our
analysis.  The bias is related to the mass via
\begin{eqnarray}
b(M,z) & = &
       1+\left.\frac{1}{\sqrt{a}\dc(z)}\right[a\nu^2\sqrt{a}+0.5\sqrt{a}(a\nu^2)^{1-c}\nonumber
       \\
       &   & \left.-\frac{(a\nu^2)^c}{(a\nu^2)^c+0.5(1-c)(1-c/2)}\right],
\label{eq:massbias}
\end{eqnarray}
where $\nu=\dc(z)/\sigma(M,z)$, $a=0.707$ and $c=0.6$.  $\dc$ is the
critical overdensity for collapse of a homogeneous spherical
perturbation. For an EdS universe $\dc=0.15(12\pi)^{2/3}\simeq1.69$.
For a general cosmology $\dc$ has a weak dependence on redshift, which
is given by Navarro, Frenk \& White (1997).  $\sigma(M)$ is the rms
fluctuation in the linear density field on a mass scale, $M$, and is
given by
\begin{equation}
\sigma^2(M)=\frac{1}{2\pi^2}\int^\infty_0 k^2P(k)w^2(kr){\rm d}k,
\end{equation}
where $P(k)$ is the power spectrum of density perturbations and
\begin{equation}
w(kr)=\frac{3(kr\sin(kr)-\cos(kr))}{(kr)^3},
\end{equation}
which is the Fourier transform of a spherical top-hat of size
\begin{equation}
r=\left(\frac{3M}{4\pi\rho_0}\right)^{1/3}.
\end{equation}
$\rho_0$ is the mean density of the universe at $z=0$ and corresponds
to $2.78\times10^{11}\Omm h^2 M_\odot~{\rm Mpc}^{-3}$.  $\sigma(M)$ at
$z=0$ is related to that at arbitrary redshift by the linear growth
factor, $D(z)$, such that
\begin{equation}
\sigma(M,z)=\sigma(M)D(z).
\end{equation}
The characteristic mass at any given redshift, $M^*(z)$, that is, the
mass scale which is just collapsing at a given redshift is defined by 
\begin{equation}
\sigma[M^*(z)]=\frac{\dc}{D(z)}.
\end{equation}

We apply Eq. \ref{eq:massbias} to estimate the typical mass of the DMHs
containing our QSOs at each redshift.  This typical mass is
plotted in Fig. \ref{fig:xibiasmass}c.  We find that the typical $\mdh$
of 2QZ QSO hosts is largely constant as a function of redshift, even
though their typical luminosity is increasing at high $z$.  There 
appears a slight tendency for low redshift QSOs to be in lower mass
DMHs, but a Spearman rank test shows no significant correlation between
redshift and $\mdh$ ($\rho=0.467$, significant at only the 83 per cent
level).  The mean mass corresponds to
$\mdh=(3.0\pm1.6)\times10^{12}h^{-1}\msun$ (rms error).  By
comparison, the characteristic mass of the Press-Schechter mass
function \cite{ps74}, $M^*$, is declining quickly at high redshift
(dotted line in Fig. \ref{fig:xibiasmass}c).  $M^*$ halos are
unbiased ($b=1$) at every redshift, with halos more massive than $M^*$
becoming progressively more biased.  We therefore see that the
increasing bias of DMHs hosting 2QZ QSOs towards higher redshift makes them
increasingly more massive than $M^*$.  However, the increase in mass
relative to $M^*$ is almost exactly cancelled out by the evolution of
$M^*$ to give an approximately constant $\mdh$.  We find that $\mdh$
for QSO hosts is in fact very similar to
$M^*(z=0)\simeq3.5\times10^{12}h^{-1}\msun$.  This is effectively the
same result discussed above, that by extrapolation $\L\qso^*$ QSOs would be
largely unbiased at $z\sim0$.  The actual mass derived is dependent on
the exact cosmology used.  Varying our assumed $\sigma_8(z=0)=0.84$
by $\pm0.08$ [the $2\sigma$ range from from analysis of WMAP and
other data \cite{wmap}] gives a range in $\mdh$ between
$(1.52\pm0.86)\times10^{12}h^{-1}\msun$ and
$(5.4\pm2.8)\times10^{12}h^{-1}\msun$ for $\sigma_8(z=0)=0.76$ and
$\sigma_8(z=0)=0.92$ respectively.  Such changes in normalization will
affect all redshift intervals equally, and also scale the value of
$M^*$ by an equal amount.  So although the derived mass might be
different our overall conclusions (in terms of constant $\mdh$ and
$b\qso\simeq1$ at $z=0$) are not affected.  Using a different form for
the relation between $b\qso$ and $\mdh$ also slightly affects out
results.  The relations described by Mo \& White (1996) and Jing
(1998) give a mean $\mdh\simeq(1.9\pm0.9)\times10^{12}h^{-1}\msun$.
These show even less dependence of $\mdh$ with redshift, as the masses
of the highest redshift halos are reduced the most.  We confirm that
similar results are found using the estimates of $\xibar(30)$, these
give a similar non-evolving $\mdh$, with a mean of
$(2.2\pm1.3)\times10^{12}h^{-1}\msun$.  Our mass estimates are
consistent with those derived by Grazian et al. (2004) based on the
QSO clustering results of Croom et al. (2001a).

\subsubsection{The lifetime of QSOs}

\begin{figure}
\centering
\centerline{\psfig{file=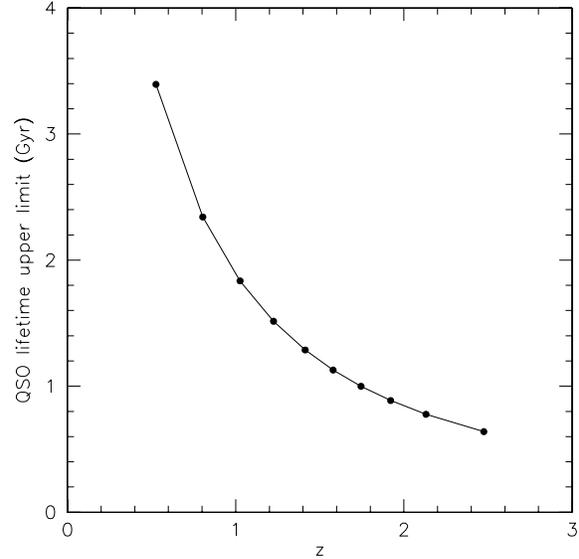,width=8cm}}
\caption{The $2\sigma$ upper limits to QSO lifetime as a function of
  redshift (connected filled circles), based on the growth in mass of
  DMHs.}
\label{fig:lifetimes}
\end{figure}

The observation that 2QZ QSOs sample the same mass DMHs at every
redshift further demonstrates that we cannot be seeing a
cosmologically long lived population.  As the mass of DMHs grow with
time through the process of accretion and merging, the low redshift
descendents of high redshift QSOs will inhabit higher mass DMHs, and
hence the QSOs we observe at high and low redshift cannot be drawn
from the same single coeval population.
We use the formalism for DMH evolution developed by Lacey \&
Cole (1993) to predict the median mass of the descendents of DMHs
hosting QSOs at
later epochs.  Eq. 2.22 of Lacey \& Cole gives the cumulative
probability that a DMH of mass $M_1$ at time $t_1$ will merge to form
a new DMH of mass greater than $M_2$ by time $t_2$.  By finding the
mass, $M_2$, that corresponds to a probability of 0.5 at a given time
$t_2$ we have the median mass of descendent DMHs.  In Fig.
\ref{fig:xibiasmass}c we plot the evolution of the median DMH mass for
a starting mass of $3.0\times10^{12}h^{-1}\msun$ (the mean QSO host
$\mdh$) at $z=0.53$, 1.41 and 2.48 (dashed lines).  At low redshift,
there is only limited time for growth, and the DMHs of QSO hosts at
$z\simeq0.5$
would only have evolved to a mass of $\simeq1\times10^{13}h^{-1}\msun$
at $z=0$.  However, the highest redshift DMHs hosting QSO have more time to
evolve and would have typical masses of $\simeq6\times10^{14}h^{-1}\msun$
at $z=0$.  It therefore appears that 2QZ QSOs at high redshift
($z\sim2$) inhabit the progenitors of low redshift galaxy clusters,
while 2QZ QSOs at lower redshift are located in the progenitors of
galaxy groups.  The growth of $\mdh$ allows us to place
constraints on the allowable lifetime of QSO activity.  Low redshift
QSOs cannot be the same population of objects as at higher redshift if
they have masses which are less than the mass of the high redshift
sources, after accounting for their expected growth over time.
Therefore calculating the time taken to reach the mean QSO host DMH
mass plus twice the measured rms gives a $\sim2\sigma$ limit on the
lifetime of QSO activity (the rms is $1.6\times10^{12}h^{-1}\msun$ and
the long dashed line in Fig. \ref{fig:xibiasmass} shows the mean plus
twice this rms).  The result 
of this is plotted in Fig. \ref{fig:lifetimes} (connected filled
circles).  At high redshift, halos merge more quickly
than at low redshift, therefore we find that the limits on QSO
lifetime using this method are smaller at high redshift than at low
redshift.  At $z=2.48$ the $2\sigma$ upper limit on QSO lifetime is
$6\times10^8$ years, while at redshifts below $z=1.7$, the upper limit
is $\geq1\times10^9$ years.  At $z=0.53$ the limit is $3\times10^9$
years.  

To further constrain QSO lifetimes, a number of authors have produced
models for QSO clustering in order to try and constrain the typical
lifetime of QSOs.  Martini \& Weinberg (2001) give fitting functions
for their models which relate $r_1$, the scale at which the rms
fluctuations in the QSO distribution is 1 (i.e. $\sigma\qso(r_1,z)=1$)
to typical QSO lifetime.  Their model makes some assumptions,
including that the brightest QSOs are always in the most massive halos
at any given redshift and that
the presence of a black hole is the only requirement for QSO activity.
This second assumption may be valid at high redshift $z\gsim2$, but
may not be at low redshift where fueling must be an
issue.  We therefore compare their models to our data for $z=2$ only
and use our two bins at $z=1.92$ and $z=2.13$ to make the
comparisons.  To convert from $\xibar(20)$ to $r_1$ we assume an
underlying CDM power spectrum with the WMAP/2dF parameters.  This
results in $r_1(z=1.92)=9.35^{+1.51}_{-1.69}\Mpc$ and
$r_1(z=2.13)=11.29^{+1.58}_{-1.76}\Mpc$.  We also need to convert
between the space density assumed by  Martini \& Weinberg
($5.27\times10^{-7}$~h$^3$~Mpc$^{-3}$ for $\Omlam=0.7$ and $\Omm=0.3$)
and the measured space density of the 2QZ at $z=2$
($5.1\times10^{-6}$~h$^3$~Mpc$^{-3}$ for the same cosmology).  This
difference increases the estimated lifetimes by a factor of 9.7
compared to those derived by Martini \& Weinberg.  We
then use the Martini \& Weinberg fitting function for lifetimes in a
$\Lambda$CDM Universe ($\sigma_8=0.9$) to find that
$t\qso=9.7^{+9.7}_{-5.8}\times10^6$ years (for the $z=1.93$ point) and
$t\qso=2.4^{+2.4}_{-1.4}\times10^7$ years (for the $z=2.13$ point).
Thus the full range of lifetimes at $z=2$ in this model is
$t\qso\simeq4-50$ Myr.  This range is lower than, but consistent
with the upper limits derived above. 

The above determination of the typical QSO lifetime is the total
period of activity for a single BH, which may be split up into several
episodes of activity.  The short lifetime indicates that there are
many generations of QSOs, and that a large fraction of galaxies pass
through an AGN phase.  The models used by Martini \& Weinberg and
others generally assume that luminosity is perfectly correlated with
host mass, thus more luminous QSOs would be in more massive DMHs and
therefore be more strongly clustered.  We will investigate this below (see
Section \ref{sec:lumdep}).  A scatter in the relation between DMH mass
and QSO luminosity, would tend to increase the effective lifetime, and
thus the estimates from the Martini \& Weinberg models become lower
limits to the QSO lifetime.

\subsubsection{Accretion efficiency and the mass of black holes}

\begin{figure}
\centering
\centerline{\psfig{file=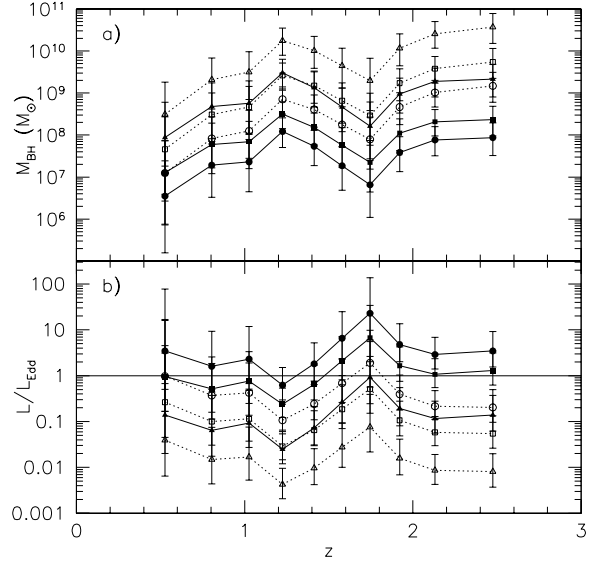,width=8cm}}
\caption{a) The estimated $\mbh$ based on the relations of Ferrarese
  (2002) (points connected by solid lines) and Wyithe \& Loeb (2004)
  (points connected by dotted lines).  We show estimates of $\mbh$
  based on Eq. \ref{eq:bhmass1} (filled circles) Eq. \ref{eq:bhmass2}
  (filled squares) and Eq. \ref{eq:bhmass3} (filled triangles) for the
  Ferrarese (2002) relations and for $\epsilon=\epsilon_{\rm SIS}$
  (open circles), $3.7\epsilon_{\rm SIS}$ (open squares) and
  $25\epsilon_{\rm SIS}$ (open triangles).  b) The derived accretion
  efficiency, $L/L_{\rm Edd}$, from the above $\mbh$ estimates,  using
  the same symbols at in the plot of $\mbh$.}
\label{fig:bhmass}
\end{figure}

There is strong evidence for a correlation between bulge velocity
dispersion, $\sigma_{\rm c}$ and central BH mass \cite{geb00a,fer00}.
This has
been extended to a correlation between $\mbh$ and $\mdh$ by
Ferrarese (2002).  The exact connection is uncertain, largely due to
uncertainty in the DMH density profile. Ferrarese suggests three
possible relations, covering the likely range of allowable assumptions:
\begin{equation}
\frac{\mbh}{10^8 M_\odot}\sim0.027\left(\frac{\mdh}{10^{12}M_\odot}\right)^{1.82}
\label{eq:bhmass1}
\end{equation}
for an isothermal dark matter profile, 
\begin{equation}
\frac{\mbh}{10^8 M_\odot}\sim0.1\left(\frac{\mdh}{10^{12}M_\odot}\right)^{1.65}
\label{eq:bhmass2}
\end{equation}
for an NFW profile \cite{nfw97} and
\begin{equation}
\frac{\mbh}{10^8 M_\odot}\sim0.67\left(\frac{\mdh}{10^{12}M_\odot}\right)^{1.82}
\label{eq:bhmass3}
\end{equation}
for a profile based on the weak lensing results of Seljak (2002)
(henceforth S02).
If we assume that these relations do not evolve with redshift, then we
can directly estimate the central BH mass of the DMHs hosting the 2QZ
QSOs.  These BH mass estimates are shown in Fig. \ref{fig:bhmass}a
(points connected by solid lines).  We assume $h=0.71$ in order to
convert from $h^{-1}\msun$ to $\msun$.  As a comparison we also plot
estimates of $\mbh$ assuming the model of Wyithe \& Loeb (2004) in
which it is the relation between velocity dispersion (or circular
velocity) and $\mbh$, $\mbh-\sigma_{\rm c}$, which is constant with
redshift \cite{shields03}.  This results in a relation between $\mdh$
and $\mbh$ of the form
\begin{equation}
\mbh=\epsilon\mdh\left(\frac{\mdh}{10^{12}M_\odot}\right)^{2/3}\left[\frac{\Delta_{\rm
  c}\Omm(0)}{18\pi^2\Omm(z)}\right]^{5/6}(1+z)^{5/2},
\end{equation}
where $\epsilon$ is a constant and 
\begin{equation}
\Delta_{\rm c}=18\pi^2+82(\Omm(z)-1)-39(\Omm(z)-1)^2.
\end{equation}
The constant $\epsilon$ depends on the density profile of the DMH and
based on the work of Ferrarese (2002) Wyithe \& Loeb suggest that for
the assumption of a singular isothermal sphere $\epsilon_{\rm
  SIS}\simeq10^{-5.1}$.  For a NFW profile
$\epsilon=3.7\epsilon_{\rm SIS}$ and for an S02 profile
$\epsilon=25\epsilon_{\rm SIS}$.
These models with, $\epsilon=\epsilon_{\rm SIS}$, $3.7\epsilon_{\rm
  SIS}$ and $25\epsilon_{\rm SIS}$ (which are direct analogues of
Eqs. \ref{eq:bhmass1}, \ref{eq:bhmass2} and \ref{eq:bhmass3} for the
case of a non-evolving $\mbh-\sigma_{\rm c}$) are plotted in
Fig. \ref{fig:bhmass}a (points connected by dotted lines).
Examination of this plot shows that models in which the
$\mbh-\sigma_{\rm c}$ is independent of redshift predict higher mass
BHs, and a significant increase in $\mbh$ with redshift for 2QZ QSOs.
The masses in this case are a factor $\sim50-100$ greater at $z=2.5$
than they are at $z=0.5$.  In contrast, for the assumption that
$\mbh-\mdh$ is independent  of redshift, there is a much weaker
trend of increasing $\mbh$. 

Given the known mean absolute magnitude of each redshift interval, we
can then calculate the accretion efficiency, $L/L\edd$, where $L$ is the
bolometric luminosity of the QSOs and $L\edd$ is the Eddington
luminosity [$L\edd=10^{39.1}(M/10^8M_\odot)$W].  To determine the
bolometric luminosity we convert from absolute magnitude in the $\bj$
band using the relation derived by McLure \& Dunlop (2004) for the $B$
band and correcting by $\bj=B-0.06$ for a mean QSO $B-V=0.22$
\cite{cv90}.  The relation is then
\begin{equation}
\mb=-2.66\log(L)+79.42
\end{equation}
for $L$ in Watts.  The resulting accretion efficiencies are shown in
Fig. \ref{fig:bhmass}b.  In some cases the mean efficiency of the
population is found to be super-Eddington.  If the Eddington limit is
a meaningful constraint on the accretion of matter onto super-massive
BHs, then the $\mbh-\mdh$ relations described by Eqs. \ref{eq:bhmass1}
and \ref{eq:bhmass2} are unlikely to hold at high redshift, as they
predict that accretion that is significantly super-Eddington.  For the
relation described by Eq. \ref{eq:bhmass3}, $L/L\edd$ evolves
little and is at $\sim0.1$ at all redshifts.  There is also little
evidence of evolution for the cases in which $\mbh-\sigma_{\rm c}$ is
independent of redshift (connected by dotted lines).  The values for
$L/L\edd$ range between $L/L\edd\sim1$ and $\sim0.01$ depending
on the value of $\epsilon$ assumed.  The more realistic values of
$\epsilon$ ($3.7\epsilon_{\rm SIS}$ and $25\epsilon_{\rm SIS}$) imply a
lower accretion efficiency.  We note that Wyithe \& Loeb (2004) have
fit models to the QSO clustering results presented by Croom et
al. (2001a).  They suggest that a model where $\mbh-\sigma_{\rm c}$ is
independent of redshift is preferred from this data, however, this
assumes that the accretion efficiency is not a function of redshift.  

An independent estimate of $\mbh$ is available by invoking the virial
theorem in the QSO broad line region and using the widths of broad
lines as a direct probe of the kinematics.  Authors have carried out
this analysis on both the 2QZ \cite{corbett03} and SDSS \cite{md04}.
There are a number of assumptions in these analysis.  The most crucial
of which is the radius-luminosity relation for broad line regions
\cite{kaspi00}.  This is generally assumed to be independent of
redshift, although this has not been demonstrated observationally.
These works provide a relatively independent comparison to the present analysis. 
Corbett et al. (2003) find little evidence of any evolution of
$L/L\edd$ in the 2QZ.  McLure \& Dunlop (2004), also find only weak
evolution in $L/L\edd$ for the SDSS.  Note that both of these samples
are flux limited so that higher luminosity QSOs are at higher
redshift, however, it is then still true that QSOs with $L\sim L^*\qso$
have little evolution in $L/L\edd$.

This implies that the evolution in luminosity of $L^*\qso$ QSOs is not
caused by a decline in fuelling, but rather, by less massive BHs
becoming active at lower redshift.  It is also possible that the
observed break in the QSO LF (see Paper XII) may be due to the 
difficulty of accreting with an efficiency above some limit
(presumably close to the Eddington limit).  However, the shape of the 
QSO LF is likely driven by a combination of accretion rate and $\mbh$.
Any spread in accretion rate for a given $\mbh$ would suppress any
luminosity dependence of QSO clustering.  We will investigate this
issue in the next Section.

\section{The luminosity dependence of QSO clustering}\label{sec:lumdep}

\begin{table*}
\baselineskip=20pt
\begin{center}
\caption{2QZ/6QZ clustering results as a function of apparent magnitude,
  $\bj$, for a WMAP/2dF cosmology. All fits are over scales
  $s=1-25\Mpc$.  We list the $\bj$ interval and the mean redshift,
  apparent magnitude and absolute magnitude for each bin together with
  the number of QSOs used.  We also give the value of $M_{\rm b_{\rm
  J}}^*$ at the mean redshift of each sample derived assuming the
  polynomial evolution model of Paper XII.  The best fit values of
  $s_0$ (in units of 
  $\Mpc$) and $\gamma$ are given with their $\chi^2$ values, number of
  dof, $\nu$ and probability of acceptance, $P(<\chi^2)$.  Lastly
  we also list the measured values of $\xibar(s)$ for $s=20$, 30 and
  $50\Mpc$.  We do not fit a power law to the brightest magnitude bin
  (6QZ data) as there are to few QSO-QSO pairs to make a reliable fit,
  we also don't list $\xibar(20)$ for this sample, as there are no
  pairs found on scales $<20\Mpc$.}
\setlength{\tabcolsep}{3pt}
\begin{tabular}{cccccccccccccc}
\hline
$\bj$ interval & $\bar{z}$ & $\bar{b}_{\rm J}$ & $\bar{M}_{\rm b_{\rm
    J}}$ & $M_{\rm b_{\rm J}}^*$ & $N\qso$ & $s_0$ & $\gamma$ &
$\chi^2$ & $\nu$ & $P(<\chi^2)$ & $\xibar(20)$ & $\xibar(30)$ & $\xibar(50)$ \\
\hline
16.00,18.25  & 1.063 & 17.81 & --25.73 & --24.48 &  275  &            --           &          --             &  --  & -- &    --    &       --        & $0.58\pm0.71$ & $-0.01\pm0.26$\\
18.25,19.45  & 1.261 & 19.02 & --25.02 & --24.84 & 3586 & $ 3.14^{+2.86}_{-3.08}$ & $-0.83^{+0.62}_{-0.55}$ &  3.2 &  6 & 7.83e-01 & $0.378\pm0.150$ & $0.140\pm0.078$ & $0.039\pm0.036$\\
19.45,19.90  & 1.336 & 19.69 & --24.53 & --24.96 & 3521 & $ 8.06^{+1.42}_{-1.53}$ & $-1.53^{+0.34}_{-0.32}$ &  3.2 &  8 & 9.23e-01 & $0.588\pm0.175$ & $0.209\pm0.084$ & $0.058\pm0.038$\\
19.90,20.25  & 1.369 & 20.09 & --24.22 & --25.01 & 3624 & $ 4.81^{+1.43}_{-1.39}$ & $-1.76^{+0.57}_{-1.05}$ &  5.5 &  7 & 6.02e-01 & $0.103\pm0.139$ & $0.121\pm0.078$ & $0.042\pm0.036$\\
20.25,20.55  & 1.384 & 20.40 & --23.93 & --25.03 & 3563 & $ 0.90^{+3.91}_{-0.84}$ & $-0.52^{+0.32}_{-0.76}$ &  1.7 &  6 & 9.43e-01 & $0.303\pm0.156$ & $0.167\pm0.083$ & $0.115\pm0.039$\\
20.55,20.85  & 1.405 & 20.70 & --23.67 & --25.06 & 3772 & $ 4.68^{+2.89}_{-4.62}$ & $-0.76^{+0.60}_{-0.46}$ &  4.4 &  7 & 7.34e-01 & $0.515\pm0.158$ & $0.167\pm0.077$ & $0.100\pm0.036$\\

\hline
\label{tab:fitparmag}
\end{tabular}
\end{center}
\end{table*}

In this section we investigate whether there is any evidence for QSO
clustering being dependent on luminosity.  There is evidence that low
redshift AGN have nuclear luminosities that are correlated with host
galaxy luminosity (e.g. Schade, Boyle \& Letawsky 2000), and in
particular with the luminosity of the bulge/spheroid component of the
host.  It has also been shown that galaxy clustering is a strong
function of luminosity brighter than $L^*\gal$ (e.g. Norberg et
al. 2001).  Thus bright QSOs, which would be expected to inhabit the
most massive galaxies, should be clustered more strongly that faint
QSOs.  Croom et al. (2002) investigated this in the first data release
of the 2QZ (Croom et al. 2001b), and found some weak evidence for
QSOs with brighter {\it apparent magnitudes} (approximately equivalent to
luminosity relative to $L\qso^*$) being more strongly clustered.
A range of physical affects could act to cancel any correlation of
clustering with luminosity.  For example, a broad range of accretion
efficiencies.

It is possible to examine the luminosity dependence of QSO clustering
in a number of ways.  Ideally, we would split the sample up into a
number of redshift and luminosity bins
and try to separate the luminosity and redshift dependencies.  This is
hard simply due to the low number density of QSOs, particular in the
most luminous intervals.  In the analysis below we follow Croom et
al. (2002) and measure the clustering of QSOs as a function of
apparent magnitude.  This has a number of advantages, as it allows us
to split the QSOs up into only a small number of sub-samples.
Apparent magnitude is also approximately equivalent to a magnitude
relative to $L\qso^*$ over the redshift range we are considering,
due to the strong evolution of the QSO LF.  This means that in a given
apparent magnitude interval, QSOs will have approximately the same
space density at every epoch.

\begin{figure}
\centering
\centerline{\psfig{file=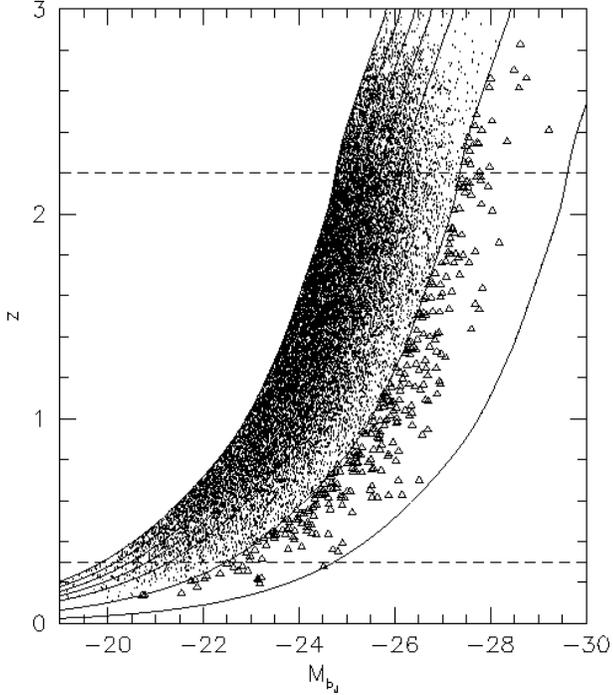,width=8cm}}
\caption{The redshift-absolute magnitude distribution of 2QZ (small
  points) and 6QZ (triangles) QSOs used in our analysis.  The solid
  lines denote the apparent magnitude limits applied to the data, while
  the dashed lines show the redshift range used.  A WMAP/2dF cosmology
  is assumed.}
\label{fig:mbz}
\end{figure}

\subsection{QSO clustering as a function of $\bj$}

We split the 2QZ QSOs into five sub-samples, on the basis of their
apparent magnitude, $\bj$.  These intervals are listed in Table
\ref{tab:fitparmag}.  To enhance the dynamic range of this analysis we
also include QSOs from the 6dF QSO Redshift Survey (6QZ; Paper XII).
This data set contains 275 QSOs at $0.3<z<2.2$ in the magnitude range
$16.0<\bj<18.25$ selected from the same photometric data as the 2QZ.
It forms a bright extension to the 2QZ, in the SGP region only (see
Paper XII).  All the QSOs in the 6QZ form a sixth magnitude interval.
The distribution of QSOs in the $z-\mb$ plane is shown in
Fig. \ref{fig:mbz}.  Even with the large sample presented here, the
steep bright-end slope of the QSO luminosity function means that we can
only cover an effective dynamic range of $\simeq3$ mag in apparent
magnitude (or a factor of $\simeq16$ in luminosity).  There is also
only a relatively small dynamic range in QSO space density, from a
mean $4.5\times10^{-6}h^3{\rm Mpc}^{-3}{\rm mag}^{-1}$ at the faintest
magnitudes to $9.2\times10^{-7}h^3{\rm Mpc}^{-3}{\rm mag}^{-1}$.  The
greatest luminosity dependence might be expected for the brightest
QSOs, as these are the rarest sources.  This is exactly the point at
which the rarity of QSOs makes clustering measurements most
difficult.  One solution to this problem is to {\it cross-correlate}
QSOs of a given luminosity with QSOs at all other luminosities.  This
approach will be discussed by Loaring et al. (in preparation).

\begin{figure*}
\centering
\centerline{\psfig{file=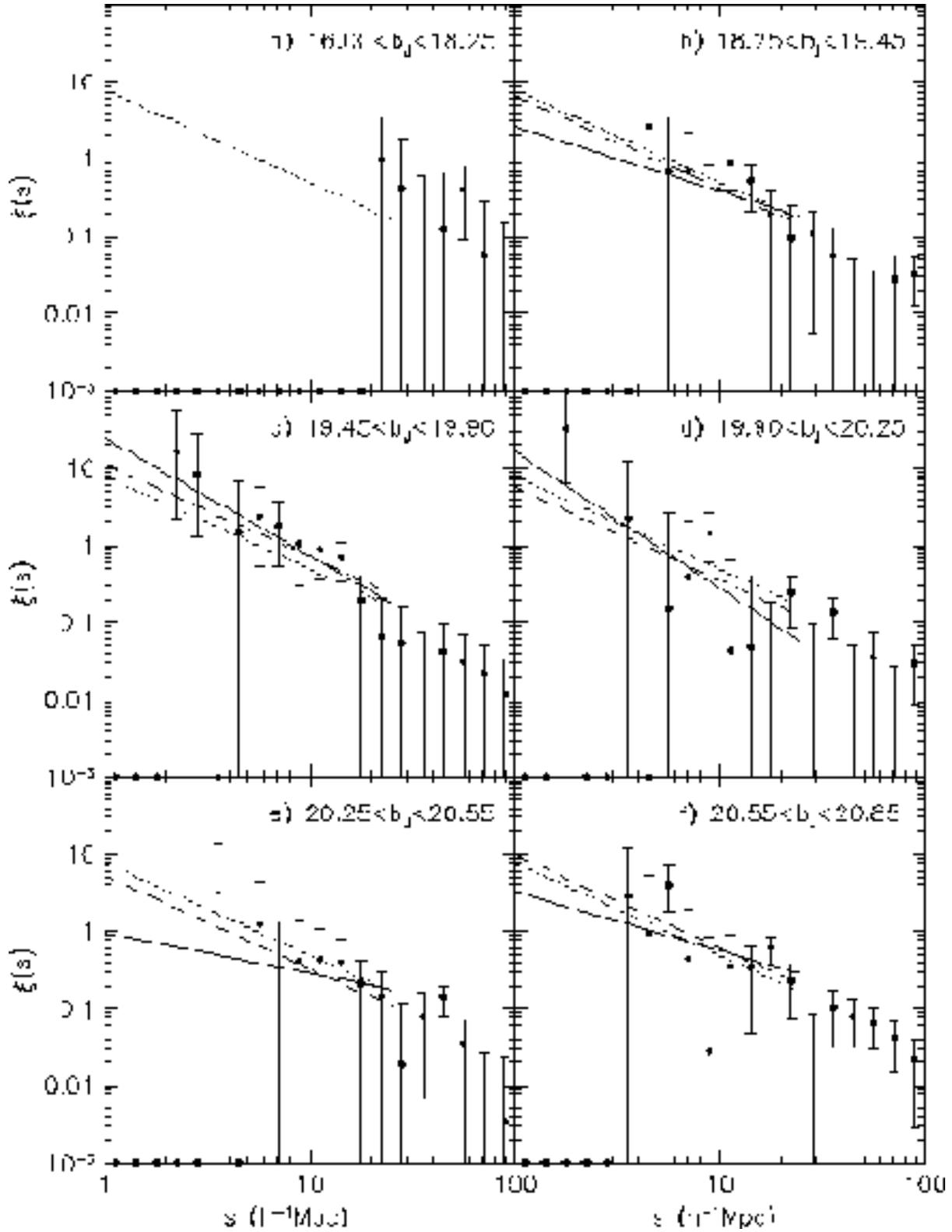,width=16cm}}
\caption{The QSO $\xi(s)$ from the 2QZ/6QZ (filled points) as a function 
  of apparent $\bj$ magnitude in 6 intervals from bright (a) to faint
  (f) magnitudes.  The best fit power law is shown in each case (solid
  line) as is the best fit power law for the full sample for
  comparison (dotted line).  We also show the best fit power law when
  fixing $\gamma$ to a value of 1.20 (dashed lines). No power law fit
  is attempted for the 6QZ data (a).  A WMAP/2dF cosmology is assumed.} 
\label{fig:xibjbins}
\end{figure*}

The measured $\bj$ dependent $\xi(s)$ are shown in
Fig. \ref{fig:xibjbins}.  At bright magnitudes
(Fig. \ref{fig:xibjbins}a) the small number and low space density of
QSOs  means that no significant signal is detected.  At fainter
magnitudes the data appear reasonably consistent with the best fit
power law for the full sample (dotted lines).  We also fit power laws
to each $\bj$ interval, showing the results as the solid lines in
Fig. \ref{fig:xibjbins}.  The values are also listed in Table
\ref{tab:fitparmag}.  The best fit parameters vary considerably, but
have large errors.  Neither the slopes or amplitudes are particularly
well constrained.  If instead we fix $\gamma=1.2$ as found above, we
find values of $s_0$ that are much closer to the mean (dashed lines in
Fig. \ref{fig:xibjbins}).  We also note
that the faintest magnitude interval (Fig. \ref{fig:xibjbins}f) shows
more structure on large scales than the other samples.  It is possible
that this is the result of increased incompleteness at the faint
limit of the sample, even though we have taken care to correct for
magnitude dependent spectroscopic completeness, as described in Paper
XII.  Estimation of $\xi(s)$ using the RA-Dec and RA-Dec-z mixing
methods described above cause some reduction in this excess at large
scales but does not completely remove it.  This suggests that some,
but not all, of this excess power could be due to residual
incompleteness affects.  Bearing this in mind we have checked whether
any of our results above are affected by removing QSOs in the faintest
bin from our sample and confirm that they have no significant impact
on our conclusions.

\begin{figure}
\centering
\centerline{\psfig{file=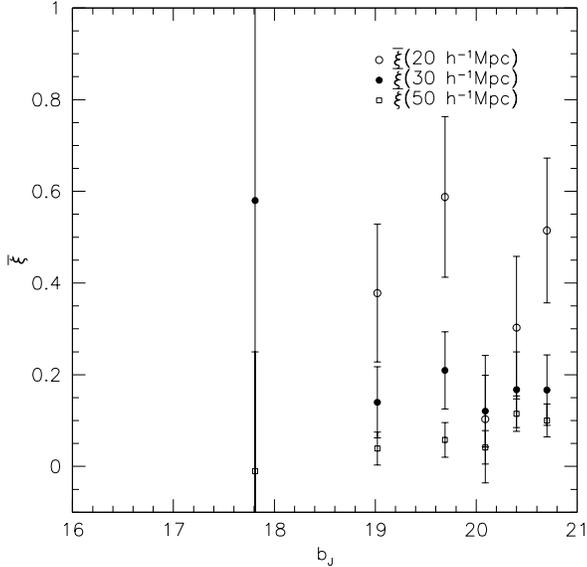,width=8cm}}
\caption{The dependence of $\xibar(s)$ on $\bj$ for three different
  values of $s=20$, 30 and $50\Mpc$ (open circles, filled circles and
  open squares respectively).  We do not plot a point at $s=20$, for
  the brightest bin, as there are no QSO pairs found.}  
\label{fig:xibarbj}
\end{figure}

In order to use a robust measure of any luminosity dependence we
calculate $\xibar(s)$ in each of the $\bj$ intervals (Table
\ref{tab:fitparmag}), which is plotted in Fig. \ref{fig:xibarbj}.  We
confirm that the estimates of $\xibar$ are not significantly changed
by using the RA-Dec mixing method to measure $\xi(s)$.  We find that
there is no significant evidence for any dependence of
clustering amplitude with $\bj$ (or equivalently luminosity relative
to $L\qso^*$).  However, given the relatively large errors found
[$\sim30$ per cent in $\xibar(20)$] this result does not rule out
models for which QSO clustering should be dependent on luminosity.  As
pointed out above, the mean space density of our brightest and
faintest samples only differs by a factor of $\sim5$.  If this
decrease in space density was solely due to higher mass (and therefore
rarer) halos acting as hosts then this would correspond to a factor of
$\sim2$ increase in $\mdh$, but only a $\sim15$ per cent increase in
bias (or $\sim30$ per cent in clustering amplitude) which is
approximately at the level of our measurement errors.  This suggests
that the increase in sensitivity provided by cross-correlating
different QSO samples may provide useful constraints on QSO models
(see Loaring et al. in preparation).

\section{conclusions}\label{section_conclusions}

We have performed a detailed analysis of the clustering of 2QZ QSOs in
redshift space as described by the two-point correlation function.
Here we now discuss our conclusions.

The QSO two-point correlation function, $\xi(s)$,
averaged over the redshift range $0.3<z<2.2$, shows a slope which
changes as a function of scale, being flatter on small scales and
steeper on large scales.  A power law is an acceptable fit on scales
less than $25\Mpc$ in a WMAP/2dF cosmology, the best fit parameters
are $s_0=5.48_{-0.48}^{+0.42}\Mpc$ and $\gamma=1.20_{-0.10}^{+0.10}$.
We demonstrate that QSO
clustering on scales $<10\Mpc$ is strongly affected by non-linear
$z$-space distortions, caused by redshift errors, shifts in QSO
broad emission lines and intrinsic peculiar velocities, which all
contribute similar amounts to the total velocity dispersion, of
$\wrms\simeq690\kms$.  A power law model which has been corrected for
both linear and non-linear $z$-space distortions is shown to be a good
description of the shape of $\xi(s)$.  Here we note that in modelling
non-linear
$z$-space distortions at high redshift it is important to include an
extra factor of $1+z$ in Eq. \ref{eq:xisigmapi} relative to the version
normally used.  

On large scales power law clustering is not an appropriate model
and we therefore compare the 2QZ $\xi(s)$ to a model CDM $\xi(s)$ in a
WMAP/2dF cosmology ($\Omm=0.27$, $\Omlam=0.73$, $\sigma_8=0.84$) 
accounting for the affects of non-linear clustering on small scales
and the affects of $z$-space distortions.  This model is well matched
to the data after allowing for a linear bias of $b\qso=2.02\pm0.07$ at
the mean redshift of the sample ($\bar{z}=1.35$).  The 2QZ $\xi(s)$
also agrees remarkably well with that measured from the low redshift
galaxies in the 2dFGRS \cite{hawk03}, in both shape and amplitude.
While the match in shape is unsurprising given that the physics (at
least on large scales) prescribing the shape should be identical, the
match in amplitude is impressive.  Given that 2dFGRS galaxies are
unbiased tracers of the mass distribution at low redshift
\cite{verdi02}, it appears that the bias of QSOs exactly cancels out
the growth of density fluctuations, to give a measured clustering
equivalent to an unbiased population at low redshift. As we find
evidence for evolution of QSO clustering in a WMAP/2dF cosmology, this
agreement must be something of a coincidence.  Also, in an EdS
Universe the 2QZ $\xi(s)$ is a factor $\sim2$ below the observed
2dFGRS clustering.

To further investigate these issues, we determine the clustering of 2QZ
QSOs as a function of redshift.  In a WMAP/2dF cosmology we find a
significant (at the 98 per cent level) correlation of clustering
amplitude with redshift as measured by the integrated correlation
function within $20\Mpc$, $\xibar(20)$.  Clustering increases with
redshift and we find $\xibar(20)=0.263\pm0.075$ at $z=0.53$, and
$\xibar(20)=0.701\pm0.174$ at $z=2.48$.  In an EdS cosmology we find
no evidence for evolution.  By assuming an underlying WMAP/2dF
cosmology we are able to directly determine the bias of QSOs, which we
find to be a strong function of redshift.  Even if there were no
evolution in the measured $\xibar(20)$ with redshift, this would still
imply a strongly evolving QSO bias.  At low redshift, 2QZ QSOs
appear largely unbiased, with $b\qso(z=0.53)=1.13\pm0.18$, while at high
redshift we find $b\qso(z=2.48)=4.24\pm 0.53$.  A complication is that
as the 2QZ is a flux limited sample, we are sampling more luminous
QSOs at high redshift.  However, the strong evolution of the QSO
population means that to good approximation we are sampling the QSO
population at the same space density at each redshift, and at the same
point relative to the evolving break in the luminosity function,
$L\qso^*$.  It thus appears that $L\qso^*$ QSOs at low redshift should
be largely unbiased, and clustered similarly to low redshift
galaxies.  This has indeed been seen by Croom et al. (2004) who
cross-correlate low redshift (and therefore low luminosity) 2QZ QSOs
with 2dFGRS galaxies and find no difference in the clustering
properties of the two populations [see also Wake et al. (2004)].  

By using the theoretical relation between $\mdh$ and bias derived
by Sheth et al. (2001) and others, it is possible for us to take the
measured bias values for 2QZ QSOs and calculate the typical masses of
their hosts' DMHs.  We find that the mass of DMHs hosting 2QZ
QSOs is approximately constant with redshift, with a mean
$\mdh=(3.0\pm1.6)\times10^{12}h^{-1}\msun$.  The fact that the hosts
of 2QZ QSOs have the same mass at all redshifts demonstrates that they
cannot be cosmologically long lived, as DMHs tend to grow and
accumulate mass over time.  Based on the formalism of Lacey \& Cole
(1993) we predict that DMHs hosting QSOs at $z\sim2.5$ would typically
have merged into DMHs of mass $\simeq6\times10^{14}h^{-1}\msun$ by the
present, and therefore exist in rich galaxy clusters (although they
would not generally be active at low redshift).  In contrast, the descendents of
lower redshift 2QZ QSOs would not have had time to form more massive
halos, and should exist in either massive galaxies or groups.  By
extrapolation it is suggested that at $z=0$, $\sim L\qso^*$ QSOs should
also sit in $\mdh=(3.0\pm1.6)\times10^{12}h^{-1}\msun$ halos, which is
very close to the characteristic mass of the Press-Schechter mass
function, $M^*(z=0)\simeq3.5\times10^{12}h^{-1}\msun$.  The predicted
growth of DMH
mass by accretion/merging allows us to place upper limits on the
lifetime of the QSO population.  Low redshift QSOs cannot be the same
population of objects as at higher redshift if they have masses which
are less then the mass of the high redshift sources, after accounting
for their expected growth over time.  Therefore calculating the time
taken to reach the mean QSO host DMH mass plus twice the measured rms
gives a $\sim2\sigma$ limit on the lifetime of QSO activity.  We find
this limit to be $t\qso<6\times10^8$ years at $z=2.48$, but weaker
at low redshift ($3\times10^9$ years at $z=0.53$).  We note that this
limit is not based on the measured number density of QSOs compared to
a Press-Schechter mass function (as many other estimates are), but is
only constrained by the clustering evolution of QSOs.  Various authors
have provided more detailed models in order to constrain the lifetime
of QSO activity.  When applied to our data, the model of Martini \&
Weinberg (2001) suggests that $z\sim2$ QSOs will have lifetimes
$t\qso\simeq4-50\times10^6$ years.  If there is scatter in the
relation between $\mdh$ and luminosity, then this is an effective
lower limit on QSO lifetimes.  The e-folding time for the evolution of
$L^*\qso$ is $\sim2\times10^9$ years (Paper XII), much less than the
ages determined from the Martini \& Weinberg model, and significantly
less than our clustering evolution upper limits at high redshift.

As a next step we determine the central BH mass, $\mbh$, of 2QZ QSOs
based on their estimated $\mdh$.  For this we use the relations
suggested by Ferrarese (2002) to estimate $\mbh$ for
different assumptions concerning the density profiles of the DMHs, and
the evolution of the correlation \cite{wl04}.  A model in which the
correlation between $\mbh$ and $\mdh$ is unchanging with redshift
predicts that BH masses should be slightly increasing with redshift,
with $\Delta\log(\mbh)\simeq1.3\pm1.1$ from the lowest to highest
redshift.  The derived BH masses are in the range $1-20\times10^7\msun$
for NFW profiles, or $0.9-20\times10^8\msun$ for S02 profiles.  The
Eddington ratio, 
$L/L\edd$,  is seen to be approximately constant as a function of
redshift when the $\mbh-\mdh$ relation is independent of redshift.
This is found to be significantly greater than 1 if isothermal DMHs
are assumed, and approximately 1 for the NFW profile, while the S02
profile gives $L/L\edd\sim0.1$.  A model in which it is the
$\mbh-\sigma_{\rm c}$ relation which is invariant with redshift gives
a much stronger evolution of $\mbh$ as DMHs of a given mass have a higher
central velocity dispersion when formed at higher redshift.  Thus the change
in $\mbh$ from low to high redshift is more significant with
$\Delta\log(\mbh)\simeq2.1\pm1.1$, and BHs of order $\sim10^{10}\msun$
being predicted at high redshift.  This increase in estimated $\mbh$ is
greater than (although not significantly) the factor $\sim30$ increase
in mean luminosity from our lowest to highest redshift interval.  As a
result there is a small (factor of a few) decline in $L/L\edd$ with
increasing redshift, although again this is not significant.  As the BH
masses predicted are higher, the accretion efficiencies are lower, in
the range $L/L\edd\sim0.01-1$ depending on DMH profile assumed.

The above suggests that any model of BH formation in which
super-massive BHs form at least as efficiently at high redshift as
they do at low redshift, will tend to have $L/L\edd$ constant or
decreasing with redshift.  This implies that it cannot be a reduction in
efficiency which is driving the fading of the QSO population to low
redshift.  Instead active BHs at high redshift
are more massive that those at low redshift, and it is this reduction
in the BH mass that causes the population of bright QSOs to disappear
in the local universe.  Because super-massive BHs cannot be destroyed,
the massive BHs active at high redshift must be largely inactive at
low redshift, otherwise we would find that low redshift QSOs would
show lower accretion efficiency, and be located in more massive DMHs.
This argument also implies that at any given redshift, the QSO
population must be dominated by objects which are active for the first
time.  Hence it is likely that each QSO passes through only one bright
active epoch (possible at the point of BH formation), although at low
redshift massive BHs may accrete at levels well below $L\edd$ without
contributing significantly to the total luminosity of the population
[see also the discussions in Corbett et al. (2003) and Croom et
al. (2004b)].

The above is valid at redshift below $z\sim2.5$, which is
approximately the point at which the space density of luminous QSOs
peaks.  Clustering measurements of QSOs at $z>2.5$ would help us to 
understand the build up of QSOs at this epoch.  However the low
surface density of $z>2.5$ QSOs currently makes any accurate clustering
measurements difficult or impossible.  The increasing number of high
redshift QSOs from the SDSS survey \cite{fan01} may remedy this
situation.

Finally, we examine our sample to look for any indication of luminosity
dependence in the clustering of 2QZ QSOs, by measuring $\xi(s)$ as a
function of apparent magnitude.  This shows no indication of any
luminosity dependence that might be expected if more luminous QSOs
inhabited more massive DMHs, but the errors are large enough that we
would not be able to detect reasonable amounts of luminosity
dependence.  More detailed investigation of this problem will be
presented by Loaring et al. (in preparation).

\section*{acknowledgments}

We warmly thank all the present and former staff of the
Anglo-Australian Observatory for their work in building and operating
both the 2dF and 6dF facilities.  The 2QZ and 6QZ are based on
observations made with the Anglo-Australian Telescope and the UK
Schmidt Telescope.  We thank Rob Sharp for useful comments.  J. da
\^{A}ngela acknowledges financial support from FCT/Portugal through
project POCTI/FNU/43753/2001 and also ESO/FNU/43753/2001.

\end{document}